\documentclass[aps,prb,reprint,groupedaddress,showpacs]{revtex4-2}
\usepackage{graphicx,graphics}
\usepackage{epstopdf}
\usepackage{dcolumn}
\usepackage{amsmath,amssymb,amsfonts}
\usepackage{latexsym,verbatim}
\usepackage{bm}
\usepackage{bbold}
\usepackage{xcolor}
\usepackage{ulem}
\usepackage[breaklinks=false,colorlinks,citecolor=blue,linkcolor=blue,urlcolor=blue]{hyperref}
\usepackage[abs]{overpic}
\usepackage{textcomp}
\usepackage{mathtools}
\usepackage{braket}
\usepackage{soul}

\begin{document}

\title{Quantum estimation and remote charge sensing with a hole-spin qubit in silicon}

\date{\today}

\author{Gaia Forghieri$^{1,2}$}
\author{Andrea Secchi$^2$}
\author{Andrea Bertoni$^2$}
\author{Paolo Bordone$^{1,2}$}
\author{Filippo Troiani$^2$}
\affiliation{$^1$Universit\`a di Modena e Reggio Emilia, I-41125 Modena, Italy\\ $^2$Centro S3, CNR-Istituto di Nanoscienze, I-41125 Modena, Italy}

\begin{abstract}
Hole-spin qubits in semiconductors represent a mature platform for quantum technological applications. Here we consider their use as quantum sensors, and specifically for inferring the presence and estimating the distance from the qubit of a remote charge. Different approaches are considered --- based on the use of single or double quantum dots, ground and out-of-equilibrium states, Rabi and Ramsey measurements --- and comparatively analyzed by means of the discrimination probability, and of the classical and quantum Fisher information. Detailed quantitative aspects result from the multiband character of the hole states, which we account for by means of the Luttinger-Kohn Hamiltonian. Furthermore, general conclusions can be drawn on the relative efficiency of the above options, and analytical expressions are derived for the Fisher information of a generic qubit within the Rabi and Ramsey schemes.
\end{abstract}

\date{\today}

\maketitle

\section{Introduction}

Spin qubits in silicon quantum dots are the subject of an intense research activity \cite{zwerver2022, Burkard23a, saraiva2022, chatterjee2021}. In recent years, high fidelities have been achieved in all fundamental primitive operations, including single-  \cite{noiri2022, yoneda2018, fernandez2022} and two-qubit gates \cite{xue2022, lawrie2021}, state preparation \cite{mills2022} and readout \cite{mills2022-2, yang2019, Zheng2019a}. 
The suitability of silicon as a host material for spin qubits \cite{boter2022, philips2022, geyer2021} is related to the long coherence times \cite{kobayashi2021,piot2022,stano2022}, resulting from the natural abundance of nonmagnetic isotopes and from the possibility of further enhancing such fraction through isotopic purification \cite{zwanenburg2013,scappucci2021}. 
% Thus, the hyperfine interaction with the nuclei is suppressed with respect to other materials, such as GaAs \cite{bosco2021}. 
Moreover, the manufacturing processes of silicon devices are well established in the modern industry \cite{zwerver2022}. Thus, dense arrays of spin qubits can be realised \cite{vandersypen2017} and possibly operated at temperatures as high as $4\,$K \cite{camenzind2022,xue2021,petit2020,petit2020-2, vandijk2020}.

\begin{figure}
\centering
\includegraphics[width=0.48\textwidth]{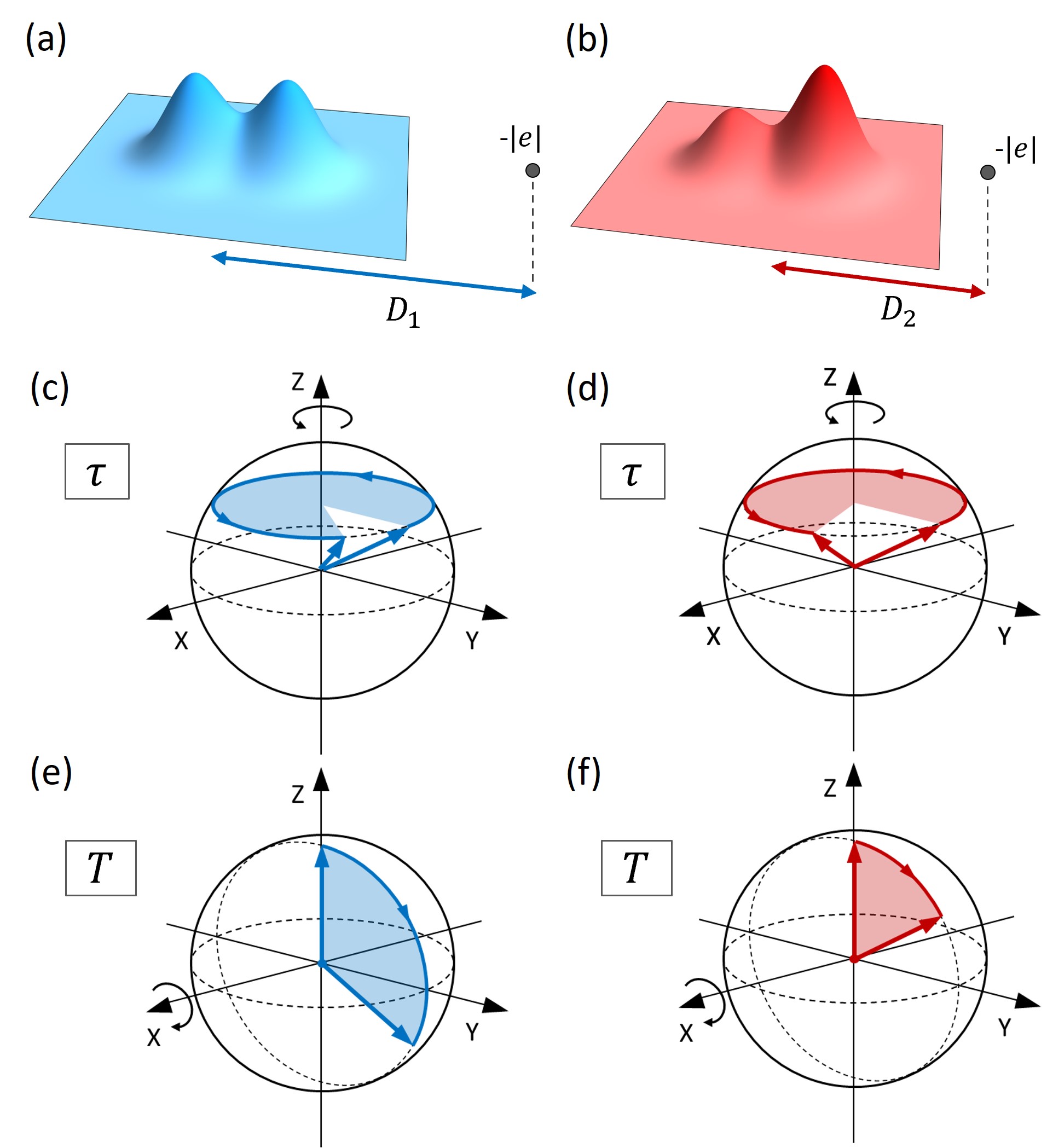}
\caption{Schematic representation of the main effects induced by a charge on a hole-spin qubit in a double quantum dot: (a,b) hole charge density; (c,d) free precession during a time $\tau$ of the qubit state vector around the $Z$ axis of the Bloch sphere; (e,f) rotation around the $X$ axis of the Bloch sphere induced by a pulse of duration $T$ . Panels (a,c,e) and (b,d,f) correspond to a distance $D_1$ and $D_2<D_1$ of the charge from the center of the double dot, respectively. The inference of the charge presence corresponds to the discrimination between two distances, one of which goes to infinity.}
\label{fig:img1}
\end{figure}
Silicon quantum dots can host both electron- and hole-spin based qubits. An appealing feature of the hole-based implementation lies in the strong atomic spin-orbit coupling, which offers the opportunity to perform all-electrical manipulation of the qubit states \cite{liles2021, froning2021, michal2021, bosco2021-2}, without the inclusion of micromagnets in the device \cite{hosseinkhani2022}. Besides, some  properties of the hole-spin qubits --- and specifically their coupling to external fields --- strongly depend on the band mixing, which can in turn be rather sensitive to the electrostatic environment. This makes hole-spin qubits highly susceptible to electric noise \cite{bosco2021-2,yoneda2018} but,  on the other hand, allows for a strong tunability and for a dynamic control of the qubit properties \cite{hsiao2020, vandiepen2018}. Here we explore the possibility of exploiting such properties in the context of remote charge sensing and quantum parameter estimation \cite{dong2022,liu2017,degen2017}. Specifically, we consider the case where one wants to infer the presence of a remote charge at a given (known) distance $D$ from the qubit, or estimate the precise value of such distance. As pictorially shown in Fig.~\ref{fig:img1}, different distances $D_1$ and $D_2$ result in different spatial distributions of the hole ground state [panels (a,b)], and in different values of the Larmor ($\omega_{\rm L}$) and Rabi ($\omega_{\rm R}$) frequencies. The absence of the remote charge can be identified with the case of a charge positioned at an infinite distance from the qubit. Different values of the Rabi and Larmor frequencies, in turn, translate into different rotation angles of the qubit state for given free precession times $\tau$ [panels (c,d)] or manipulation pulses [panels (e,f)]. The inference of the unknown parameter $D$ can thus be based either on the properties of the hole ground state ({\it static approach}), or on the final out-of-equilibrium state $|\psi\rangle$ of the spin qubit resulting from a given pulse sequence ({\it dynamic approach}). 
We assess and compare the quality of the remote charge sensing and of the parameter estimation based on these two approaches by computing the discrimination probability and the (classical and quantum) Fisher information, respectively. In fact, the static approach is essentially meant to provide a general and quantitative characterization of the qubit susceptibility to the external charge, and a benchmark for the dynamic approach, whose implementation is based on the standard readout of the qubit state.

Remote charge sensing has been demonstrated in recent experiments \cite{Duan2020,Patomaki}, where the properties (charge occupation) of a target dot are inferred from the positions of the Coulomb peaks in the conductance of a sensor dot during a transport measurement. Our investigation moves along these lines, but considers two significant elements of novelty. On the one hand, as in the quantum sensing paradigm, the information on the remote charge is encoded in the quantum state of the qubit, rather than on the positions of peaks observed in continuous transport measurements. On the other hand, as in the typical quantum estimation scenario, the quantity to be inferred can be a continuous one, namely the distance of a charge from the qubit.

Inferring the presence of an external charge and estimating its distance from the qubit are related and yet distinct tasks. In fact, the suitable measurement schemes are different in the two cases, and the optimal values of the control parameters need to be identified through different figures of merit. In case both inferences are required, they should thus be carried out sequentially, through different and independently optimized measurement schemes.

The eigenstates of the hole-spin qubit and its dynamics induced by the application of pulsed electric fields result from a complex interplay between the shape of the electrostatic potential, the atomic spin-orbit coupling, the external magnetic field, and the Coulomb interaction with the remote charge. 
Such interplay is fully captured by the six-band $\mathit{k\cdot p}$ approach \cite{Bellentani2021a,Secchi2021a,Secchi2021b,Venitucci18}, which is here used for the calculation of the hole eigenstates. At a qualitative level, the role of the confinement potential emerges from the comparison between single (SQDs) and double quantum dots (DQDs). 
The latter ones are characterized by a wider polarizability of the hole ground state, and thus by a stronger dependence of its properties on the distance of the remote charge. Moreover, the Rabi and Ramsey frequencies of DQDs have a stronger dependence on $D$ than those of SQDs, eventually resulting in a larger discrimination probability and in a more precise parameter estimation. Overall, the dynamic approach leads to higher values of all the considered figures of merit. More specifically, estimation procedures based on the Ramsey measurement with a DQD allows one to achieve values of the Fisher information that are several orders of magnitude larger than those obtained with the other considered approaches. This emerges as a robust conclusion, irrespective of the detailed properties of the quantum dot systems. 

The rest of the paper is organized as follows. In Section II we introduce the multiband model of the hole-spin qubit, along with the Hubbard model that qualitatively reproduces the main features of the DQD implementation. In Section III we discuss the static estimation strategy, based on the dependence of the hole ground state on the charge distance. Section IV is devoted to the dynamic strategy, where the remote charge sensing and the parameter estimation are based on the Rabi and Ramsey measurements. Finally, in Section V we report the conclusions and the outlook.

\section{The model system}

The multiband character of the hole eigenstates plays a crucial role in determining the relevant qubit properties, specifically the Larmor and Rabi frequencies. In fact, these result from the interplay between the atomic spin-orbit coupling, the confining potential, and the external magnetic field. In order to account for these features, we model the hole-spin qubit by means of a six-band Luttinger-Kohn Hamiltonian \cite{Luttinger55} (Subsec.~\ref{subsec:A}), with an external confining potential that models either a SQD or a DQD. In the DQD case, the interplay between interdot tunneling and charge-induced bias can also be described by means of a biased Hubbard model (Subsec.~\ref{subsec: Hubbard}), where the bias between the dots is due to their different distances from the remote charge. This simplified model allows one to obtain analytical expressions that qualitatively reproduce several trends and dependencies of the numerical results on the parameters of the system; such agreement corroborates the generality of our findings.

\subsection{Multiband model of the single and double quantum dots\label{subsec:A}}

The six-band Luttinger-Kohn envelope-function Hamiltonian is appropriate for the description of the low-energy hole states close to the maximum of the valence band in silicon ($\bf \Gamma$ point) \cite{Voon09, Chao92, Secchi2021b}. There, the atomic spin-orbit coupling provides an energy splitting $\Delta_{\rm SO} = 44$ meV between the $j = 3/2$ quartet (heavy and light hole bands) and the $j = 1/2$ doublet (split-off band). The single-hole Hamiltonian defined in such combined position-spin space and with the inclusion of external (electric and magnetic) fields, is diagonalized numerically in order to derive the hole energies and eigenstates \cite{Secchi2021a, Secchi2021b, Bellentani2021a}. 

\paragraph{Single and double quantum dots.}

The hole confinement is induced by a combination of band offset and electrostatic coupling to the metal gates. In the presence of an additional external charge, the total potential acting on the hole is given by: $V({\bf r}) = V_{\rm conf}({\bf r}) + V_{\rm c}({\bf r})$, where $V_{\rm conf}({\bf r})$ is the confinement potential, and $V_{\rm c}({\bf r})$ accounts for the Coulomb interaction with the charge. In the cases of the single (SQD) and double quantum dots (DQD) considered hereafter, $V_{\rm conf}({\bf r})$ coincides with 
\begin{equation}
     V_{\rm SQD}({\bf r}) = \frac{\kappa}{2} \left( x^2 + y^2 \right) + V_{\parallel}(z) \,,
    \label{V SQD}
\end{equation}
and
\begin{equation}
    V_{\rm DQD}({\bf r}) = \frac{\kappa}{2} \left[ \frac{(x^2-a^2)^2}{4a^2} +y^2\right] + V_{\parallel}(z) \,,
    \label{V DQD}
\end{equation}
respectively. 
In both cases, the last contribution accounts for the confinement in the vertical direction and reads
\begin{align}
V_{\parallel}(z) = \left\{ \begin{matrix} - E_{\parallel} z & |z| < L_z / 2 \\ V_{\rm bo} & |z| > L_z / 2 \end{matrix} \right.\,, 
\label{V par}
\end{align}
where $L_z$ is the channel width, $V_{\rm bo}$ is the band offset between the well and the barriers along the $z$ direction (here $V_{\rm bo} = 4000\,$meV and $L_z = 5\,$nm), and $E_{\parallel}$ is an electric field applied along the $z$ direction \cite{Bellentani2021a} ($E_{\parallel}$ is set to zero in Sec. \ref{section:SA} and to $-50\,$mV/nm in Sec. \ref{sec:DA}). 
The strength of the in-plane confinement is determined by $\kappa$ and, close to the minima of $V_{\rm DQD}({\bf r})$, it approaches that of the SQD, both in the $x$ and $y$ directions (here $\kappa = 7.6566 \times 10^{-2}\,$meV$/$nm$^2$). The interdot distance is identified with the distance $2a$ between the minima of $V_{\rm DQD}({\bf r})$. 

The Coulomb interaction between the confined hole and the external charge (assumed to be equal to $- |e|$) is
\begin{equation}\label{eq:vc}
    V_{\rm c}({\bf r}) = -  \frac{e^2}{\varepsilon \sqrt{(x-x_{\rm c})^2+(y-y_{\rm c})^2+(z-z_{\rm c})^2}} \,,
\end{equation}
where $\varepsilon = 11.68$ is the dielectric constant of bulk silicon \cite{Dunlap53}. In the following, we specifically consider the case where the charge is positioned along the $x$ axis, ${\bf r}_{\rm c} = (D,0,0)$, in order to highlight the interplay between the interdot tunneling and the Coulomb interaction. Different values of $\varepsilon$ will also be considered, in order to effectively account for screening effects, resulting from the presence of metal gates within different gate geometries \cite{Hogg23a}.

Finally, the coupling between the hole and the static magnetic field is accounted for by the Zeeman-Bloch Hamiltonian and the substitution ${\bf k} \rightarrow {\bf k} + (e / \hbar) {\bf A}$ in the $\mathit{k\cdot p}$ Hamiltonian, with ${\bf A}$ being the vector potential associated to the magnetic field \cite{Venitucci18, Bellentani2021a}.

\paragraph{Larmor and Rabi frequencies.}
In order to realize a hole-spin qubit, it is crucial to split the Kramers degeneracy in the ground doublet by means of a static magnetic field ${\bf B}=B\,(\sin\theta\,\cos\phi,\sin\theta\,\sin\phi,\cos\theta)$. 
The resulting energy splitting between the lowest energy eigenvalues $e_1$ and $e_2$ defines the Larmor angular frequency
\begin{equation}
    \omega_{\rm L} \equiv \frac{1}{\hbar} (e_2-e_1)   \,, 
    \label{omega L}
\end{equation}
whose value depends on the field intensity ($B$) and orientation ($\theta$ and $\phi$), on the detailed composition of the hole eigenstates, and (therefore) on the parameters that define the confinement potential.

Hole-spin qubits can be manipulated electrically, by applying voltage pulses to a nearby metal gate. Here, we assume that such voltage induces a time-dependent potential perturbation $ \delta V({\bf r},t) = - \delta E_0 \,\cos (\omega t)\, z $, corresponding to an oscillating, homogeneous electric field (here $\delta E_0 = 3\,$meV$/$nm). The resulting Rabi angular frequency is given by:
\begin{equation}
    \omega_{\rm R} \equiv  \frac{1}{\hbar} \,\delta E_0 \,  \left|\langle\psi_1|  \hat{z} |\psi_2\rangle \right|  \,,
    \label{omega R}
\end{equation}
where $|\psi_1\rangle$ and $|\psi_2\rangle$ are the hole eigenstates belonging to the ground doublet.

Both $\omega_{\rm L}$ and $\omega_{\rm R}$ are affected by the Coulomb interaction between the hole and the external charge [Eq. \eqref{eq:vc}], and depend on the distance $D$ between this and the center of the (double) quantum dot. The figures of merit that determine the accuracy in the estimation of $D$ (Section \ref{sec:DA}) are functions of both these angular frequencies and their derivatives with respect to $D$, $\omega_{\rm L}'\equiv\partial_D \omega_{\rm L}$ and $\omega_{\rm R}'\equiv\partial_D \omega_{\rm R}$. These are here calculated through the extended Hellmann-Feynman theorem, which allows to avoid the finite-differences approach and the related numerical noise (see Appendix \ref{app:QFI} for further details). 

\subsection{Hubbard model of the double quantum dot}
\label{subsec: Hubbard}

The coupling of the hole to the external (electric and magnetic) fields, and the resulting accuracy in the parameter estimation achieved through the dynamic approach (Sec.~\ref{sec:DA}) depends on the detailed composition of the hole eigenstates, and thus on the specific geometry of the host device and on the resulting shape of the confining potential. The properties of the hole ground state that determine the results obtained within the static approach (Sec.~\ref{section:SA}) are less dependent on such details. It is therefore possible to identify general trends and properties, which can be effectively described in terms of a simpler model. In fact, as shown below, the simplest model of a DQD system, namely, the biased two-site single-band Hubbard Hamiltonian, qualitatively accounts for several features of the more realistic multiband system, including the dependence on $D$ of the quantum Fisher information.

The biased DQD Hubbard Hamiltonian, for a single hole, reads:
\begin{equation}
    \mathcal{\hat H}^{\rm Hub} \!=\! \sum_{\alpha = \Uparrow, \Downarrow} \left( \epsilon_1 \hat{n}_{1, \alpha} \!+\! \epsilon_2 \hat{n}_{2, \alpha} \!-\! t  \hat{c}_{1, \alpha}^\dagger \hat{c}_{2, \alpha}  \!-\! t^* \hat{c}_{2, \alpha}^\dagger \hat{c}_{1, \alpha}  \right) .
    \label{Hubbard 1 particle}
\end{equation}
Here, $\alpha \in \lbrace \Uparrow, \Downarrow \rbrace$ is a pseudospin quantum number, distinguishing time-reversal conjugated single-particle states \cite{Secchi2021a}; $\hat{c}^{\dagger}_{i, \alpha}$ creates a hole in a state localized in dot $i$ with pseudospin $\alpha$, and $\hat{n}_{i,\alpha} = \hat{c}^{\dagger}_{i, \alpha} \hat{c}_{i, \alpha}$; $t$ is the inter-dot tunneling parameter, and $\epsilon_i$ is the onsite energy for site $i$. The eigenenergies are independent of $\alpha$, and read 
\begin{align}
e^{\rm Hub}_{\pm} = \frac{1}{2} \left[ \epsilon_1 + \epsilon_2 \pm \sqrt{(\epsilon_1 - \epsilon_2)^2 + 4 |t|^2 } \right] \,.
\label{Hubbard energies}
\end{align}

The parameters introduced in Eq.~\eqref{Hubbard 1 particle} can be related to the energies characterizing the multiband system. In particular, the two dots are assumed to be symmetric, so that the difference between the onsite energies $\epsilon_i$ is entirely due to the presence of the external charge. As a simple approximation, we then take $\epsilon_i$ to coincide with $V_{\rm c}({\bf r}_i)$, where ${\bf r}_{1/2} = (\pm a,0,0)$ correspond to the minima of $V_{\rm DQD}({\bf r})$:
\begin{equation}\label{eq:bias}
    \epsilon_1 = V_c ({\bf r}_1) = -\frac{e^2}{\varepsilon (D+a)}, \,\, \epsilon_2 = V_c ({\bf r}_2) = -\frac{e^2}{\varepsilon (D-a)} \,.
\end{equation} 
The bias $\delta_{\rm c}$ between the two dots, induced by the charge, is then given by
\begin{equation}
    \delta_{\rm c} = \epsilon_1 - \epsilon_2 = \frac{2ae^2}{\varepsilon(D^2-a^2)}.
\end{equation}

As to the energy separation $\delta_{3 , 1 } \equiv e_3 - e_1$ between the ground and first excited doublets, obtained from the multiband calculations at zero magnetic field and for $V_{\rm c} = 0$, this is identified with $2 |t|$, corresponding to the energy gap $e^{\rm Hub}_{+}-e^{\rm Hub}_{-}$ of the Hubbard model in the absence of external charges. As discussed in Subsec.~\ref{subsec:QFI} and Appendix \ref{app: tunnelfit}, the hopping parameter in the DQD depends on $a$ according to $|t| = t_0 {\rm e}^{- \gamma a^2}$ to a very good approximation.

\section{Static approach to the estimate of the charge distance\label{section:SA}}

The presence of a charge in the vicinity of the DQD affects the eigenstates of the system. In particular, the dependence of the hole ground state on the remote charge can be used to infer its presence or estimate its distance [Fig.~\ref{fig:img1} (a,b)]. The maximal probability of discriminating between the presence and the absence of the remote charge (at a given position) is quantified through the Helstrom bound (Subsec. \ref{subsecB}). The highest achievable precision in the estimate of the distance is quantified by means of the quantum Fisher information (Subsec. \ref{subsec:QFI}). The physical origin of its maxima, corresponding to optimal working points, is investigated through the dependence on the charge position of the hole density distribution and of the energy gap between the ground and first excited doublets (Subsec. \ref{subsecA}). 

\begin{figure}
\centering
\includegraphics[width=0.48 \textwidth]{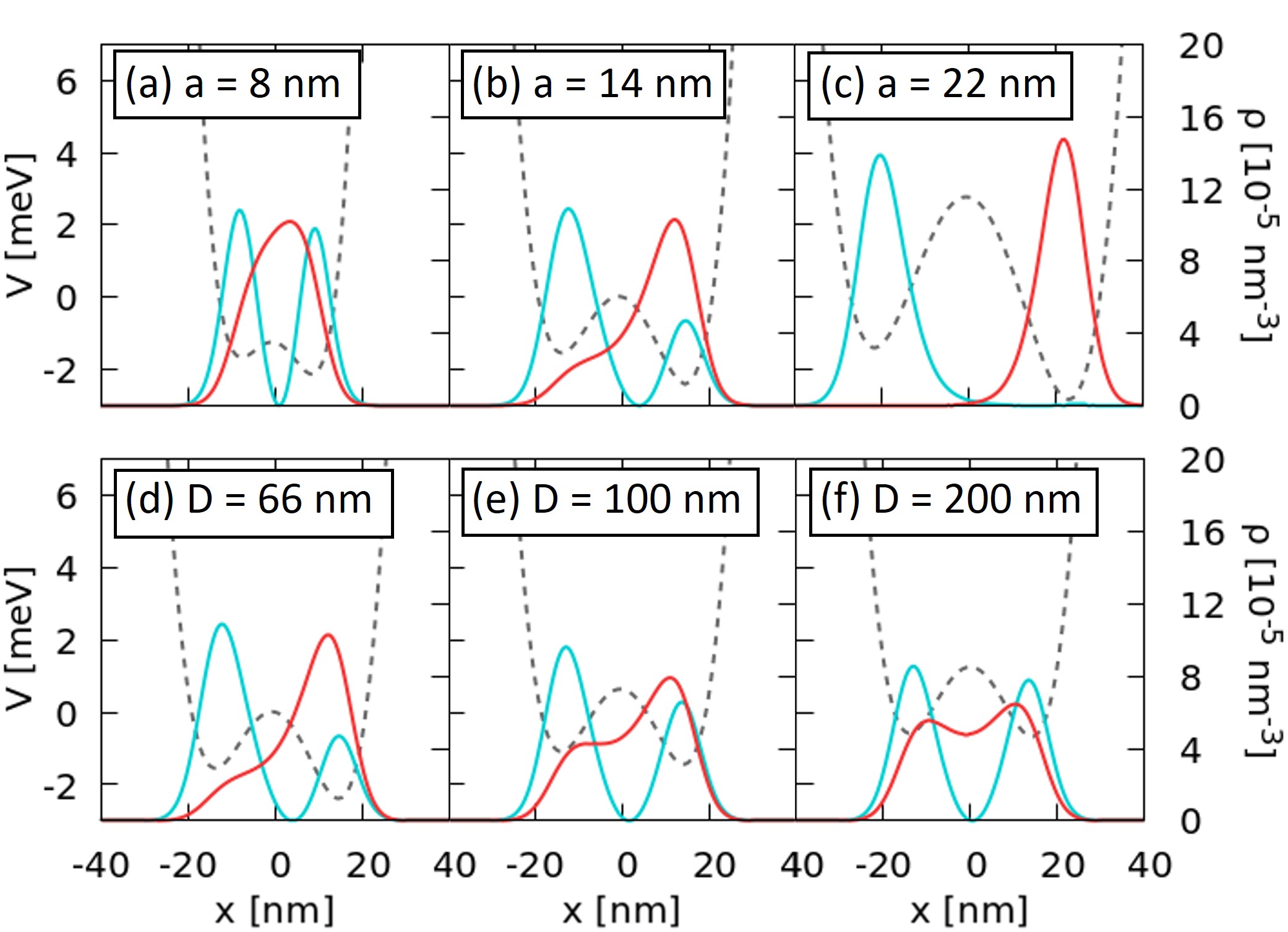}
\caption{Charge densities (solid lines) for the ground (red) and first-excited (blue) hole states of a silicon DQD in the presence of an external charge. The grey dashed curves are the profiles of the confining potential for $y = z = 0$. (a,b,c) The calculations are performed for $D = 66\,$nm and for different values of the half interdot distance $a$. (d,e,f) The calculations are performed for $a = 14\,$nm and for different values of the charge distance $D$. }
\label{Fig:2}
\end{figure}

\subsection{Charge distribution and energy eigenvalues\label{subsecA}}
If the confinement potential $V_{\rm DQD}({\bf r})$ is symmetric with respect to the plane $x=0$, the same applies to the hole density of the eigenstates. This symmetry is broken by the presence of the charge positioned away from the $yz$ plane (Fig.~\ref{Fig:2}). In particular, a negative charge at ${\bf r}_{\rm c} =(D,0,0)$ (with $D>0$) tends to localize the hole density $n_k({\bf r})=|\langle {\bf r}|\psi_k\rangle|^2$ for the ground ($k=1,2$, red curves) and first excited doublets ($k=3,4$, cyan curves) respectively in the right ($x>0$) and left ($x<0$) dots.
The degree of localization results from the competition between the Coulomb interaction with the charge, which induces a bias between the two dots and thus tends to localize the hole, and the interdot tunneling, which tends to delocalize the eigenstates over the DQD. 
A transition between a delocalized and a localized-hole regime is observed for increasing values of $a$ at a given charge distance $D$ [panels (a-c)], and for decreasing values of $D$ at a given half interdot distance $a$ [panels (d-f)].

\begin{figure}
\centering
\includegraphics[width=0.45\textwidth]{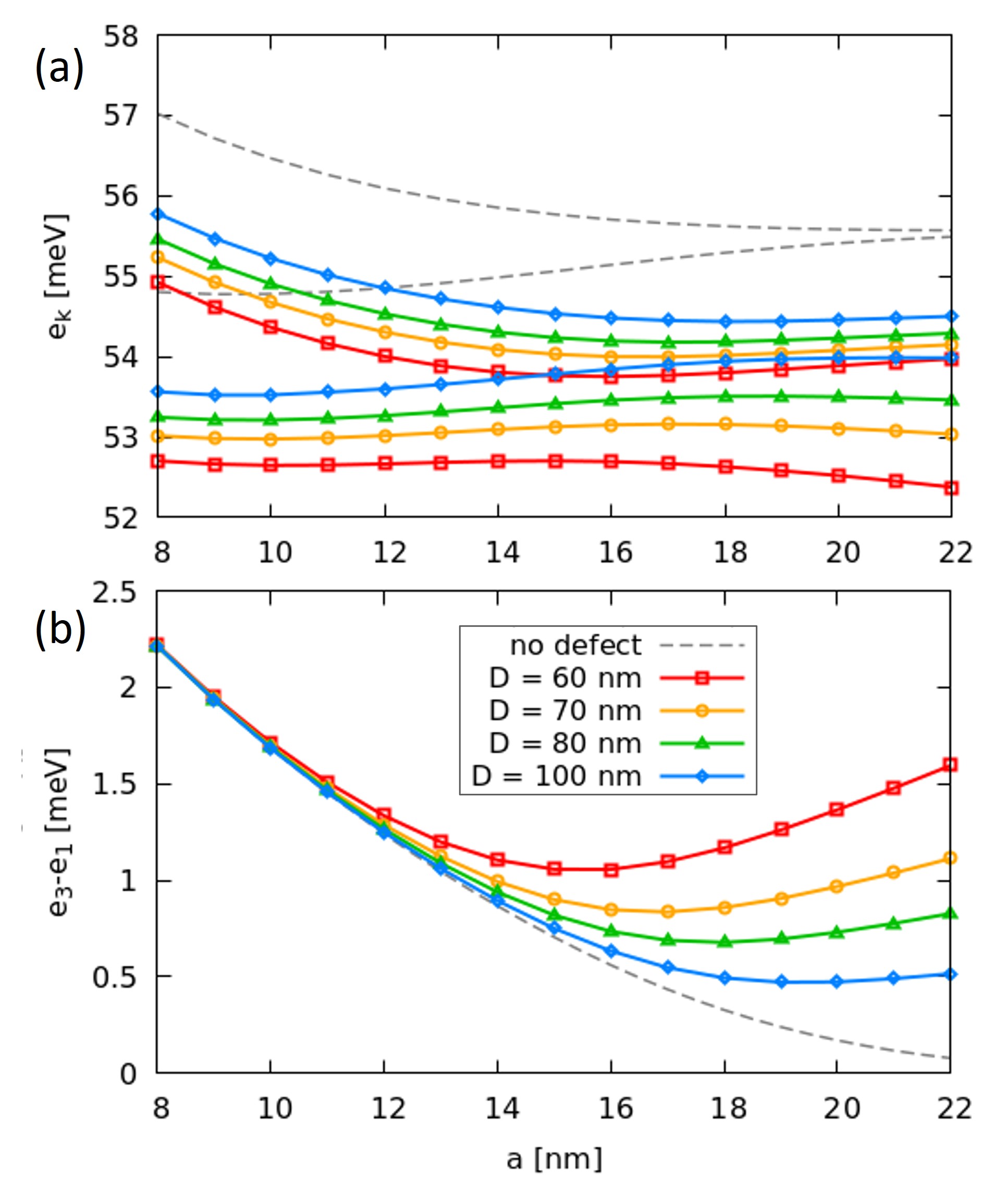}
\caption{(a) Energies $e_1=e_2$ and $e_3= e_4$ of the hole ground and first-excited doublets in a DQD as a function of the half interdot distance $a$ (and for $B=0$), in the absence (dashed line) and in the presence of the external charge, for $D$ equal to $60$ (red curves), $70$ (orange), $80$ (green) and $100$ nm (blue). (b) Energy gap $e_3 - e_1$, with the same color code as above.}
\label{Fig:3}
\end{figure}

Such transition clearly shows up in the behavior of the hole energy eigenvalues at zero magnetic field. In particular, we consider the energies $e_1=e_2$ and $e_3=e_4$ of the Kramers-degenerate ground and first-excited doublets as functions of $a$ [Fig.~\ref{Fig:3}(a)]. In the absence of an external charge, the energy gap $\delta_{3,1} \equiv e_3-e_1$ decreases monotonically in the considered range of values [Fig.~\ref{Fig:3}(b), dashed grey line]. This is because $\delta_{3,1}$ is related to the tunneling amplitude between the two dots, which decreases approximately as $ {\rm e}^{-\gamma a^2}$ (for further details, see Appendix \ref{app: tunnelfit}). This behavior changes in the presence of the external charge (solid lines), where  $\delta_{3,1}$ displays a minimum for $a\approx a_{\rm min}$. There, a transition takes place from a regime where the interdot tunneling is larger than the charge-induced bias and the hole state is delocalized, to the opposite for $a\gtrsim a_{\min}$. Correspondingly, the energy levels $e_1$ and $e_3$ undergo an avoided crossing, as can be seen by comparing the curves corresponding to the same value of $D$ in Fig.~\ref{Fig:3}(a). Higher energy levels lie at least $1.5\,$meV's above $e_3$ for all the considered values of $a$ and $D$. 

This interplay is qualitatively reproduced by the Hubbard model described in Subsec.~\ref{subsec: Hubbard}. In fact, the bias between the two dots, induced by the external charge, concurs with quantum tunneling to determine the avoided crossing between the energy levels $e_-^{\rm Hub}$ and $e_+^{\rm Hub}$ [Eq.~\eqref{Hubbard energies}], corresponding to $e_1$ and $e_3$, respectively (see Appendix \ref{app: tunnelfit}).

\subsection{Remote charge sensing\label{subsecB}}

In this section, we consider the case where one has to infer the presence (or absence) of the remote charge at a given distance from the qubit. The inference is based on the fact that, in the two cases, the ground state of the hole corresponds to two different --- though non-orthogonal --- states, $|\psi_c\rangle$ (presence of the charge) and $|\psi_{nc}\rangle$ (no charge present). This problem can be formalized within the framework of quantum state discrimination \cite{Helstrom1969,Bae2015a}. In particular, the probability of discriminating between the two qubit states and thus correctly inferring the presence of the remote charge, is upper bounded by the Helstrom bound, defined as:
\begin{gather}
p_{d,max} = \frac{1}{2} + \frac{1}{2} \left[1-4p_c\,p_{nc}|\langle\psi_c|\psi_{nc}\rangle|^2\right]^{1/2},
\end{gather}
where $p_c$ and $p_{nc}$ are the \textit{a priori} probabilities assigned to the two hypotheses, which in the following are assumed to be $1/2$. 

\begin{figure}
\centering
\includegraphics[width=0.47\textwidth]{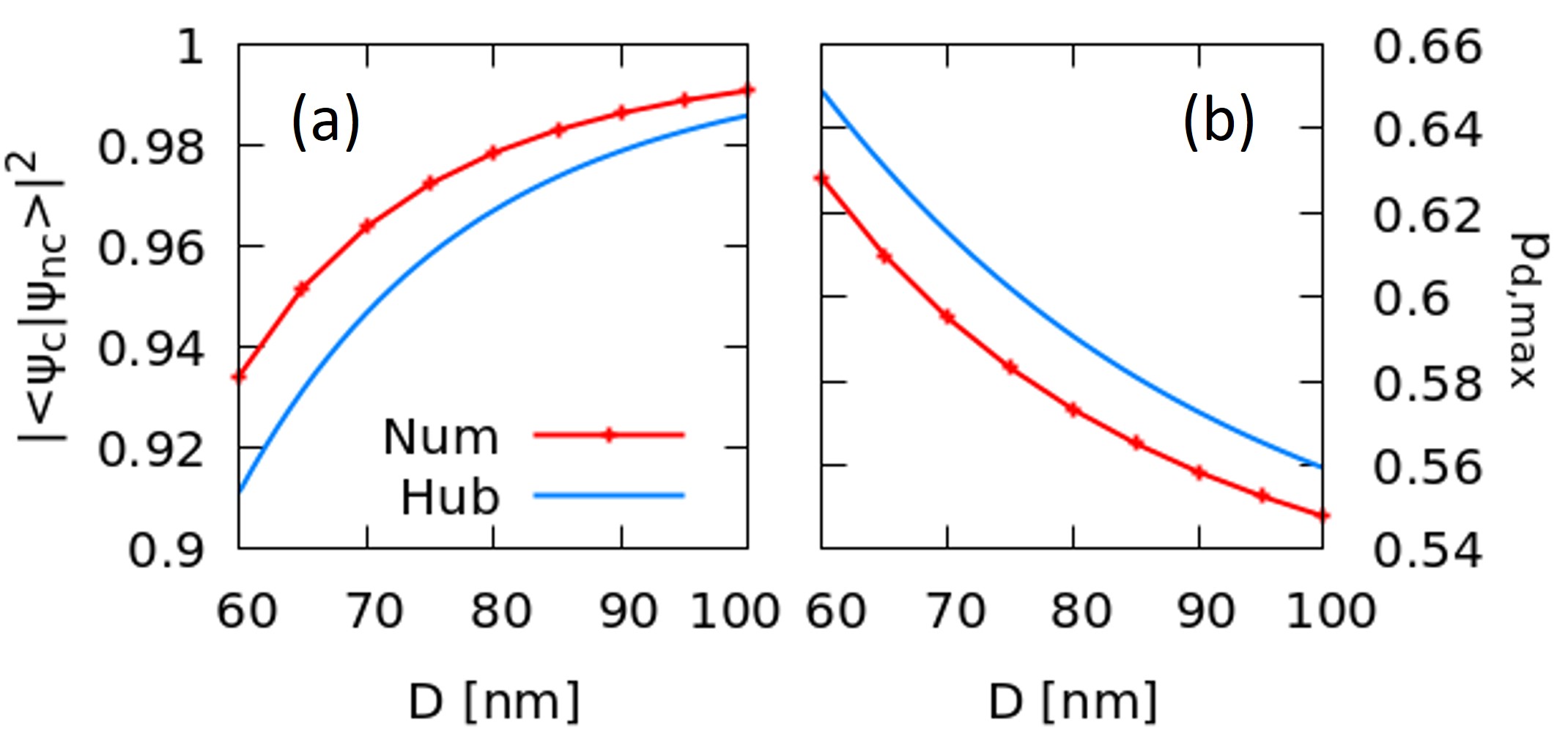}
\caption{(a) Squared overlap between the two hole states $|\psi_{c}\rangle$ and $|\psi_{nc}\rangle$ and (b) corresponding Helstrom bound as a function of the distance $D$ of the remote charge. The blue curves correspond to the case of the Hubbard model, the red symbols are obtained by numerically solving the Luttinger-Kohn Hamiltonian for the hole state, with half interdot distance $a=12\,$nm.}
\label{Fig:A4}
\end{figure}

The dependence of $p_{d,max}$ on the distance $D$ of the remote charge is reported in Fig.~\ref{Fig:A4} for a representative case, corresponding to an interdot distance of $2a=24\,$nm. The presence of the remote charge at the distance $D= 60\,$nm slightly modifies the hole wave function [panel (a)], yielding a maximum discrimination probability around $0.63$ [panel (b)]. For larger values of $D$, interdot tunneling dominates and the ground state tends to be delocalized in spite of the Coulomb interaction between the hole and the external charge. As a result, the two states $|\psi_{c}\rangle$ and $|\psi_{nc}\rangle$ tend to coincide, and $p_{d,max}$ approaches its theoretical minimum, equal to $1/2$. In accordance with the discussion in the previous Subsection, the value of $D$ at which such transition takes place decreases with the half interdot distance $a$. 

Finally, the Hubbard model (blue curves) slightly overestimates the values of $p_{d,max}$ obtained from the Luttinger-Kohn Hamiltonian (red curves); the relative discrepancy between the values obtained from the model and those obtained from the numerical calculations is between $2 \%$ and $3 \%$ in the whole investigated range of $D$, and the curves are qualitatively similar in the two cases. In the limit of a close charge, where the bias between the two dots is much larger than the tunneling parameter, the squared overlap between the states $|\psi_{c}\rangle$ and $|\psi_{nc}\rangle$ in the Hubbard model would be $1/2$, corresponding to a Helstrom bound $p_{d,max}=\frac{1}{2}(1+1/\sqrt{2})\approx 0.854$. We finally note that these values of the maximal discrimination probability corresponds to single shot measurements performed on the sensor qubit, and can be increased by performing repeated measurements \cite{Troiani20a}.

\subsection{Quantum Fisher information of the hole ground state\label{subsec:QFI}}
The dependence on the interdot distance of the hole localization and the energy gap anticipates and explains that of the Quantum Fisher Information (QFI). The QFI is a key quantity in quantum parameter estimation \cite{Paris09}, for it provides --- via the quantum Cram\'er-Rao bound --- the highest achievable precision in the estimate of the unknown value of the parameter. For a generic state $|\psi\rangle$ depending on a parameter $\lambda$, the QFI reads as
\begin{equation}
    H = 4 \left( \langle \partial_\lambda\psi|\partial_\lambda\psi\rangle + \langle \partial_\lambda\psi|\psi\rangle^2 \right) .
    \label{QFI formula}
\end{equation}
The QFI related to the parameter $\lambda=D$ is computed hereafter in order to further characterize the dependence of the hole ground state on the position of the charge and to provide a benchmark for the dynamic approach to the estimate of the charge distance.

\begin{figure}
\centering
\includegraphics[width=0.45\textwidth]{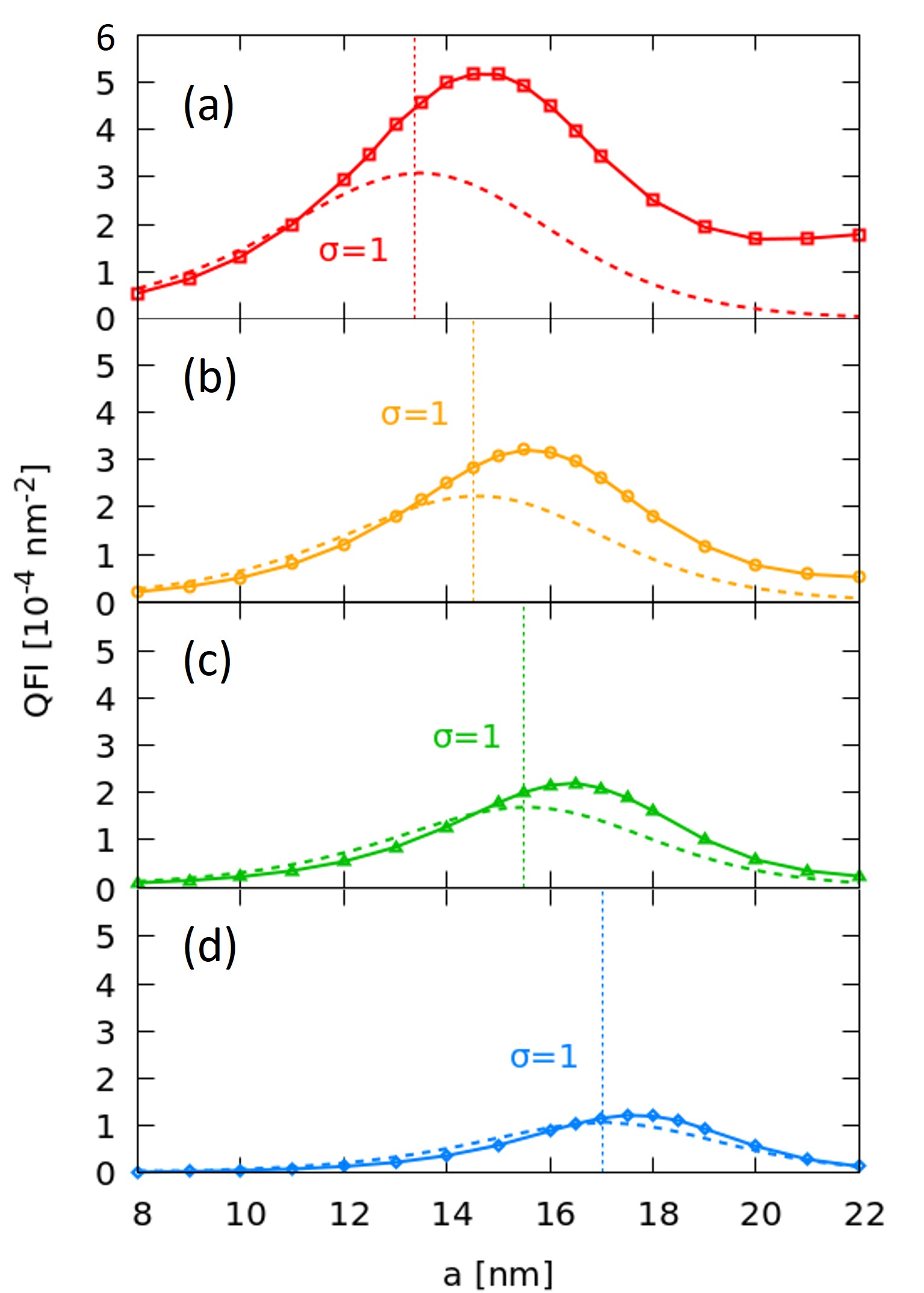}
\caption{Quantum Fisher information of the hole ground state as a function of the half interdot distance $a$ and for different values of the charge distance $D$: (a) $60$ , (b) $70$, (c) $80$, and (d) $100$ nm. Each panel compares the numerical results obtained from the multiband simulations (solid line) with the analytical results from the Hubbard model (dashed). The vertical dotted lines indicate the value of $a$ corresponding to $\sigma = 1$, for each value of $D$.}
\label{Fig:4}
\end{figure}

The dependence of $H$ on the half interdot distance $a$, for different distances $D$ of the charge from the DQD, is characterized by a maximum for $a\approx a_{\rm min}$, where $\delta_{3,1}$ is minimal (Fig.~\ref{Fig:4}, solid lines). In agreement with what was previously observed for the energy eigenvalues, the maxima occur at values of the interdot distance that increase with the distance $D$ between the charge and the DQD.

The behavior obtained from the six-band $\mathit{k\cdot p}$ Hamiltonian is qualitatively reproduced by the Hubbard model (dashed lines), especially for values of $D$ that are sufficiently larger than $2a$. Besides, in the case of the Hubbard model, one can derive the analytical expression of the QFI:
\begin{align}\label{eq:analytical}
    H^{\rm Hub} = \left[ \frac{2e^2 \, a \, |t| \,\varepsilon \, D }{a^2 e^4+ |t|^2\varepsilon^2(D^2-a^2)^2} \right]^2  ,
\end{align}
which depends on the half interdot distance $a$ also through the tunneling parameter $t$ (see Appendix \ref{app: tunnelfit}). At a given value of $a$, the maximum value of $H^{\rm Hub}$ as a function of $D$ (in the case of $D > a$) is obtained at 
\begin{align}
D_{\max H}^{\rm Hub} =  \sqrt{\frac{a}{3} \left( a + \sqrt{4 a^2 + \frac{3 e^4}{\varepsilon^2 |t|^2} } \, \right) } \,.
\end{align}
The above expression thus gives the value of $D$ at which the hole ground state is most sensitive to small variations in $D$. The problem of identifying the value of $a$ that maximizes $H^{\rm Hub}$, for a given $D$, is not analytically solvable. However, we find that such optimal value of $a$ is close to satisfying the condition $\sigma \approx 1$ (see Fig.~\ref{Fig:4}), where $\sigma \equiv 2 |t| / \delta_{\rm c}$ is the ratio between the tunneling energy gap and the charge-induced bias. The Hubbard model thus provides a clear indication that the maximal sensitivity of the hole ground state to the position of the external charge is achieved when the competition between interdot tunneling and charge-induced bias is balanced. 

\begin{figure}
\centering
\includegraphics[width=0.45\textwidth]{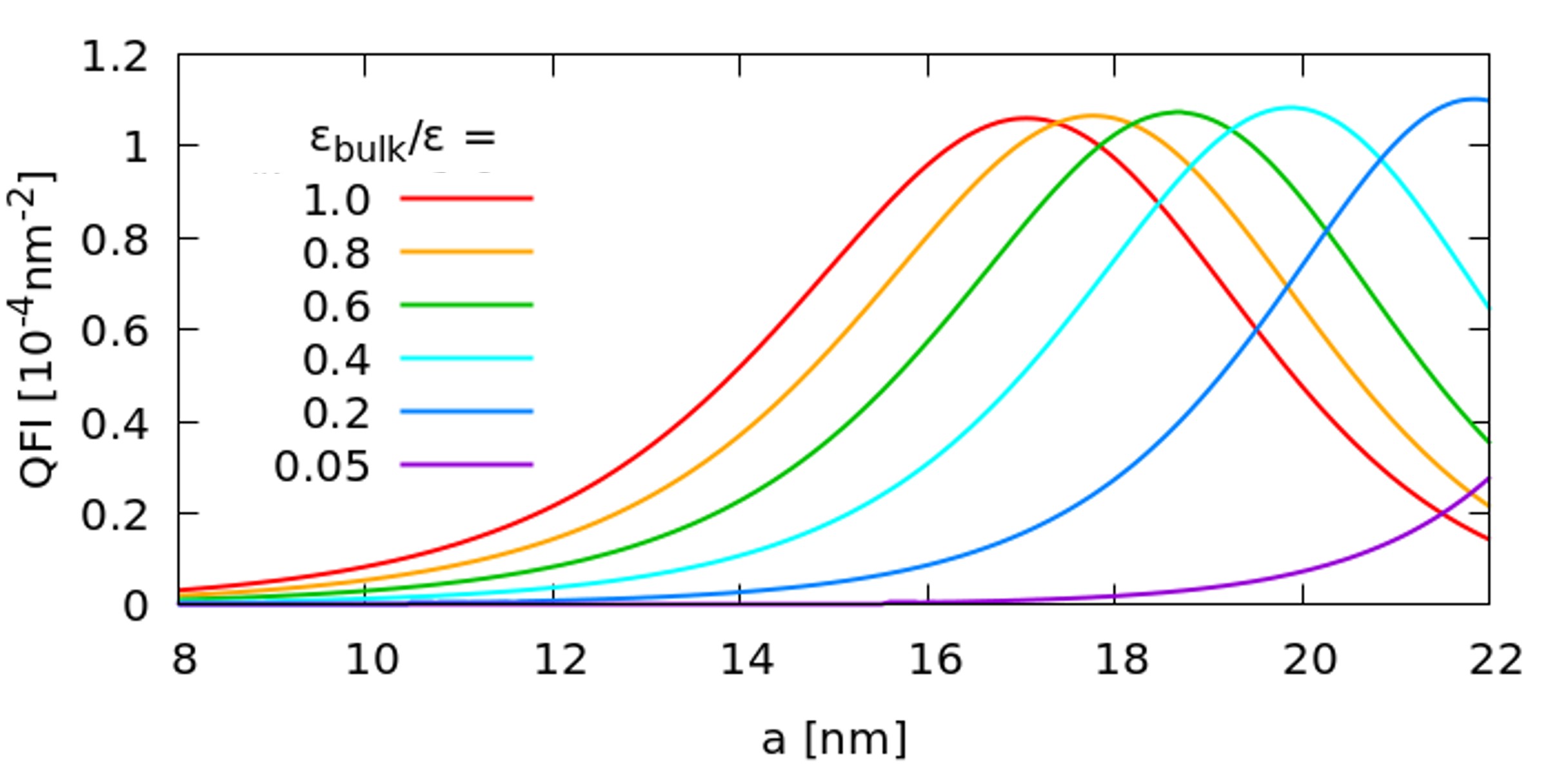}
\caption{Dependence of the quantum Fisher information on the half interdot distance $a$ at fixed distance $D=100\,$nm between the qubit and the remote charge. Different colors correspond to different values of the screening constant $\varepsilon$, normalized to the value in bulk silicon $\varepsilon_{\rm bulk}$.}
%\caption{{\color{red} Dependence of the quantum Fisher information on: (a) the half interdot distance $a$, for different values of the screening constant $\epsilon$, normalized to the value of bulk silicon $\epsilon_{bulk}$; (b) $\epsilon$ for different values of $a$.}}
\label{Fig:A1}
\end{figure}

The effect of the remote charge on the hole-spin qubit depends not only on its position, but also on the screening of the Coulomb interaction between the two charges resulting, for example, from the presence of neighboring metal gates \cite{Hogg23a}. A detailed analysis of such dependence is beyond the scope of this paper. However, a preliminary understanding of such effect can be gained by assuming that such screening results in a renormalization of the dielectric constant $\varepsilon$, whose value is known before the quantum estimation or state discrimination procedures are implemented. As is apparent from the expressions of the Coulomb interaction [Eq.~(\ref{eq:vc})] and of the QFI for the Hubbard model [Eq.~(\ref{eq:analytical})], the dependence of the figures of merit on the dielectric constant cannot be absorbed in a redefinition of the distance $D$, due to the extended character of the hole wave function. As shown in Fig.~\ref{Fig:A1}, the dependence of the QFI on $a$ essentially undergoes a shift in the position of the maximum, whose value remains approximately constant for a significant range of values of $\varepsilon$.

\section{Dynamic approach to the estimate of the charge distance \label{sec:DA}}

In the dynamic approach, the presence of the charge or its position are inferred from the outcome of the qubit readout, performed after a coherent control scheme. When the hole is operated as a qubit, the only degree of freedom which is accessed is the particle spin, defined by the lowest-energy doublet: $\{|0\rangle\equiv|\psi_1\rangle,|1\rangle\equiv|\psi_2\rangle\}$. 
The particular qubit state $|\psi\rangle$ that is generated after manipulating the hole depends on the Larmor and Rabi frequencies and, through these, on the presence of the charge and on its distance $D$ from the (double) quantum dot. 

\subsection{Larmor and Rabi frequencies\label{subsec:lrf}}

\begin{figure}
\centering
\includegraphics[width=0.45\textwidth]{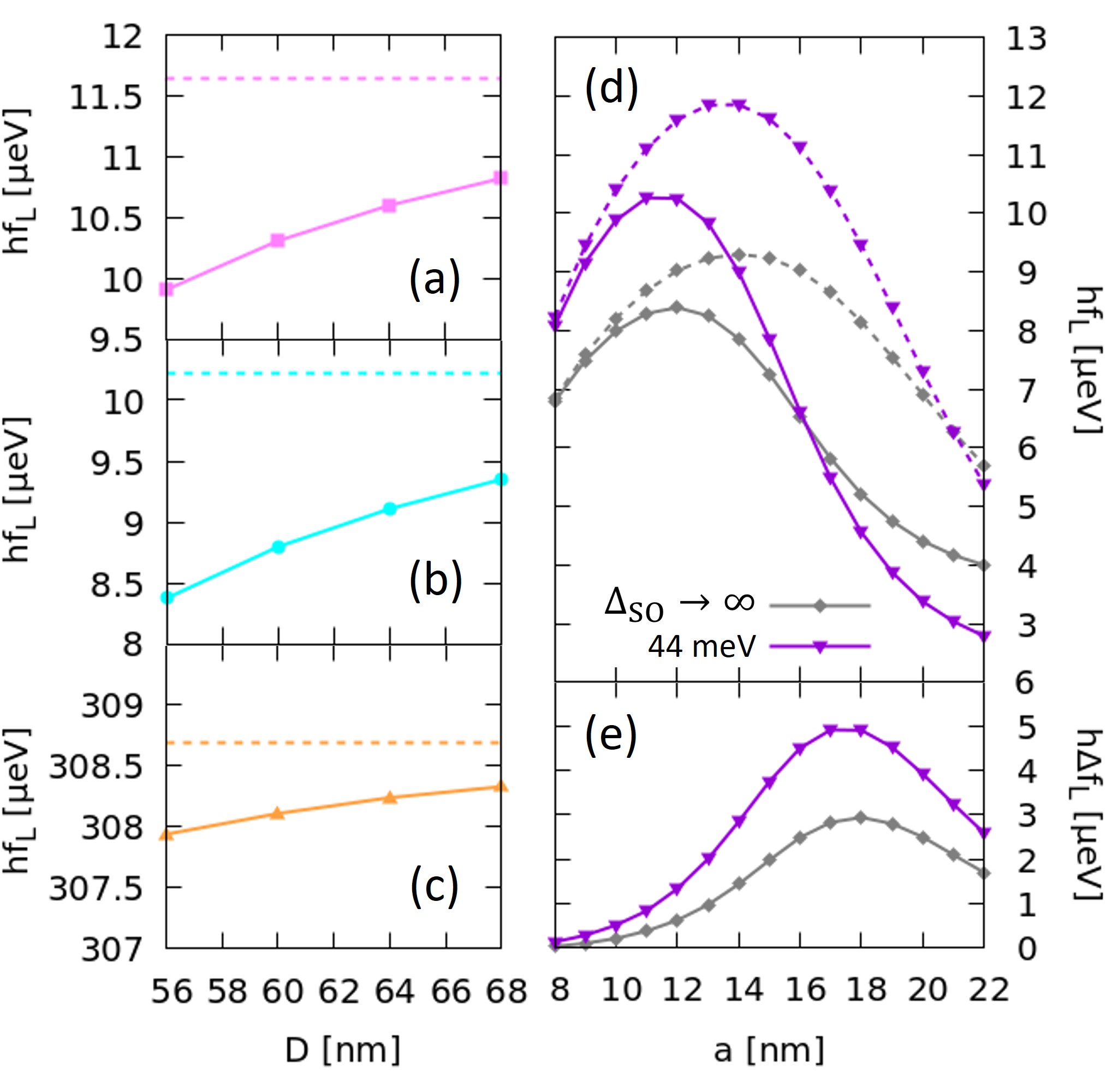}
\caption{(a-c) Energy gap between the qubit states $|0 \rangle$ and $|1 \rangle$ of a DQD, induced by a magnetic field of $1\,$T and for $a = 12$ nm. The gap is plotted as a function $D$ (solid lines), for $\bf B$ oriented along (a) the $x$, (b) $y$, and (c) $z$ axes. The dashed horizontal lines indicate the value of the gap in the absence of an external charge. (d) Same energy gap, as a function of the half interdot distance $a$, with the charge at $D = 60\,$nm (solid lines) and without the charge (dashed), for a field $B=1\,$T, oriented along the $x$ axis. The difference between the two cases is reported in panel (e). }
\label{Fig:5}
\end{figure}

The presence of the external charge affects the Larmor frequency in ways that depend on the orientation of the magnetic field $\mathbf{B}$. As the charge distance $D$ from the DQD increases, $f_{\rm{L}}\equiv\omega_{\rm L}/2\pi$ monotonically increases [Fig.~\ref{Fig:5}(a-c), solid lines], and eventually tends to the value obtained in the absence of the charge (dashed lines). While this behavior is qualitatively independent of the magnetic-field orientation, the sensitivity of $f_{\rm{L}}$ to $D$ is maximal when $\bf B$ is oriented along either the $x$ or $y$ directions. The situation is analogous for a SQD, although in this case the curves corresponding to $\mathbf{B} \parallel \mathbf{x}$ and $\mathbf{B} \parallel \mathbf{y}$ have opposite curvatures (i.e. opposite values of $f_{\rm L}'$). 

Both the value of the Larmor frequency ($\omega_{\rm L}$) and its change ($\Delta\omega_{\rm L}$) in the presence of an external charge display a non-monotonic dependence on the interdot distance [panels (d-e), $\mathbf{B} \parallel \mathbf{x}$]. The position of the maxima of both $\omega_{\rm L}$ and $\Delta\omega_{\rm L}$ depends on the distance $D$. The role played by spin-orbit coupling, and specifically by the value of the splitting between the $j=3/2$ and the $j=1/2$ bands, is demonstrated by the comparison between the above mentioned results and those obtained by a four-band calculation (corresponding to the limit where $\Delta_{\rm SO}\rightarrow\infty$). In the opposite limit with no spin-orbit coupling ($\Delta_{\rm SO}=0$) the Larmor frequency is entirely independent of the orbital degrees of freedom and, thus, of the electrostatic potential. In this case $f_{\rm L}$ is given by the Zeeman splitting, $hf^{\rm Z}_{\rm{L}}=g_{\rm s} \mu_{\rm{B}}B=115.77$~$\mu$eV, where $\mu_{\rm{B}}$ is the Bohr magneton and $g_{\rm s} \approx 2$ is the electron-spin $g$-factor. This implies that the presence of $\Delta_{\rm SO}\neq 0$ is fundamental, since the dependence of the Larmor frequency on $D$ plays a crucial role in the estimation of the parameter $D$ performed both in the Rabi and in the Ramsey measurement schemes (see Subsec. \ref{subsec:qpe}). 

\begin{figure}
\centering
\includegraphics[width=0.48\textwidth]{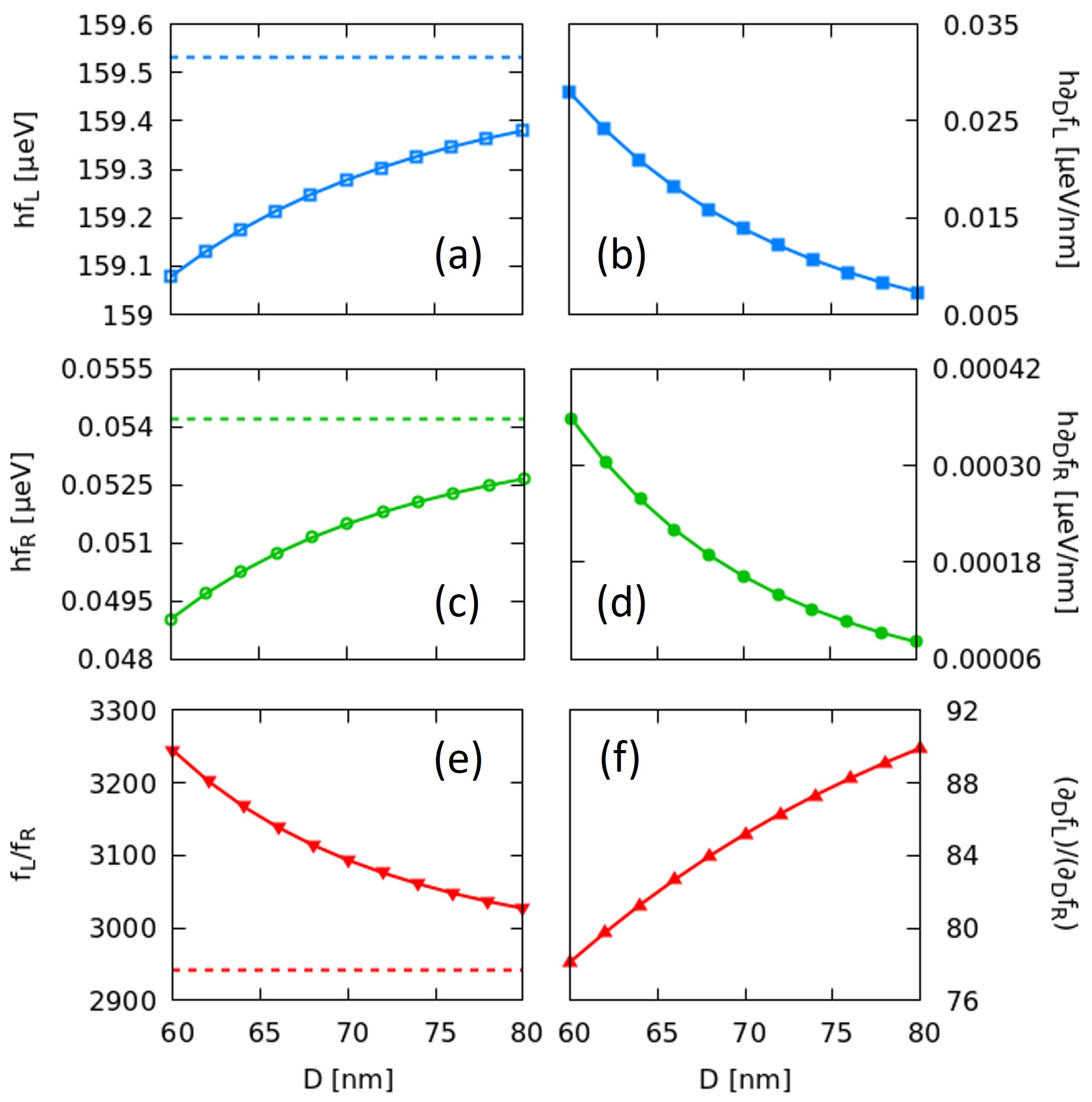}
\caption{Dependence of the Larmor and Rabi frequencies on the position of the charge, for a specific orientation of the magnetic field ($\theta=60^{\circ}$ and $\phi=90^{\circ}$) with intensity $B=1\,$T and fixed half interdot distance $a = 12\,$nm. The panels show (a) the Larmor frequency $f_{\rm L}$ and (b) its derivative, (c) the Rabi frequency $f_{\rm R}$ and (d) its derivative, and the ratios (e) $f_{\rm L}/f_{\rm R}$ and (f) $\partial_{D}f_{\rm R}/\partial_{D} f_{\rm L}$. The dashed lines represent the values of $f_{\rm L}$, $f_{\rm R}$ and their ratio in the absence of a charge.}
\label{Fig:6}
\end{figure}

The Rabi frequency for a hole-spin qubit in a Si SQD has been previously studied for different devices and confining potentials. A crucial role in the tuning of $\omega_{\rm R}$ is played by the orientation of the magnetic field with respect to the crystal and dot axes. The dependence of the Rabi frequency on the angles $\theta$ and $\varphi$, that we obtain here, qualitatively coincides with the results reported elsewhere for SQDs~\cite{Bellentani2021a}. In the following, we focus of the field orientation defined by $\theta=60^{\circ}$ and $\phi=90^{\circ}$, which represents a good compromise between the conflicting requirements of obtaining large Larmor and Rabi frequencies. Both $\omega_{\rm R}$ and its derivative with respect to $D$ decrease monotonically with the distance $D$ [Fig.~\ref{Fig:6}(c,d)], in analogy to what is obtained for the Larmor frequency [panels (a,b)]. The comparison between the two quantities shows that the Larmor frequency, aside from being much larger than the Rabi frequency in absolute terms, displays a much stronger dependence on the position of the charge [panels (e,f)]. This will have important implications in the identification of the optimal parameter estimation protocols.

\subsection{Quantum state discrimination\label{subsec:dqsd}}

As shown in the previous Subsection, both Larmor and Rabi frequencies are modified by the interaction between the hole and the remote charge, localized in a given (known) position. In analogy to what was discussed in Subsection \ref{subsecB} for the static approach, this results in the generation of two different qubit states in the two cases, hereafter denoted as $|\psi_{c}\rangle $ (charge) and $ |\psi_{nc}\rangle $ (no charge), after a given pulse sequence. The problem of inferring the presence of the charge is thus reduced to that of discriminating between these two qubit states. In the case where $ | \langle 0 |\psi_{c}\rangle |^2 > | \langle 0 |\psi_{nc}\rangle |^2$, from the outcome $0$ ($1$) of the qubit readout one infers the presence (absence) of the remote charge. The discrimination probability $p_d = p_c\, p(0|c) + p_{nc}\, p(1|nc)$ in the case of a coherent qubit state evolution thus reads:
\begin{gather}
    p_{d,\rm coh} = p_c | \langle 0 |\psi_{c}\rangle |^2 + p_{nc} | \langle 1 |\psi_{nc}\rangle |^2,
\end{gather}
where $p_c$ and $p_{nc}$ are the \textit{a priori} probabilities, which are assumed to coincide with $1/2$. 

\begin{figure}
\centering
\includegraphics[width=0.45\textwidth]{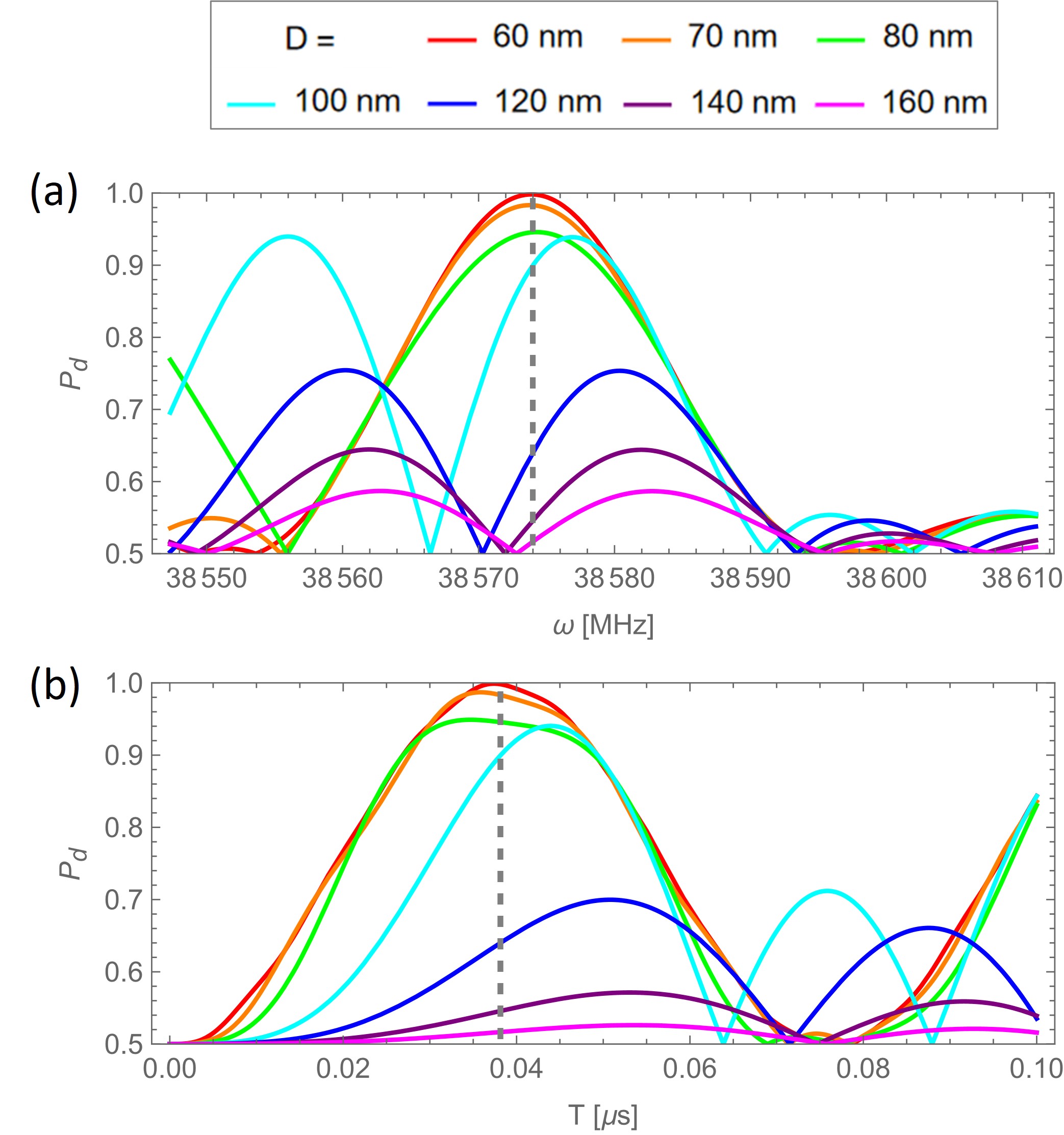}
\caption{Discrimination probability based on the Rabi measurement as a function of the pulse (a) frequency and (b) duration, in the absence of decoherence, for a fixed half interdot distance of $a=12\,$nm and $B=1\,$T ($\theta=60^{\circ}$ and $\phi=90^{\circ}$). Different colors correspond to different values of the distance $D$ between the qubit and the remote charge. In panel (a) the pulse duration is fixed to $T=\pi/\omega_{\rm R}^{nc}$, where $\omega_{\rm R}^{nc}$ is the Rabi frequency in the absence of a remote charge. Panel (b) fixes the pulse frequency in resonance to the Larmor frequency in the absence of a remote charge, $\omega=\omega_{\rm L}^{nc}$. The grey dashed line corresponds to the implementation of $\pi$ rotations in the absence of charge.}
\label{Fig:A3}
\end{figure}

In the Rabi measurement scheme the qubit readout takes place after a pulse of duration $T$ and frequency $\omega$ (Fig. \ref{Fig:A3}). For $D\lesssim 100\,$nm, the maxima in the discrimination probabilities $p_{d,\rm coh}$ are obtained with pulses that implement a $\pi$ pulse in the absence of remote charge (dotted grey lines). As $D$ increases, the Larmor and Rabi frequencies become more and more similar to those characterizing the qubit in the absence of the remote charge, and consequently the maxima of the discrimination probability decrease. The reported results are well above those obtained, for the same values of $D$, within the static approach (Fig.~\ref{Fig:A4}). 

Decoherence tends to erase the information on the presence of the external charge encoded in the qubit state. In the following, we consider for simplicity the case of an environment that acts as a depolarizing channel (see Ref. \cite{nielsen} and Appendix \ref{app:ramseycoh}), so that the final density matrix corresponds to a mixture of the target state, with probability $q$, and of the fully mixed state. The  resulting discrimination probability reads:
\begin{gather}\label{eq:dpwd}
    p_d = q\,p_{d,\rm coh} + \frac{1}{2} (1-q) .
\end{gather}
The probability $q$ is in general a decreasing function of the measurement duration $T_m$, with a functional dependence on $T_m$ and a characteristic time scale $T_d$ that depend on the dominant decoherence mechanism \cite{yoneda2018,stano2022,piot2022,Connors22a} and on the measurement scheme \cite{Burkard23a} (see discussion in Sec. \ref{subsec:qpe}).  
As a result, the oscillations between $1/2$ and $1$ that characterize the discrimination probability in the coherent case [Fig. \ref{Fig:A3}(b)] would be multiplied by a decaying envelope $q(T)$ (being $T_m=T$).

\subsection{Fisher information of the qubit state\label{subsec:qpe}}

Once the dependence of the Larmor and Rabi frequencies on $D$ has been derived, one can also devise a strategy for precisely estimating the charge distance, in the case where the presence of the charge is known. This consists in identifying a manipulation protocol that results in a strong dependence on $D$ of the qubit state $|\psi\rangle$, and specifically of the measurement-outcome probabilities $p_{0}=\langle 0 | \rho | 0\rangle $ and $p_{1}=\langle 1 | \rho | 1 \rangle$. In the following, we focus on the Rabi and Ramsey measurement schemes.
In both cases, the key figure of merit is the Fisher information (FI) associated with the observable $\sigma_Z$. For a qubit in a generic state $|\psi\rangle$, the FI related to the parameter $D$ reads:
\begin{equation}
    F(\rho,\sigma_Z) = \sum_{k=0,1} \frac{(p_k')^2}{p_k} = \frac{(p_1')^2}{(1-p_1) p_1} \ ,
    \label{FI qubit}
\end{equation}
where the prime denotes differentiation with respect to the unknown parameter $D$.

Decoherence tends to erase the information on the value of $D$ encoded in the qubit state, and thus to reduce the FI. In order to account for this effect, we compare in the following the values of $F(\rho,\sigma_Z)$ obtained in the presence and in the absence of decoherence. If decoherence is mainly induced by the hyperfine interactions \cite{yoneda2018,stano2022,piot2022}, the coherent component of the density operator undergoes a Gaussian decay: $q(T_m)=e^{-(T_m/T_d)^2}$, being $T_m$ and $T_d$ the overall duration of the qubit manipulation and the relevant decoherence time, respectively (see Appendix \ref{app:ramseycoh}). 
The FI in the presence of decoherence can be expressed as a function of that obtained in the coherent case:
\begin{align}
F &=F_{\rm coh} \frac{1-(1-2p_{1,\rm coh})^2}{q^{-2}-(1-2p_{1,\rm coh})^2}\, ,
\end{align}
where $p_{0,\rm coh}=|\langle 0 | \psi\rangle|^2$ and $p_{1,\rm coh}=|\langle 1 | \psi\rangle|^2$.
This shows how the ratio $F / F_{\rm coh}$ varies from 1 in the limit where $T_m$ is much smaller than the decoherence time $T_d$, to $q^{2}$ in the opposite limit.

\paragraph{Rabi measurement.}

In the Rabi scheme and under the rotating-wave approximation, the occupation probability of state $|1\rangle$, for a qubit initialized in $|0\rangle$ and addressed by a rectangular pulse of frequency $\omega$ and duration $T$, is given by \cite{Sakurai}:
\begin{equation}
    p_{1,\rm coh}=\frac{\omega_{ \rm R}^2}{\Omega^2} \sin^2\left(\frac{T \Omega}{2} \right) \,,
    \label{p1 Rabi}
\end{equation}
where $\Omega = \sqrt{\omega_{\rm R}^2 + \Delta^2}$, and $\Delta \equiv \omega - \omega_{\rm L}$. Combining Eqs.~(\ref{FI qubit}-\ref{p1 Rabi}), one obtains the expression of the FI, which reads:
%\begin{widetext}
%\begin{align}
%F_{\rm Rabi, coh} = \frac{\left[ T \cos\left(\frac{ \Omega T}{2} \right) \omega_{\rm R} \Omega \left( \omega_{\rm R} \omega'_{\rm R} - \Delta  \omega'_{\rm L}  \right) + 2 \Delta \sin\left(\frac{  \Omega T }{2} \right) \left( \omega_{\rm R}  \omega'_{\rm L}  + \Delta   \omega'_{\rm R} \right)   \right]^2}{\Omega^4 \left[ \Delta^2 + \omega_{\rm R}^2 \cos^2\left( \frac{ \Omega T}{2}\right) \right]}  \,.
%\label{FI Rabi general}
%\end{align}
%\end{widetext}
\begin{align}
F_{\rm Rabi, coh} = \frac{\left[ C_1 \cos\left(\frac{ \Omega T}{2} \right) + C_2 \sin\left(\frac{  \Omega T }{2} \right) \right]^2}{\Omega^4 \left[ \Delta^2 + \omega_{\rm R}^2 \cos^2\left( \frac{ \Omega T}{2}\right) \right]}  \,.
\label{FI Rabi general}
\end{align}
where the terms $C_1$ and $C_2$ are given by
\begin{align}
C_1 & = T \omega_{\rm R} \Omega \left( \omega_{\rm R} \omega'_{\rm R} - \Delta  \omega'_{\rm L}  \right) \\ 
C_2 & = 2 \Delta \left( \omega_{\rm R}  \omega'_{\rm L}  + \Delta   \omega'_{\rm R} \right) .
\end{align}

For each pulse duration ($T$) and central frequency ($\omega$), the qubit read out thus allows the estimate of $D$ with a precision whose upper bound is fixed by the above $F_{\rm Rabi}$. We stress that the above expression is valid for a generic Rabi measurement, irrespective of the involved physical system. 

\begin{figure}
\centering
\includegraphics[width=0.49\textwidth]{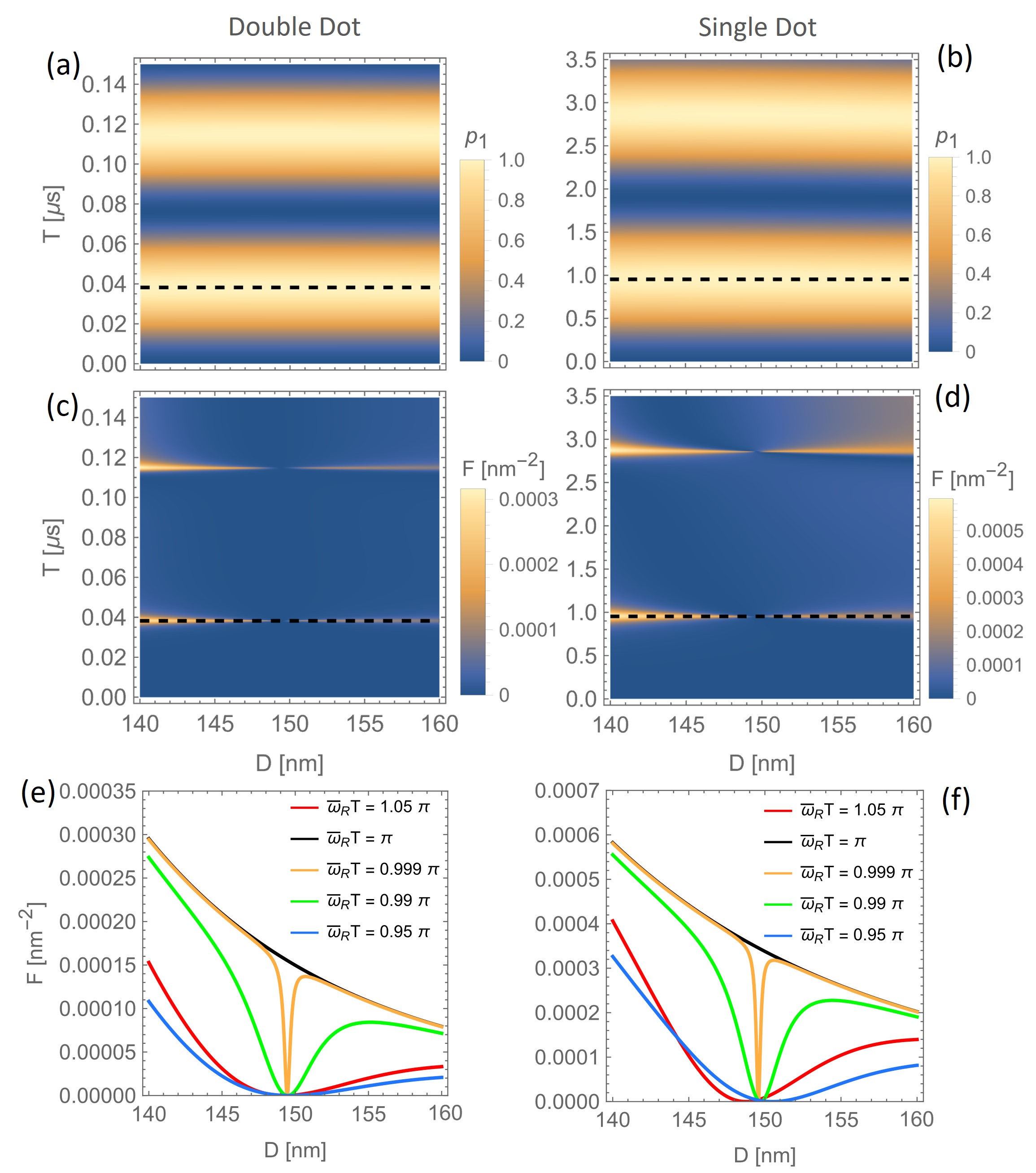}
\caption{Color plots, related to the Rabi measurement, of the probability $p_1$ [panels (a), (b)] and of the Fisher Information $F$ [panels (c), (d)] for a DQD with half interdot distance $a=12\,$nm [panels (a), (c)] and a SQD [panels (b), (d)], as functions of the defect coordinate $D$ and the pulse duration $T$. The dashed line in panels (a-d) show the values of $T_{\rm Rabi}$ (black line) for both systems, as defined in the text. Panels (e) and (f) show the profile of $F$ as a function of $D$ for selected values of $T$ close to $T_{\rm Rabi}$, respectively for the DQD and the SQD.}
\label{Fig:Rabi}
\end{figure}

In local quantum parameter estimation, the goal is the precise estimate of a parameter, whose possible values are already restricted to a limited range by an \textit{a priori} knowledge. In order to account for this, we consider hereafter values of the charge distance from the (double) quantum dot ranging from $D_{\rm min} = 140\,$nm to $D_{\rm max} = 160\,$nm. A suitable choice of the pulse is crucial in order to optimize the estimation process. The uncertainty on the value of $D$ also implies an uncertainty on $\omega_{\rm R}$ and $\omega_{\rm L}$, and thus on the pulse duration and frequency that are required to generate a given qubit state $|\psi\rangle$. However, if the variation of the Larmor and Rabi frequencies within the range of possible values of $D$ is sufficiently small, then a given pulse leads to the implementation of well-defined rotations. More specifically, we consider a value of the pulse frequency equal to $\bar{\omega}_{\rm L}$, defined as the integral average of the Larmor frequency in the interval [$D_{\rm min}$;$D_{\rm max}$]. By doing so, we effectively show in Fig.~\ref{Fig:Rabi} that the final occupation probability of the qubit state $|1\rangle$ displays a negligible dependence on the exact value of $D$ for both the double [panel (a)] and single dot [panel (b)]. We also evaluate the integral average of the Rabi frequency, $\bar{\omega}_{\rm R}$, which allows us to define a reference time for the duration of the pulse, $T_{\rm Rabi}=\pi/\bar{\omega}_R$. We note that the value of $\bar{\omega}_{\rm R}$ is approximately $25$ times larger for the DQD than for the SQD, which determines a much shorter pulse duration $T_{\rm Rabi}$ in the former case ($0.038$ vs $0.95\,\mu$s).

As to the FI, the regions of interest clearly coincide with those where $\bar{\omega}_{\rm R}T\approx (2n+1)\pi$ [panels (c,d)]. In order to gain further insight into the parameter dependence of $F_{\rm Rabi}$, we focus hereafter on the case $n=0$ [panels (e,f)], where the assumption that $T_m = T_{\rm Rabi} \ll T_d$ (and thus $q \approx 1$) is better satisfied. The FI, plotted as a function of $D$ for slightly different values of the pulse duration, displays a non-trivial behavior. It generally tends to decrease with the distance between the hole and the charge. However, the curves also display a marked minimum in correspondence of the resonance condition ($\Delta=0$), where the FI approaches zero. Besides, such dip becomes increasingly narrow as $\bar{\omega}_{\rm R}T$ approaches the value of $\pi$.
While such complex behavior can only be described by the full expression of the FI [Eq. \eqref{FI Rabi general}], the trend obtained away from the resonance condition at $T=T_{\rm Rabi}$ is reflected by the simplified expression:
\begin{align}
    F_{\rm Rabi}\approx \frac{4 \omega_{\rm R}^2(\omega_{\rm L}')^2}{\Omega^4}\, .
    \label{FI Rabi approx}
\end{align}
This is obtained from Eq. \eqref{FI Rabi general} if $\Omega T_{\rm Rabi}\approx \pi$ for any charge distance within the considered range, and exploiting the fact that $|\omega_{\rm L}'|\gg |\omega_{\rm R}'|$ (see Subsec.~\ref{subsec:lrf}).

Overall, the values of the FI are comparable to those obtained for the QFI in the static approach, and larger for the single than for the double quantum dot. On the other hand, the pulse duration $T_{\rm Rabi}$ required for the DQD is about 25 times shorter than that for the SQD, which might represent a significant advantage in dots with shorter decoherence times.

\paragraph{Ramsey measurement.}
In the Ramsey measurement, one ideally applies two $\pi/2$ pulses to the system, separated by a waiting time $\tau$. The final occupation probability of the state $|1\rangle$ depends on the phase difference between the two basis states accumulated during the waiting time $\tau$, and thus on $\omega_{\rm L}$. The inference of $D$ is based on such dependence, but also on the dependence on $\omega_{\rm L}$ and $\omega_{\rm R}$ of the rotations implemented by the two pulses. 
 Within the RWA, and in the absence of decoherence, the expression of the final $p_1$ reads as \cite{ramsey1950}:
\begin{align}
 p_{1,{\rm coh}}  &  =  4 \frac{ \left| \omega_{\rm R} \right|^2 }{\Omega^2}     \sin^2\left( \frac{ \Omega T }{2} \right)    \Bigg[   
\cos\left( \frac{ \Omega T }{2} \right)    \cos\left( \frac{\Delta \tau}{2} \right)  \nonumber \\
&  \quad - \frac{\Delta}{\Omega} \sin\left( \frac{ \Omega T }{2} \right)    \sin\left( \frac{\Delta \tau}{2} \right)   \Bigg]^2  \,,
\label{p1 Ramsey}
\end{align} 
where $T$ is the duration of each pulse.

We report hereafter the analytical expression of the FI. Even though it is quite complex, it applies to any qubit within the Ramsey measurement scheme. The FI is given by
\begin{align}\label{eq:fisherramsey}
F_{\rm Ramsey, coh}=&\frac{4 B^2}{1-X^2 A^2}\, ,
\end{align}
where the terms $A$ and $B$ read as
\begin{equation}
A= \cos \left(\frac{\Delta  \tau}{2} \right) \sin \left(\Omega T \right)-Y\sin \left(\frac{\Delta  \tau}{2} \right) \left[1-\cos \left(\Omega T \right)\right]\, ,
\end{equation}
and
\begin{align}\label{eq:bramsey}
B=& A \frac{\omega_{\rm R}'-X\Omega'}{\Omega} 
+ A_1 \frac{\omega_{\rm L}'\tau}{2} + A_2 \Omega' T
+ A_3 \frac{\omega_{\rm L}'+Y\Omega'}{\Omega}\, .
\end{align}
In the above equations, we have introduced the dimensionless quantities $X\equiv\omega_{\rm R}^X/\Omega$ and $Y\equiv\Delta/\Omega$, defined such that $X^2+Y^2=1$. As usual the prime denotes the derivative with respect to the unknown parameter $\lambda = D$.
Finally, the three terms $A_1$, $A_2$, and $A_3$, which appear in the expression of $B$, are given by:
\begin{align}
A_1 =& \sin \left(\frac{\Delta\tau}{2} \right) \sin \left(\Omega T \right)+Y\cos \left(\frac{\Delta\tau}{2} \right) \left[1-\cos \left(\Omega T \right)\right]\, , \\
A_2 =& \cos \left(\frac{\Delta\tau}{2} \right) \cos \left(\Omega T \right)-Y\sin \left(\frac{\Delta\tau}{2} \right) \sin \left(\Omega T \right)\, , \\
A_3 =& \sin \left(\frac{\Delta\tau}{2} \right) \left[1-\cos \left(\Omega T \right)\right]\, .
\end{align}

The sensitivity allowed by the Ramsey measurement can ideally increase with the duration of the waiting time $\tau$, but it is in practice limited by decoherence. In order to account for such effect, and being in general $\tau\gg T$, we assume that the effect of decoherence is mainly manifest during the free evolution phase [$T_m=\tau +2T\approx\tau$, and thus $q=q(\tau)$; see Appendix \ref{app:ramseycoh} for details]. The FI in the presence of decoherence thus reads:
\begin{align}
F_{\rm Ramsey}&=\frac{16X^2A^2B^2}{(q^{-2}-1)+4X^2A^2(1-X^2A^2)}\, .
\end{align}

\begin{figure}
\centering
\includegraphics[width=0.49\textwidth]{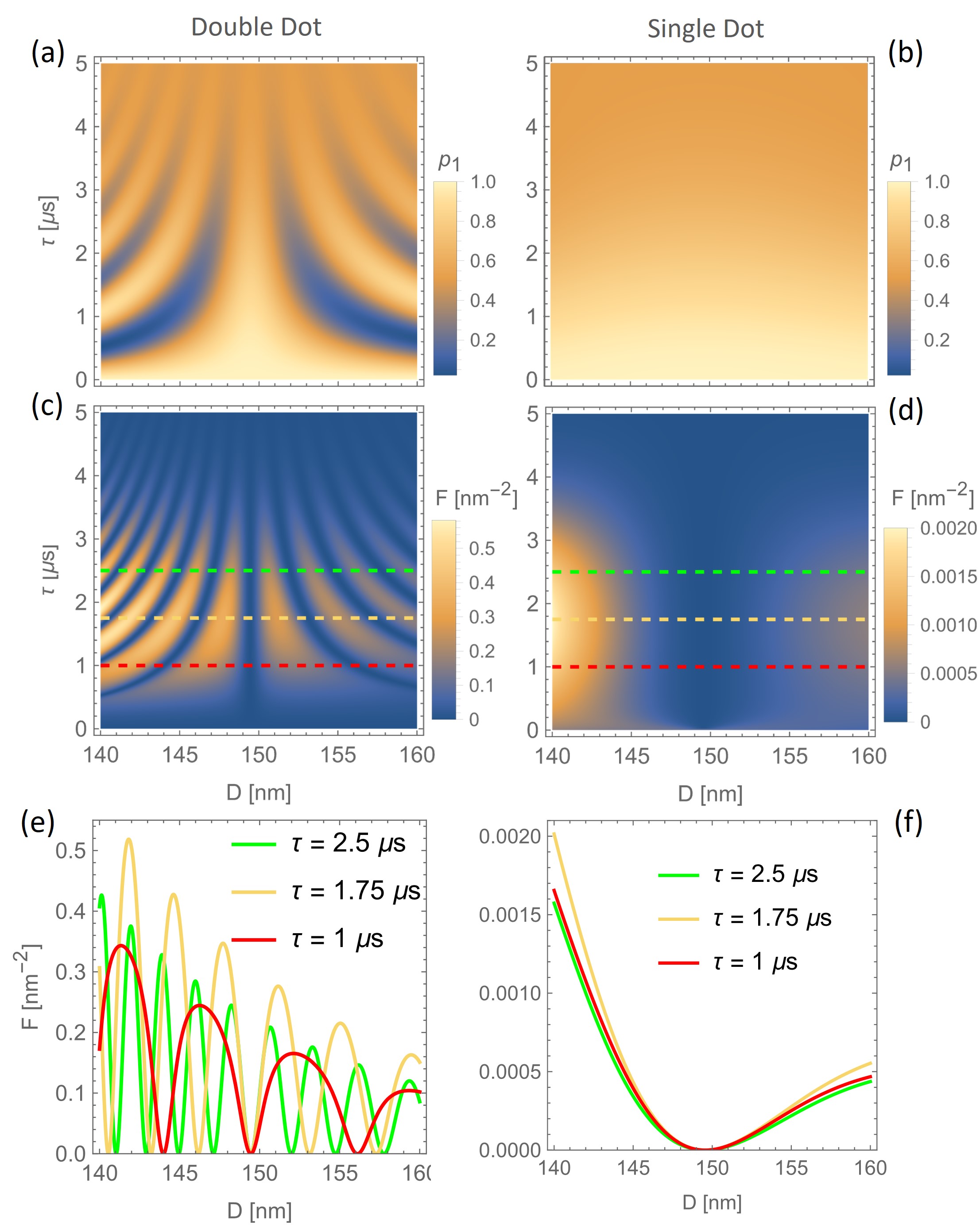}
\caption{Color plots related to the Ramsey measurement in the presence of dephasing. The panels show the probability $p_1$ (a,b) and the Fisher information $F$ (c,d) as functions of the defect coordinate $D$ and the waiting time $\tau$, for a DQD with half interdot distance $a=12\,$nm (a,c) and for a SQD (b,d). Panels (e) and (f) show the dependence of $F$ on $D$ in the case of DQD and SQD, respectively. The curves refer to given values of $\tau$, also highlighted by the dashed lines in the panels (c,d): $1\,\mu$s (red), $1.75\,\mu$s (yellow) and $2.5\,\mu$s (green).}
\label{Fig:Ramsey deph}
\end{figure}

The general expressions of the qubit occupation probabilities and of the corresponding FI are finally applied to the case of the hole-spin qubit in the single and double quantum dots, where we include the specific values of the Larmor and Rabi frequencies and of their derivatives with respect to $\lambda=D$ obtained from our $\mathit{k\cdot p}$ simulations.
In particular, we plot $p_1$ and $F_{\rm Ramsey}$ as a function of $D$ and of the waiting time $\tau$, for $T_d=T_2^*=2.5\,\mu$s and $q(\tau) = e^{-(\tau/T_2^*)^2}$ (Fig. \ref{Fig:Ramsey deph}). We fix the pulse duration to a reference time $T=T_{\rm Ramsey}=\pi/2\bar{\omega}_{\rm R}$. In the case of the DQD, the dependence of $p_1$ on $\tau$ and on $D$ [panel (a)] recalls the one typically obtained in Ramsey spectroscopy \cite{wang2022}, with $D$ modulating $\omega_{\rm L}$ and thus being the cause of detuning. In both cases, the oscillations as a function of $D$ are damped for values of the waiting time $\tau$ that are much larger than $T_2^*$. 
The FI displays analogous oscillations as a function of $D$ and $\tau$, with a shorter periodicity and a similar damping at long waiting times [panel (c)]. Further details can be appreciated from the sections of the contour plots, corresponding to given values of $\tau$ [panel (e)]. The dependence of the frequency in the oscillations on $\tau$ reflects the fact that the sensitivity of the accumulated phase on the exact value of $\omega_{\rm L}$ (and thus of $D$) increases with $\tau$. This should also imply a monotonic increase of the local maxima with $\tau$, which however does not occur in the presence of decoherence.
The effect of decoherence can be appreciated by comparing these plots with those obtained for larger values of $T_2^*$ and for the coherent case (Appendix~\ref{app:ramseycoh}). 

The difference with respect to the SQD case [panels (b,d,f)] is striking. On the one hand, in the SQD case one hardly observes an oscillatory behavior for either $p_1$ and $F$, because $\omega_{\rm L}$ varies much less than in the case of the DQD in the considered range of values of $D$. On the other hand, the maximal values of the Fisher information in the DQD are two orders of magnitude larger than in the SQD. For optimal values of $\tau$, $F_{\rm Ramsey}$ is roughly 10 times larger than $F_{\rm Rabi}$ for the SQD, while it is 1000 times larger for the DQD. In fact, since $F_{\rm Rabi}$ is similar in both systems, the discriminating factor in $F_{\rm Ramsey}$ is the value of $\omega_{\rm R}$, which we found to be an order of magnitude larger in the DQD with respect to the SQD.
This conclusion results from a simplified expression of the Fisher information, derived for $\Omega T_{\rm Ramsey}\approx \pi/2$ and $| \omega_{\rm L}' \tau| \gg |X'|, |Y'|$. In this case, one has that
\begin{align} 
   F_{\rm Ramsey} & \approx C \, \frac{4\omega_R^2(\omega_L')^2}{\Omega^4} \, (\omega_R\tau)^2  \,.
\label{FI Ramsey approx}
\end{align}
where $C$ is a factor that oscillates as a function of $\Delta$ (and thus of $D$), whose expression in reported in Appendix \ref{app:ramseycoh}. The second factor in the above formula coincides with the approximate expression of $F_{\rm Rabi}$ given in Eq. \eqref{FI Rabi approx}. The third factor increases quadratically with the free evolution time $\tau$. Since $\omega_{\rm R} T_{\rm Ramsey} \approx \pi/2$, when $\tau \gg T_{\rm Ramsey}$ the factor $(\omega_{\rm R}\tau)^2$ might become very large, allowing the maximal values of $F_{\rm Ramsey}$ to be much larger than those of $F_{\rm Rabi}$. On the other hand, $\tau$ cannot be chosen to be large at will, since $C$ also includes an exponentially decreasing dependence on $\tau/T_2^*$, which suppresses $F_{\rm Ramsey}$ at $\tau \gg T_2^*$. Therefore, the optimal value of $\tau$ should represent a compromise between these conflicting requirements.

The strongly oscillating behavior of $F_{\rm Ramsey}$ with respect to both $D$ and $\tau$ should be taken into account in devising the measurement strategy. The oscillations as a function of the charge distance imply no problem if their period is larger than the \textit{a priori} uncertainty on the value of $D$. If this is not the case, one can either enhance the period by reducing the free evolution time $\tau$, or make use of a composite measure, which combines two Ramsey measurements, obtained with different values of the pulse frequency $\omega$, and thus with different positions of the minima. This approach would lead to a smearing of the oscillating features, and thus to an enhancement of the minimal precision, at the cost of reducing the maxima.

\section{Conclusions and outlook}

The purpose of this paper is to explore the potentialities of hole-spin qubits for performing quantum sensing and remote charge sensing. These potentialities ultimately lie in the multiband character of the eigenstates, which can be significantly influenced by the electrostatic environment. This, in turn, affects the coupling of the qubit to the external (electric and magnetic) fields, and thus the qubit state $|\psi\rangle$ that is generated by a given pulse sequence. We have considered different approaches for inferring the presence of a remote charge and, in the case where such presence is known in advance, for precisely estimating its distance $D$ from the qubit. The computed figures of merit are, in the two cases, the discrimination probability and the classical Fisher information, upper bounded by the Helstrom bound and by the quantum Fisher information, respectively.

Beyond the dependence of these quantities on the specific properties of the quantum dots, some general aspects emerge from our investigation. First, the use of double --- rather than single --- quantum dots allows for a significant enhancement of the precision in the estimate. In fact, at the static level, the hole ground state is more polarizable, and thus more sensitive to the position of the external charge, especially in the range where the charge-induced bias is comparable to the interdot tunneling amplitude. Besides, at the dynamic level, when the hole is localized in the double quantum dot, its Larmor and Rabi frequencies, $\omega_{\rm L}$ and $\omega_{\rm R}$, are more sensitive to the charge position. Second, a dynamic approach, where the value of $D$ is encoded in the statistics of a Rabi or a Ramsey measurement, allows for better estimates than any possible static approach where the distance of the charge determines the statistics of an arbitrary measurement performed on the hole ground state. Third, among the dynamic approaches, the Ramsey measurement performs better than the Rabi scheme for quantum estimation, because it allows for more efficient exploitation of the dependence of $\omega_{\rm L}$ on the charge distance, which is stronger than that of $\omega_{\rm R}$. Comparable values of the discrimination probability are obtained instead for the Rabi and Ramsey schemes.

Finally, while the dependence on $D$ of the ground state, $\omega_{\rm L}$, and $\omega_{\rm R}$ is specific to the hole-spin qubit, the analytical expressions of the Fisher information corresponding to the Rabi and Ramsey measurements that we have derived are completely general and apply to any physical implementation of the qubit.

% Possible generalizations of the presented results can be envisaged. These include the inclusion of a multidimensional estimate, where more than one parameter needs to be simultaneously inferred, and less constrained spatial distributions of the point charge. On the other hand, the downside of the sensitivity of the hole-spin qubit to the electrostatic environment is represented by its fragility to electrical noise. The present approach and figures of merit can in principle also be used to investigate and quantify such fragility, and devise strategies for reducing or minimizing the effects of electrical-noise-induced decoherence. 

\acknowledgements

The authors acknowledge financial support from the European Commission, through the project IQubits (Call: H2020--FETOPEN--2018--2019--2020--01, Project ID: 829005), and from the PNRR MUR project PE0000023-NQSTI. Fruitful discussions with M. G. A. Paris are also acknowledged.

\appendix

\section{Numerical evaluation of the QFI via the extended Hellmann-Feynman theorem}
\label{app:QFI}

The evaluation of the QFI for a parameter-dependent state $| \psi_\lambda \rangle$ requires the evaluation of the state derivative, $| \partial_\lambda \psi_\lambda \rangle$. Here, the unknown parameter is the charge distance from the origin ($\lambda=D$), but the discussion that follows is more general. Each state $| \psi_\lambda \rangle$ is obtained by numerically diagonalizing the Hamiltonian for a given value of $\lambda$. The state $| \psi_\lambda \rangle$ is thus given with a global phase factor (gauge) that is fixed randomly, i.e. without any defined or user-controlled relation with the phase assigned to the state $| \psi_{\lambda'} \rangle$, resulting from another diagonalization. In order to compute the derivative $| \partial_\lambda \psi_\lambda \rangle$ via, e.g., the finite-differences method, one should perform two numerical calculations at the values $\lambda \pm \delta \lambda / 2$, for sufficiently small values of $\delta\lambda$, and then fix the gauge in order to remove the random non-derivable global phase. One possible choice would consist in requiring that the scalar product of the state of interest with a fixed reference (basis) state is real. Then, one can compute the derivative
\begin{align}
| \partial_\lambda \psi_\lambda \rangle \approx \frac{ \left| \psi_{\lambda+\delta\lambda / 2} \right> - \left| \psi_{\lambda-\delta\lambda / 2} \right>}{\delta\lambda} \,.
\label{finite difference}
\end{align} 
Although well-defined, this procedure has two drawbacks: (1) it requires two numerical calculations for each value of $\lambda$, and (2) in practice, the result might depend on $\delta \lambda$ and, ultimately, be inaccurate. In fact, if $\delta \lambda$ is too large, then Eq.~\eqref{finite difference} is not a good approximation of the derivative. Instead, if it is too small, the derivative is given by the ratio between two very small quantities, which might be prone to significant numerical noise.

To avoid this problem, we apply the extended Hellmann-Feynman theorem \cite{Singh89}, whose statement can be summarized as follows. Let us consider the Schr\"odinger equation $\hat{\mathcal{H}} | \psi_n \rangle = e_n | \psi_n \rangle$, where the Hamiltonian, its eigenstates and eigenvalues depend on the parameter $\lambda$. Then, the following relations hold:
\begin{align}
\partial_\lambda e_n = \mathcal{H}'_{n,n} \,,
\label{HF1}
\end{align}
and
\begin{align}
 | \partial_\lambda \psi_n \rangle = {\rm i} w_{n} | \psi_n \rangle + \sum_{m \neq n} \frac{\mathcal{H}'_{m,n}}{e_n - e_m}  | \psi_m \rangle \,,
\label{HF2}
\end{align}
where $\mathcal{H}'_{m,n} \equiv \langle \psi_m | ( \partial_\lambda \hat{\mathcal{H}} ) | \psi_n \rangle $, while $w_{n}$ is an undetermined real number. By an appropriate choice of the overall phase of the state $| \psi_n \rangle$, $w_n$ can be set to zero, even though this is not required in what follows.

The advantage of the present approach lies in the fact that it does not require to perform and combine the results of two independent numerical calculations, and avoids the above mentioned drawbacks. In fact, the derivative with respect to $\lambda$ is only applied to the Hamiltonian, which is a known analytical function of the parameter, and can thus be computed by performing calculations for a single value of $\lambda$. In particular, the numerical evaluation of the QFI for the ground state ($n=1$) can be performed by combining Eq.~\eqref{HF2} and Eq. \eqref{QFI formula}. The result is
\begin{align}
H = 4 \sum_{m \neq 1} \frac{\mathcal{H}'_{1,m} \mathcal{H}'_{m,1}}{(e_m-e_1)^2} \,,
\label{QFI after HF}
\end{align}
independent of $w_n$. 
In calculating the QFI with Eq.~\eqref{QFI after HF}, one should only take care that the sum converges. % While it might seem that the convergence of the sum requires convergence on the calculation of the excited states (and thus the expansion of the Hamiltonian on a large basis set), we have verified that in practice only the convergence of the ground state is required. 

\begin{figure}[b]
\centering
\includegraphics[width=0.4\textwidth]{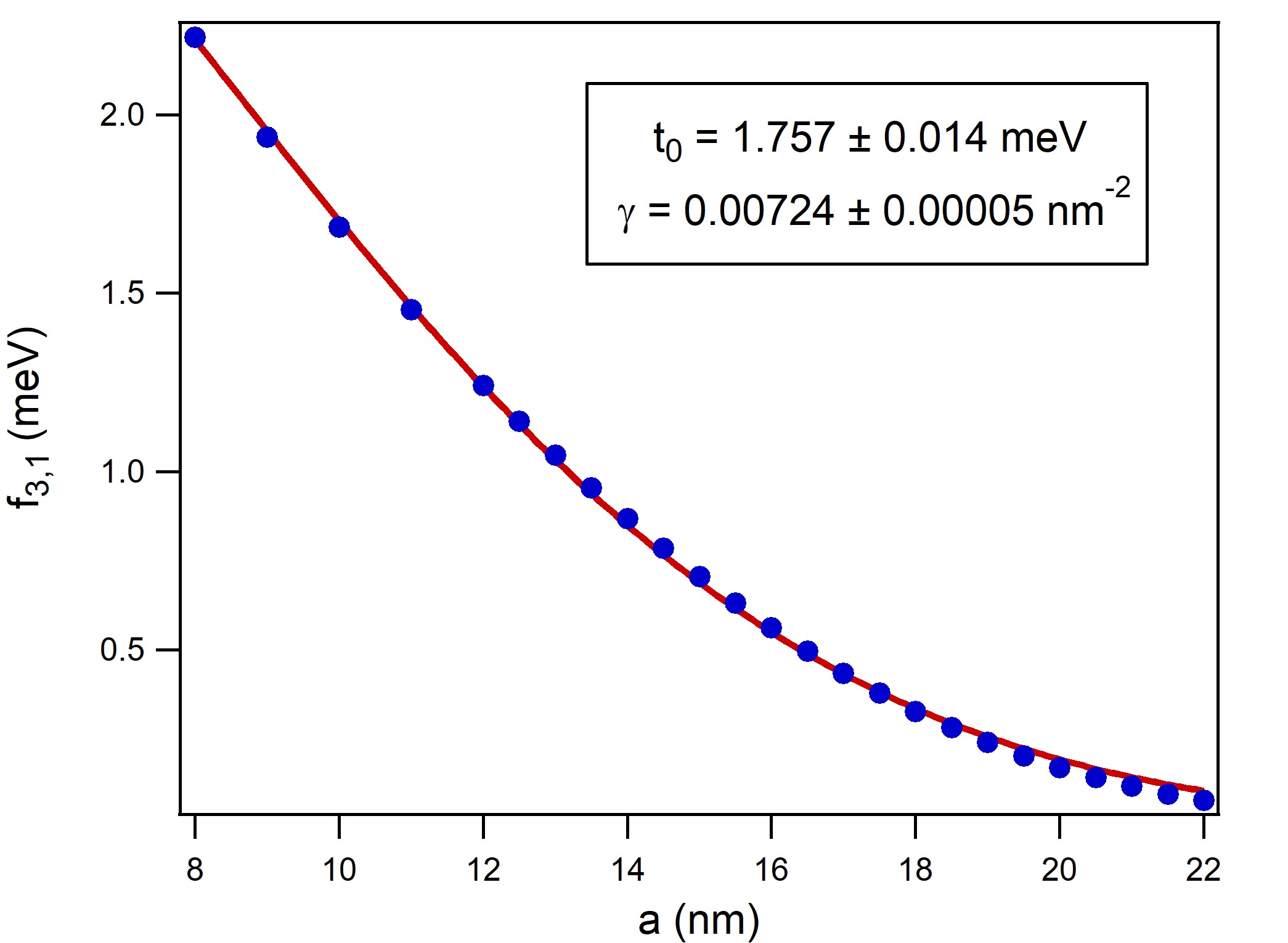}
\caption{Energy gap between the lowest Kramers doublets as a function of the half-interdot distance $a$. The blue dots are the values obtained from the multiband calculations in the absence of an external charge and for zero magnetic field. The red curve corresponds to the fitting function $f_{3,1}(a;t_0,\gamma)$, with values of the parameters specified in the inset.}
\label{Fig:app1}
\end{figure}

In order to compute the QFI for the state generated by the Rabi and Ramsey experiments, one needs to numerically evaluate the derivatives with respect to $D$ of the Larmor and Rabi frequencies. We use the extended Hellmann-Feynman theorem for this purpose as well. The derivative of the Larmor frequency is obtained by applying Eq.~\eqref{HF1}:
\begin{align}
\partial_D \omega_{\rm L} = \frac{1}{\hbar} \left(  \mathcal{H}'_{2,2}   - \mathcal{H}'_{1,1} \right) \,.
\label{larmorder}
\end{align}
The derivative of the Rabi frequency, instead, requires the application of Eq.~\eqref{HF2}, which yields:
\begin{align}
\partial_D \omega_{\rm R} = \frac{\left| \delta E_0 \right| }{\hbar |z_{1,2}|} & {\rm Re} \left( z_{1,2} \sum_{m \neq 2} \frac{\mathcal{H}'_{2,m} }{e_2 - e_m} z_{m,1} \right. \nonumber \\
& \left. \quad \quad + \, z_{2,1} \sum_{m \neq 1} \frac{\mathcal{H}'_{1,m} }{e_1 - e_m} z_{m,2}  \right)  ,
\label{rabider}
\end{align}
where
$
z_{m,n} \equiv \langle \psi_m | \hat{z} | \psi_n \rangle 
$.
This procedure can be applied to the numerical calculation of analogous differential quantities, corresponding to the derivative with respect to $\lambda$ of diagonal or off-diagonal matrix elements.

\section{Tunneling parameter and energy levels in the Hubbard model}
\label{app: tunnelfit}

\begin{figure}[b]
\centering
\includegraphics[width=0.4\textwidth]{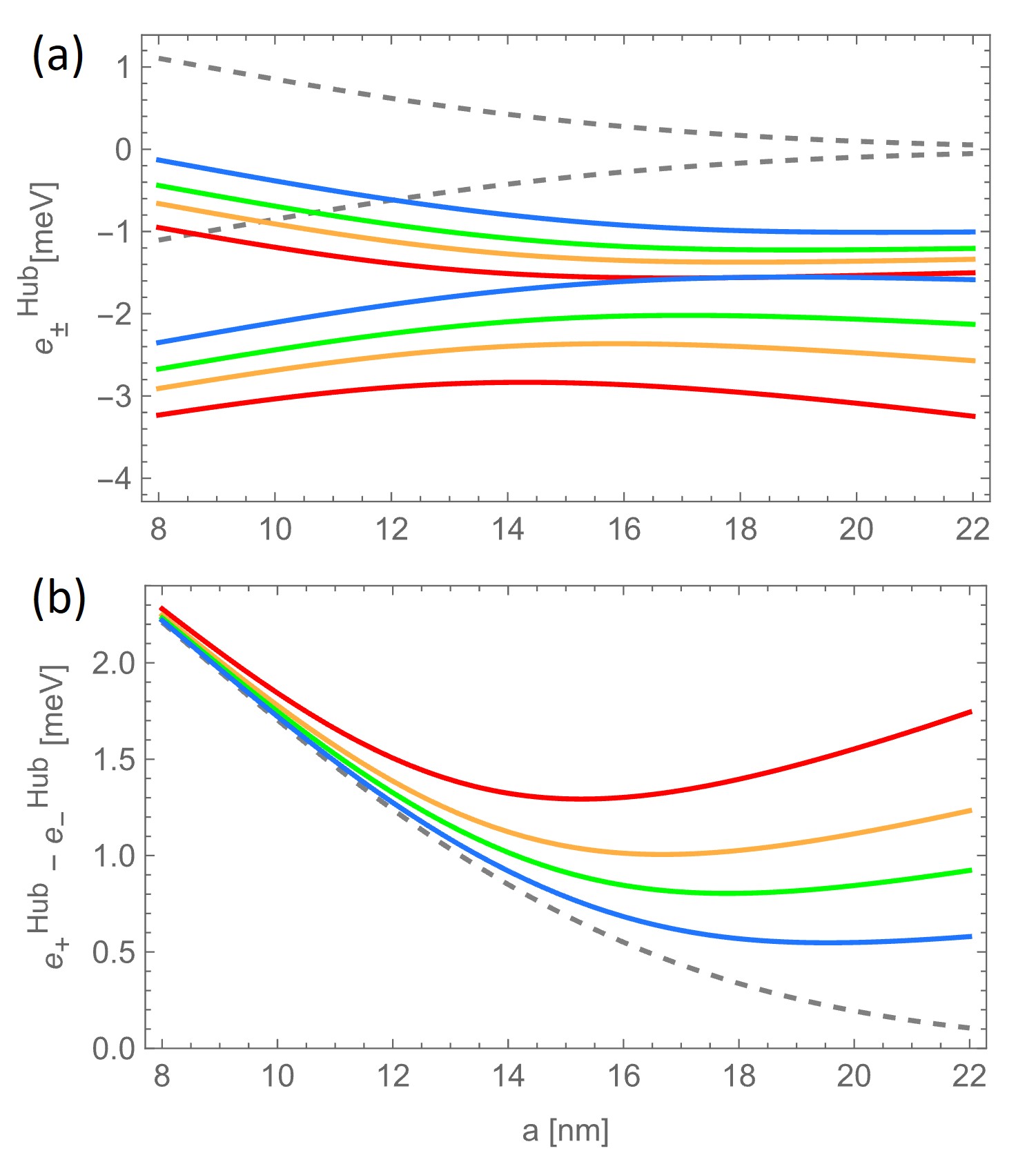}
\caption{(a) Hole energies $e_\pm^{\rm{Hub}}$ and (b) energy gap $e_+^{\rm{Hub}}-e_-^{\rm{Hub}}$ as a function of the half interdot distance $a$. The energies are computed in the absence of an external charge (dashed lines) and with a charge at a distance $D$ of $ 60$ (solid red curves), $70$ (orange), $80$ (green), and $100$ nm (blue).}
\label{Fig:app2}
\end{figure}

In the absence of external charges and magnetic field, the gap $\delta_{3,1} = e_3 - e_1$ between the lowest Kramers doublets is identified with the corresponding energy gap in the solutions of the Hubbard model, given by $2 |t|$ (Subsec.~\ref{subsec: Hubbard}). In general, the tunneling parameter is expected to decrease exponentially with height of the barrier between the two dots, due to the exponential decay of the dot-localized envelope functions on each side of the barrier. For the potential defined in Eq.~\eqref{V DQD}, the barrier height is $V_{\rm DQD}({\bf 0}) = \kappa a^2 / 8$, and thus quadratic in $a$. It is thus reasonable to try and fit the dependence on $a$ of the energy gap obtained from the diagonalization of the $\mathit{k\cdot p}$ Hamiltonian by means of the function
\begin{align}
f_{3,1}(a;t_0,\gamma) = 2 t_0 {\rm e}^{- \gamma a^2} \,,
\label{fit equation}
\end{align}  
where the parameters $t_0\in\mathbb{R}^+$ and $\gamma$ are determined with the nonlinear least-squares method. As shown in Fig.~\ref{Fig:app1}, the fitting can indeed reproduce the dependence of the gap on the half interdot distance with a high degree of accuracy for a significant range of values.

Given the functional dependence of the tunneling parameter on the half interdot distance, one can also analytically express the energy levels and the quantum Fisher information as a function of $a$. The QFI derived from the Hubbard model is shown to reproduce qualitatively the behavior of the one derived from the multiband approach (Sec. \ref{section:SA}). The same applies to the energy levels and to the gap between the ground and first excited levels. 
Figure~\ref{Fig:app2}(a) shows the energies $e_{\pm}^{\rm{Hub}}$ obtained from Eq.~\ref{Hubbard energies} in the presence of an external charge, after identifying the hopping parameter $\delta_{3,1}$ with the $t_0 {\rm e}^{- \gamma a^2}$. The corresponding energy gap $e_+^{\rm{Hub}}-e_-^{\rm{Hub}}$ is shown in Fig.~\ref{Fig:app2}(b), for different values of $D$. The comparison between these results and those reported in Fig.~\ref{Fig:3} shows that the Hubbard model qualitatively reproduces also the behavior of the energy levels obtained from the six-band calculations.

\section{Effect of qubit decoherence on the Rabi and Ramsey measurement}
\label{app:ramseycoh}

The precision that one can achieve both in the Rabi and in the Ramsey scheme is limited by the environment-induced decoherence that affects the time evolution of the qubit state. For the sake of simplicity, we assume here that the environment acts as a depolarizing channel \cite{nielsen}, with depolarization probability $p=1-q$. 
Therefore, if $| \psi \rangle$ is the qubit state in the absence of decoherence, the state density matrix at the end of the pulse sequence reads
\begin{align}\label{eq:DC}
\hat{\rho}  = q\, | \psi \rangle \langle \psi |  + \frac{(1-q)}{2}\, \hat{1} \,,
\end{align}
where the occupation probabilities corresponding to the two basis states are related to those in the absence of decoherence by the relations
\begin{gather}\label{eq:prdec}
    p_k = q\, p_{k,\rm coh} + \frac{1}{2} (1-q)\ \ \ (k=1,2). 
\end{gather}
As a result, the discrimination probability is given by Eq. (\ref{eq:dpwd}). As to the Fisher information, its expression can be derived by combining the above expression with Eq. (\ref{FI qubit}):
\begin{equation}
    F  =  \frac{\left( \chi' \right)^2}{1  - \chi^2   } = 
    \frac{\left( q\,\chi'_{\rm coh} \right)^2}{1  - (q\,\chi_{\rm coh})^2   } \,,
    \label{FI qubit with dephasing}
\end{equation}
where $\chi = p_0 - p_1 = q\,\chi_{\rm coh} $, and the prime denotes the differentiation with respect to $D$.

\begin{figure}
\centering
\includegraphics[width=0.47\textwidth]{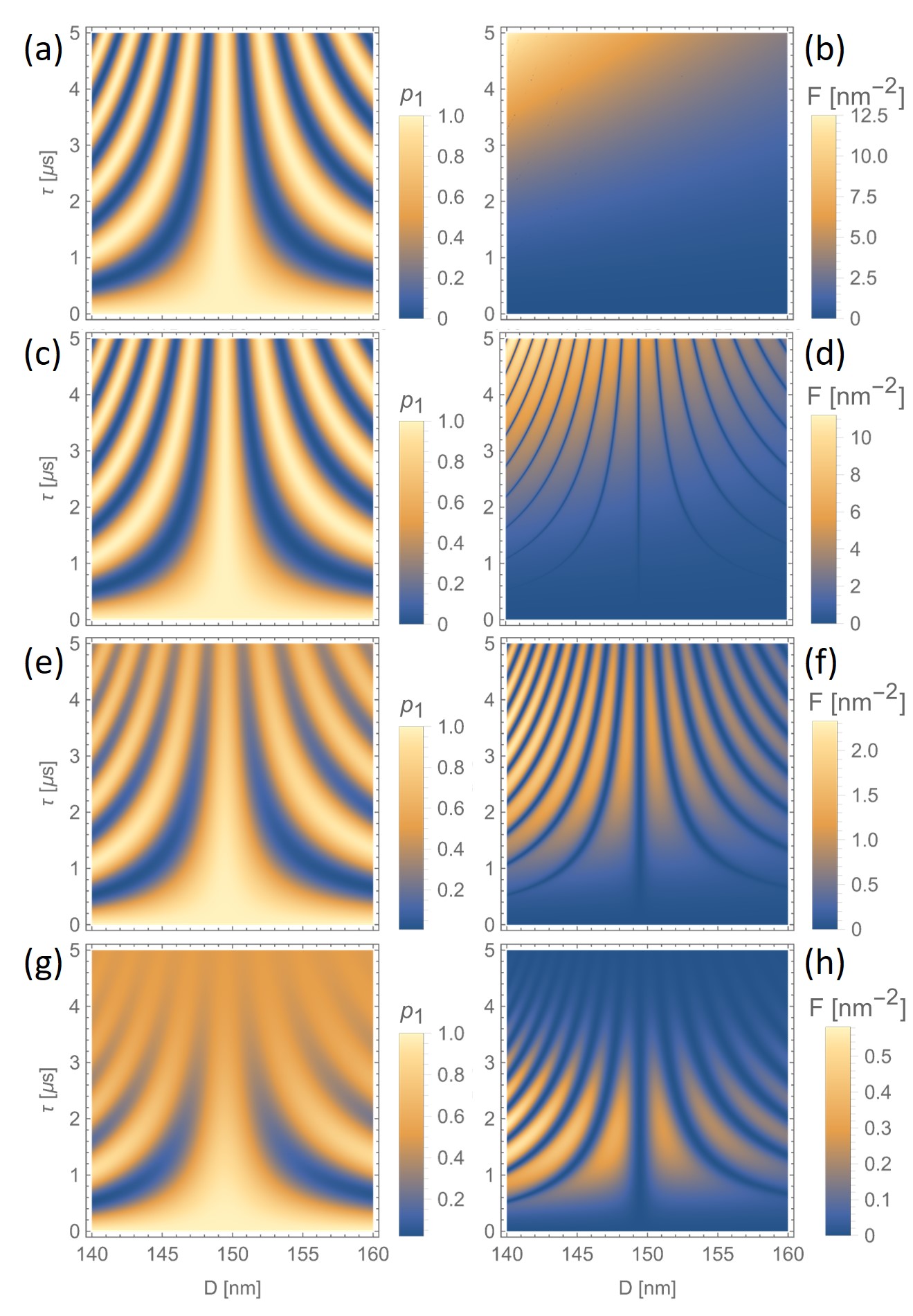}
\caption{Occupation probability $p_1$ (left panels) and corresponding Fisher information $F_{\rm Ramsey}$ (right) as a function of $D$ and $\tau$, in the absence of decoherence (a,b) and for $T_2^*$ equal to $25$ (c,d), $5$ (e,f), and $2.5\,\mu$s (g,h). All the calculations refer to a DQD with half-distance $a=12\,$nm.}
\label{Fig:app3}
\end{figure}

In order to highlight the effect of decoherence on the final occupation probability $p_1$ and on the Fisher information within the Ramsey measurement scheme, we plot these quantities in the case of a fully coherent evolution and for decreasing values of $T_2^*$, for $q=e^{-(T_m/T_d)^2} \approx e^{-(\tau/T_2^*)^2}$ (Fig. \ref{Fig:app3}). All quantities are shown as a function of the charge distance $D$ and the waiting time $\tau$, for a DQD with half-distance $a=12\,$nm, while the pulse duration $T$ is fixed in each case to the corresponding value of $T_{\rm Ramsey}$ (defined in the main text). The probability $p_{1,\rm coh}$ [panel (a)] displays features similar to those obtained in the presence of decoherence [panels (c,e,g)], with a contrast in the oscillations which increases with $T_2^*$. As to the Fisher information, the maxima progressively move from the region of small $\tau$, for the smallest value of the decoherence times [panel (h)], to the one of large $\tau$ in the coherent case [panel (b)]. Correspondingly, the contrast in the oscillations as a function of $D$ decreases, resulting in a qualitative difference between the behaviors in the presence and in the absence of decoherence.

We finally report the expression of the oscillating prefactor that appears in the approximate expression of $F_{\rm Ramsey}$, given in Eq. \eqref{FI Ramsey approx}:
\begin{align}
    C = \frac{ \left[\cos\left(\frac{\Delta \tau}{2}\right) - \sin\left(\frac{\Delta \tau}{2}\right) Y \right]^2 \left[ \cos\left(\frac{\Delta \tau}{2}\right)   Y     + \sin\left(\frac{ \Delta \tau}{2}\right)         \right]^2       }{ {q}^{-2} - 
\left\{ 1-2 X^2 \left[  \cos\left(\frac{\Delta \tau}{2}\right) - \sin\left(\frac{\Delta \tau}{2}\right) Y \right]^2\right\}^2 },
\end{align}
where $\Delta$ is the detuning, $X \equiv \omega_R^X/\Omega$, and $Y \equiv \Delta/\Omega$.

\bibliography{paper}

\begin{thebibliography}{58}%
\makeatletter
\providecommand \@ifxundefined [1]{%
 \@ifx{#1\undefined}
}%
\providecommand \@ifnum [1]{%
 \ifnum #1\expandafter \@firstoftwo
 \else \expandafter \@secondoftwo
 \fi
}%
\providecommand \@ifx [1]{%
 \ifx #1\expandafter \@firstoftwo
 \else \expandafter \@secondoftwo
 \fi
}%
\providecommand \natexlab [1]{#1}%
\providecommand \enquote  [1]{``#1''}%
\providecommand \bibnamefont  [1]{#1}%
\providecommand \bibfnamefont [1]{#1}%
\providecommand \citenamefont [1]{#1}%
\providecommand \href@noop [0]{\@secondoftwo}%
\providecommand \href [0]{\begingroup \@sanitize@url \@href}%
\providecommand \@href[1]{\@@startlink{#1}\@@href}%
\providecommand \@@href[1]{\endgroup#1\@@endlink}%
\providecommand \@sanitize@url [0]{\catcode `\\12\catcode `\$12\catcode
  `\&12\catcode `\#12\catcode `\^12\catcode `\_12\catcode `\%12\relax}%
\providecommand \@@startlink[1]{}%
\providecommand \@@endlink[0]{}%
\providecommand \url  [0]{\begingroup\@sanitize@url \@url }%
\providecommand \@url [1]{\endgroup\@href {#1}{\urlprefix }}%
\providecommand \urlprefix  [0]{URL }%
\providecommand \Eprint [0]{\href }%
\providecommand \doibase [0]{https://doi.org/}%
\providecommand \selectlanguage [0]{\@gobble}%
\providecommand \bibinfo  [0]{\@secondoftwo}%
\providecommand \bibfield  [0]{\@secondoftwo}%
\providecommand \translation [1]{[#1]}%
\providecommand \BibitemOpen [0]{}%
\providecommand \bibitemStop [0]{}%
\providecommand \bibitemNoStop [0]{.\EOS\space}%
\providecommand \EOS [0]{\spacefactor3000\relax}%
\providecommand \BibitemShut  [1]{\csname bibitem#1\endcsname}%
\let\auto@bib@innerbib\@empty
%</preamble>
\bibitem [{\citenamefont {Zwerver}\ \emph {et~al.}(2022)\citenamefont
  {Zwerver}, \citenamefont {Kr{\"a}henmann}, \citenamefont {Watson},
  \citenamefont {Lampert}, \citenamefont {George}, \citenamefont
  {Pillarisetty}, \citenamefont {Bojarski}, \citenamefont {Amin}, \citenamefont
  {Amitonov}, \citenamefont {Boter}, \citenamefont {Caudillo}, \citenamefont
  {Correas-Serrano}, \citenamefont {Dehollain}, \citenamefont {Droulers},
  \citenamefont {Henry}, \citenamefont {Kotlyar}, \citenamefont {Lodari},
  \citenamefont {L{\"u}thi}, \citenamefont {Michalak}, \citenamefont {Mueller},
  \citenamefont {Neyens}, \citenamefont {Roberts}, \citenamefont {Samkharadze},
  \citenamefont {Zheng}, \citenamefont {Zietz}, \citenamefont {Scappucci},
  \citenamefont {Veldhorst}, \citenamefont {Vandersypen},\ and\ \citenamefont
  {Clarke}}]{zwerver2022}%
  \BibitemOpen
  \bibfield  {author} {\bibinfo {author} {\bibfnamefont {A.~M.~J.}\
  \bibnamefont {Zwerver}}, \bibinfo {author} {\bibfnamefont {T.}~\bibnamefont
  {Kr{\"a}henmann}}, \bibinfo {author} {\bibfnamefont {T.~F.}\ \bibnamefont
  {Watson}}, \bibinfo {author} {\bibfnamefont {L.}~\bibnamefont {Lampert}},
  \bibinfo {author} {\bibfnamefont {H.~C.}\ \bibnamefont {George}}, \bibinfo
  {author} {\bibfnamefont {R.}~\bibnamefont {Pillarisetty}}, \bibinfo {author}
  {\bibfnamefont {S.~A.}\ \bibnamefont {Bojarski}}, \bibinfo {author}
  {\bibfnamefont {P.}~\bibnamefont {Amin}}, \bibinfo {author} {\bibfnamefont
  {S.~V.}\ \bibnamefont {Amitonov}}, \bibinfo {author} {\bibfnamefont {J.~M.}\
  \bibnamefont {Boter}}, \bibinfo {author} {\bibfnamefont {R.}~\bibnamefont
  {Caudillo}}, \bibinfo {author} {\bibfnamefont {D.}~\bibnamefont
  {Correas-Serrano}}, \bibinfo {author} {\bibfnamefont {J.~P.}\ \bibnamefont
  {Dehollain}}, \bibinfo {author} {\bibfnamefont {G.}~\bibnamefont {Droulers}},
  \bibinfo {author} {\bibfnamefont {E.~M.}\ \bibnamefont {Henry}}, \bibinfo
  {author} {\bibfnamefont {R.}~\bibnamefont {Kotlyar}}, \bibinfo {author}
  {\bibfnamefont {M.}~\bibnamefont {Lodari}}, \bibinfo {author} {\bibfnamefont
  {F.}~\bibnamefont {L{\"u}thi}}, \bibinfo {author} {\bibfnamefont {D.~J.}\
  \bibnamefont {Michalak}}, \bibinfo {author} {\bibfnamefont {B.~K.}\
  \bibnamefont {Mueller}}, \bibinfo {author} {\bibfnamefont {S.}~\bibnamefont
  {Neyens}}, \bibinfo {author} {\bibfnamefont {J.}~\bibnamefont {Roberts}},
  \bibinfo {author} {\bibfnamefont {N.}~\bibnamefont {Samkharadze}}, \bibinfo
  {author} {\bibfnamefont {G.}~\bibnamefont {Zheng}}, \bibinfo {author}
  {\bibfnamefont {O.~K.}\ \bibnamefont {Zietz}}, \bibinfo {author}
  {\bibfnamefont {G.}~\bibnamefont {Scappucci}}, \bibinfo {author}
  {\bibfnamefont {M.}~\bibnamefont {Veldhorst}}, \bibinfo {author}
  {\bibfnamefont {L.~M.~K.}\ \bibnamefont {Vandersypen}},\ and\ \bibinfo
  {author} {\bibfnamefont {J.~S.}\ \bibnamefont {Clarke}},\ }\bibfield  {title}
  {\bibinfo {title} {Qubits made by advanced semiconductor manufacturing},\
  }\href {https://doi.org/10.1038/s41928-022-00727-9} {\bibfield  {journal}
  {\bibinfo  {journal} {Nature Electronics}\ }\textbf {\bibinfo {volume} {5}},\
  \bibinfo {pages} {184} (\bibinfo {year} {2022})}\BibitemShut {NoStop}%
\bibitem [{\citenamefont {Burkard}\ \emph {et~al.}(2023)\citenamefont
  {Burkard}, \citenamefont {Ladd}, \citenamefont {Pan}, \citenamefont
  {Nichol},\ and\ \citenamefont {Petta}}]{Burkard23a}%
  \BibitemOpen
  \bibfield  {author} {\bibinfo {author} {\bibfnamefont {G.}~\bibnamefont
  {Burkard}}, \bibinfo {author} {\bibfnamefont {T.~D.}\ \bibnamefont {Ladd}},
  \bibinfo {author} {\bibfnamefont {A.}~\bibnamefont {Pan}}, \bibinfo {author}
  {\bibfnamefont {J.~M.}\ \bibnamefont {Nichol}},\ and\ \bibinfo {author}
  {\bibfnamefont {J.~R.}\ \bibnamefont {Petta}},\ }\bibfield  {title} {\bibinfo
  {title} {Semiconductor spin qubits},\ }\href
  {https://doi.org/10.1103/RevModPhys.95.025003} {\bibfield  {journal}
  {\bibinfo  {journal} {Rev. Mod. Phys.}\ }\textbf {\bibinfo {volume} {95}},\
  \bibinfo {pages} {025003} (\bibinfo {year} {2023})}\BibitemShut {NoStop}%
\bibitem [{\citenamefont {Saraiva}\ \emph {et~al.}(2022)\citenamefont
  {Saraiva}, \citenamefont {Lim}, \citenamefont {Yang}, \citenamefont {Escott},
  \citenamefont {Laucht},\ and\ \citenamefont {Dzurak}}]{saraiva2022}%
  \BibitemOpen
  \bibfield  {author} {\bibinfo {author} {\bibfnamefont {A.}~\bibnamefont
  {Saraiva}}, \bibinfo {author} {\bibfnamefont {W.~H.}\ \bibnamefont {Lim}},
  \bibinfo {author} {\bibfnamefont {C.~H.}\ \bibnamefont {Yang}}, \bibinfo
  {author} {\bibfnamefont {C.~C.}\ \bibnamefont {Escott}}, \bibinfo {author}
  {\bibfnamefont {A.}~\bibnamefont {Laucht}},\ and\ \bibinfo {author}
  {\bibfnamefont {A.~S.}\ \bibnamefont {Dzurak}},\ }\bibfield  {title}
  {\bibinfo {title} {Materials for silicon quantum dots and their impact on
  electron spin qubits},\ }\href
  {https://doi.org/https://doi.org/10.1002/adfm.202105488} {\bibfield
  {journal} {\bibinfo  {journal} {Advanced Functional Materials}\ }\textbf
  {\bibinfo {volume} {32}},\ \bibinfo {pages} {2105488} (\bibinfo {year}
  {2022})}\BibitemShut {NoStop}%
\bibitem [{\citenamefont {Chatterjee}\ \emph {et~al.}(2021)\citenamefont
  {Chatterjee}, \citenamefont {Stevenson}, \citenamefont {De~Franceschi},
  \citenamefont {Morello}, \citenamefont {de~Leon},\ and\ \citenamefont
  {Kuemmeth}}]{chatterjee2021}%
  \BibitemOpen
  \bibfield  {author} {\bibinfo {author} {\bibfnamefont {A.}~\bibnamefont
  {Chatterjee}}, \bibinfo {author} {\bibfnamefont {P.}~\bibnamefont
  {Stevenson}}, \bibinfo {author} {\bibfnamefont {S.}~\bibnamefont
  {De~Franceschi}}, \bibinfo {author} {\bibfnamefont {A.}~\bibnamefont
  {Morello}}, \bibinfo {author} {\bibfnamefont {N.~P.}\ \bibnamefont
  {de~Leon}},\ and\ \bibinfo {author} {\bibfnamefont {F.}~\bibnamefont
  {Kuemmeth}},\ }\bibfield  {title} {\bibinfo {title} {Semiconductor qubits in
  practice},\ }\href {https://doi.org/10.1038/s42254-021-00283-9} {\bibfield
  {journal} {\bibinfo  {journal} {Nature Reviews Physics}\ }\textbf {\bibinfo
  {volume} {3}},\ \bibinfo {pages} {157} (\bibinfo {year} {2021})}\BibitemShut
  {NoStop}%
\bibitem [{\citenamefont {Noiri}\ \emph {et~al.}(2022)\citenamefont {Noiri},
  \citenamefont {Takeda}, \citenamefont {Nakajima}, \citenamefont {Kobayashi},
  \citenamefont {Sammak}, \citenamefont {Scappucci},\ and\ \citenamefont
  {Tarucha}}]{noiri2022}%
  \BibitemOpen
  \bibfield  {author} {\bibinfo {author} {\bibfnamefont {A.}~\bibnamefont
  {Noiri}}, \bibinfo {author} {\bibfnamefont {K.}~\bibnamefont {Takeda}},
  \bibinfo {author} {\bibfnamefont {T.}~\bibnamefont {Nakajima}}, \bibinfo
  {author} {\bibfnamefont {T.}~\bibnamefont {Kobayashi}}, \bibinfo {author}
  {\bibfnamefont {A.}~\bibnamefont {Sammak}}, \bibinfo {author} {\bibfnamefont
  {G.}~\bibnamefont {Scappucci}},\ and\ \bibinfo {author} {\bibfnamefont
  {S.}~\bibnamefont {Tarucha}},\ }\bibfield  {title} {\bibinfo {title} {Fast
  universal quantum gate above the fault-tolerance threshold in silicon},\
  }\href {https://doi.org/10.1038/s41586-021-04182-y} {\bibfield  {journal}
  {\bibinfo  {journal} {Nature}\ }\textbf {\bibinfo {volume} {601}},\ \bibinfo
  {pages} {338} (\bibinfo {year} {2022})}\BibitemShut {NoStop}%
\bibitem [{\citenamefont {Yoneda}\ \emph {et~al.}(2018)\citenamefont {Yoneda},
  \citenamefont {Takeda}, \citenamefont {Otsuka}, \citenamefont {Nakajima},
  \citenamefont {Delbecq}, \citenamefont {Allison}, \citenamefont {Honda},
  \citenamefont {Kodera}, \citenamefont {Oda}, \citenamefont {Hoshi},
  \citenamefont {Usami}, \citenamefont {Itoh},\ and\ \citenamefont
  {Tarucha}}]{yoneda2018}%
  \BibitemOpen
  \bibfield  {author} {\bibinfo {author} {\bibfnamefont {J.}~\bibnamefont
  {Yoneda}}, \bibinfo {author} {\bibfnamefont {K.}~\bibnamefont {Takeda}},
  \bibinfo {author} {\bibfnamefont {T.}~\bibnamefont {Otsuka}}, \bibinfo
  {author} {\bibfnamefont {T.}~\bibnamefont {Nakajima}}, \bibinfo {author}
  {\bibfnamefont {M.~R.}\ \bibnamefont {Delbecq}}, \bibinfo {author}
  {\bibfnamefont {G.}~\bibnamefont {Allison}}, \bibinfo {author} {\bibfnamefont
  {T.}~\bibnamefont {Honda}}, \bibinfo {author} {\bibfnamefont
  {T.}~\bibnamefont {Kodera}}, \bibinfo {author} {\bibfnamefont
  {S.}~\bibnamefont {Oda}}, \bibinfo {author} {\bibfnamefont {Y.}~\bibnamefont
  {Hoshi}}, \bibinfo {author} {\bibfnamefont {N.}~\bibnamefont {Usami}},
  \bibinfo {author} {\bibfnamefont {K.~M.}\ \bibnamefont {Itoh}},\ and\
  \bibinfo {author} {\bibfnamefont {S.}~\bibnamefont {Tarucha}},\ }\bibfield
  {title} {\bibinfo {title} {A quantum-dot spin qubit with coherence limited by
  charge noise and fidelity higher than 99.9{\%}},\ }\href
  {https://doi.org/10.1038/s41565-017-0014-x} {\bibfield  {journal} {\bibinfo
  {journal} {Nature Nanotechnology}\ }\textbf {\bibinfo {volume} {13}},\
  \bibinfo {pages} {102} (\bibinfo {year} {2018})}\BibitemShut {NoStop}%
\bibitem [{\citenamefont {Fern\'andez-Fern\'andez}\ \emph
  {et~al.}(2022)\citenamefont {Fern\'andez-Fern\'andez}, \citenamefont {Ban},\
  and\ \citenamefont {Platero}}]{fernandez2022}%
  \BibitemOpen
  \bibfield  {author} {\bibinfo {author} {\bibfnamefont {D.}~\bibnamefont
  {Fern\'andez-Fern\'andez}}, \bibinfo {author} {\bibfnamefont
  {Y.}~\bibnamefont {Ban}},\ and\ \bibinfo {author} {\bibfnamefont
  {G.}~\bibnamefont {Platero}},\ }\bibfield  {title} {\bibinfo {title} {Quantum
  control of hole spin qubits in double quantum dots},\ }\href
  {https://doi.org/10.1103/PhysRevApplied.18.054090} {\bibfield  {journal}
  {\bibinfo  {journal} {Phys. Rev. Appl.}\ }\textbf {\bibinfo {volume} {18}},\
  \bibinfo {pages} {054090} (\bibinfo {year} {2022})}\BibitemShut {NoStop}%
\bibitem [{\citenamefont {Xue}\ \emph {et~al.}(2022)\citenamefont {Xue},
  \citenamefont {Russ}, \citenamefont {Samkharadze}, \citenamefont {Undseth},
  \citenamefont {Sammak}, \citenamefont {Scappucci},\ and\ \citenamefont
  {Vandersypen}}]{xue2022}%
  \BibitemOpen
  \bibfield  {author} {\bibinfo {author} {\bibfnamefont {X.}~\bibnamefont
  {Xue}}, \bibinfo {author} {\bibfnamefont {M.}~\bibnamefont {Russ}}, \bibinfo
  {author} {\bibfnamefont {N.}~\bibnamefont {Samkharadze}}, \bibinfo {author}
  {\bibfnamefont {B.}~\bibnamefont {Undseth}}, \bibinfo {author} {\bibfnamefont
  {A.}~\bibnamefont {Sammak}}, \bibinfo {author} {\bibfnamefont
  {G.}~\bibnamefont {Scappucci}},\ and\ \bibinfo {author} {\bibfnamefont
  {L.~M.~K.}\ \bibnamefont {Vandersypen}},\ }\bibfield  {title} {\bibinfo
  {title} {Quantum logic with spin qubits crossing the surface code
  threshold},\ }\href {https://doi.org/10.1038/s41586-021-04273-w} {\bibfield
  {journal} {\bibinfo  {journal} {Nature}\ }\textbf {\bibinfo {volume} {601}},\
  \bibinfo {pages} {343} (\bibinfo {year} {2022})}\BibitemShut {NoStop}%
\bibitem [{\citenamefont {Lawrie}\ \emph {et~al.}(2023)\citenamefont {Lawrie},
  \citenamefont {Rimbach-Russ}, \citenamefont {Riggelen}, \citenamefont
  {Hendrickx}, \citenamefont {Snoo}, \citenamefont {Sammak}, \citenamefont
  {Scappucci}, \citenamefont {Helsen},\ and\ \citenamefont
  {Veldhorst}}]{lawrie2021}%
  \BibitemOpen
  \bibfield  {author} {\bibinfo {author} {\bibfnamefont {W.~I.~L.}\
  \bibnamefont {Lawrie}}, \bibinfo {author} {\bibfnamefont {M.}~\bibnamefont
  {Rimbach-Russ}}, \bibinfo {author} {\bibfnamefont {F.~v.}\ \bibnamefont
  {Riggelen}}, \bibinfo {author} {\bibfnamefont {N.~W.}\ \bibnamefont
  {Hendrickx}}, \bibinfo {author} {\bibfnamefont {S.~L.~d.}\ \bibnamefont
  {Snoo}}, \bibinfo {author} {\bibfnamefont {A.}~\bibnamefont {Sammak}},
  \bibinfo {author} {\bibfnamefont {G.}~\bibnamefont {Scappucci}}, \bibinfo
  {author} {\bibfnamefont {J.}~\bibnamefont {Helsen}},\ and\ \bibinfo {author}
  {\bibfnamefont {M.}~\bibnamefont {Veldhorst}},\ }\bibfield  {title} {\bibinfo
  {title} {Simultaneous single-qubit driving of semiconductor spin qubits at
  the fault-tolerant threshold},\ }\href
  {https://doi.org/10.1038/s41467-023-39334-3} {\bibfield  {journal} {\bibinfo
  {journal} {Nature Communications}\ }\textbf {\bibinfo {volume} {14}},\
  \bibinfo {pages} {3617} (\bibinfo {year} {2023})}\BibitemShut {NoStop}%
\bibitem [{\citenamefont {Mills}\ \emph
  {et~al.}(2022{\natexlab{a}})\citenamefont {Mills}, \citenamefont {Guinn},
  \citenamefont {Gullans}, \citenamefont {Sigillito}, \citenamefont {Feldman},
  \citenamefont {Nielsen},\ and\ \citenamefont {Petta}}]{mills2022}%
  \BibitemOpen
  \bibfield  {author} {\bibinfo {author} {\bibfnamefont {A.~R.}\ \bibnamefont
  {Mills}}, \bibinfo {author} {\bibfnamefont {C.~R.}\ \bibnamefont {Guinn}},
  \bibinfo {author} {\bibfnamefont {M.~J.}\ \bibnamefont {Gullans}}, \bibinfo
  {author} {\bibfnamefont {A.~J.}\ \bibnamefont {Sigillito}}, \bibinfo {author}
  {\bibfnamefont {M.~M.}\ \bibnamefont {Feldman}}, \bibinfo {author}
  {\bibfnamefont {E.}~\bibnamefont {Nielsen}},\ and\ \bibinfo {author}
  {\bibfnamefont {J.~R.}\ \bibnamefont {Petta}},\ }\bibfield  {title} {\bibinfo
  {title} {Two-qubit silicon quantum processor with operation fidelity
  exceeding 99\%},\ }\href {https://doi.org/10.1126/sciadv.abn5130} {\bibfield
  {journal} {\bibinfo  {journal} {Science Advances}\ }\textbf {\bibinfo
  {volume} {8}},\ \bibinfo {pages} {eabn5130} (\bibinfo {year}
  {2022}{\natexlab{a}})}\BibitemShut {NoStop}%
\bibitem [{\citenamefont {Mills}\ \emph
  {et~al.}(2022{\natexlab{b}})\citenamefont {Mills}, \citenamefont {Guinn},
  \citenamefont {Feldman}, \citenamefont {Sigillito}, \citenamefont {Gullans},
  \citenamefont {Rakher}, \citenamefont {Kerckhoff}, \citenamefont {Jackson},\
  and\ \citenamefont {Petta}}]{mills2022-2}%
  \BibitemOpen
  \bibfield  {author} {\bibinfo {author} {\bibfnamefont {A.}~\bibnamefont
  {Mills}}, \bibinfo {author} {\bibfnamefont {C.}~\bibnamefont {Guinn}},
  \bibinfo {author} {\bibfnamefont {M.}~\bibnamefont {Feldman}}, \bibinfo
  {author} {\bibfnamefont {A.}~\bibnamefont {Sigillito}}, \bibinfo {author}
  {\bibfnamefont {M.}~\bibnamefont {Gullans}}, \bibinfo {author} {\bibfnamefont
  {M.}~\bibnamefont {Rakher}}, \bibinfo {author} {\bibfnamefont
  {J.}~\bibnamefont {Kerckhoff}}, \bibinfo {author} {\bibfnamefont
  {C.}~\bibnamefont {Jackson}},\ and\ \bibinfo {author} {\bibfnamefont
  {J.}~\bibnamefont {Petta}},\ }\bibfield  {title} {\bibinfo {title}
  {High-fidelity state preparation, quantum control, and readout of an
  isotopically enriched silicon spin qubit},\ }\href
  {https://doi.org/10.1103/PhysRevApplied.18.064028} {\bibfield  {journal}
  {\bibinfo  {journal} {Phys. Rev. Appl.}\ }\textbf {\bibinfo {volume} {18}},\
  \bibinfo {pages} {064028} (\bibinfo {year} {2022}{\natexlab{b}})}\BibitemShut
  {NoStop}%
\bibitem [{\citenamefont {Yang}\ \emph {et~al.}(2019)\citenamefont {Yang},
  \citenamefont {Chan}, \citenamefont {Harper}, \citenamefont {Huang},
  \citenamefont {Evans}, \citenamefont {Hwang}, \citenamefont {Hensen},
  \citenamefont {Laucht}, \citenamefont {Tanttu}, \citenamefont {Hudson},
  \citenamefont {Flammia}, \citenamefont {Itoh}, \citenamefont {Morello},
  \citenamefont {Bartlett},\ and\ \citenamefont {Dzurak}}]{yang2019}%
  \BibitemOpen
  \bibfield  {author} {\bibinfo {author} {\bibfnamefont {C.~H.}\ \bibnamefont
  {Yang}}, \bibinfo {author} {\bibfnamefont {K.~W.}\ \bibnamefont {Chan}},
  \bibinfo {author} {\bibfnamefont {R.}~\bibnamefont {Harper}}, \bibinfo
  {author} {\bibfnamefont {W.}~\bibnamefont {Huang}}, \bibinfo {author}
  {\bibfnamefont {T.}~\bibnamefont {Evans}}, \bibinfo {author} {\bibfnamefont
  {J.~C.~C.}\ \bibnamefont {Hwang}}, \bibinfo {author} {\bibfnamefont
  {B.}~\bibnamefont {Hensen}}, \bibinfo {author} {\bibfnamefont
  {A.}~\bibnamefont {Laucht}}, \bibinfo {author} {\bibfnamefont
  {T.}~\bibnamefont {Tanttu}}, \bibinfo {author} {\bibfnamefont {F.~E.}\
  \bibnamefont {Hudson}}, \bibinfo {author} {\bibfnamefont {S.~T.}\
  \bibnamefont {Flammia}}, \bibinfo {author} {\bibfnamefont {K.~M.}\
  \bibnamefont {Itoh}}, \bibinfo {author} {\bibfnamefont {A.}~\bibnamefont
  {Morello}}, \bibinfo {author} {\bibfnamefont {S.~D.}\ \bibnamefont
  {Bartlett}},\ and\ \bibinfo {author} {\bibfnamefont {A.~S.}\ \bibnamefont
  {Dzurak}},\ }\bibfield  {title} {\bibinfo {title} {Silicon qubit fidelities
  approaching incoherent noise limits via pulse engineering},\ }\href
  {https://doi.org/10.1038/s41928-019-0234-1} {\bibfield  {journal} {\bibinfo
  {journal} {Nature Electronics}\ }\textbf {\bibinfo {volume} {2}},\ \bibinfo
  {pages} {151} (\bibinfo {year} {2019})}\BibitemShut {NoStop}%
\bibitem [{\citenamefont {Zheng}\ \emph {et~al.}(2019)\citenamefont {Zheng},
  \citenamefont {Samkharadze}, \citenamefont {Noordam}, \citenamefont {Kalhor},
  \citenamefont {Brousse}, \citenamefont {Sammak}, \citenamefont {Scappucci},\
  and\ \citenamefont {Vandersypen}}]{Zheng2019a}%
  \BibitemOpen
  \bibfield  {author} {\bibinfo {author} {\bibfnamefont {G.}~\bibnamefont
  {Zheng}}, \bibinfo {author} {\bibfnamefont {N.}~\bibnamefont {Samkharadze}},
  \bibinfo {author} {\bibfnamefont {M.~L.}\ \bibnamefont {Noordam}}, \bibinfo
  {author} {\bibfnamefont {N.}~\bibnamefont {Kalhor}}, \bibinfo {author}
  {\bibfnamefont {D.}~\bibnamefont {Brousse}}, \bibinfo {author} {\bibfnamefont
  {A.}~\bibnamefont {Sammak}}, \bibinfo {author} {\bibfnamefont
  {G.}~\bibnamefont {Scappucci}},\ and\ \bibinfo {author} {\bibfnamefont
  {L.~M.~K.}\ \bibnamefont {Vandersypen}},\ }\bibfield  {title} {\bibinfo
  {title} {Rapid gate-based spin read-out in silicon using an on-chip
  resonator},\ }\href {https://doi.org/10.1038/s41565-019-0488-9} {\bibfield
  {journal} {\bibinfo  {journal} {Nature Nanotechnology}\ }\textbf {\bibinfo
  {volume} {14}},\ \bibinfo {pages} {742} (\bibinfo {year} {2019})}\BibitemShut
  {NoStop}%
\bibitem [{\citenamefont {Boter}\ \emph {et~al.}(2022)\citenamefont {Boter},
  \citenamefont {Dehollain}, \citenamefont {van Dijk}, \citenamefont {Xu},
  \citenamefont {Hensgens}, \citenamefont {Versluis}, \citenamefont {Naus},
  \citenamefont {Clarke}, \citenamefont {Veldhorst}, \citenamefont
  {Sebastiano},\ and\ \citenamefont {Vandersypen}}]{boter2022}%
  \BibitemOpen
  \bibfield  {author} {\bibinfo {author} {\bibfnamefont {J.~M.}\ \bibnamefont
  {Boter}}, \bibinfo {author} {\bibfnamefont {J.~P.}\ \bibnamefont
  {Dehollain}}, \bibinfo {author} {\bibfnamefont {J.~P.}\ \bibnamefont {van
  Dijk}}, \bibinfo {author} {\bibfnamefont {Y.}~\bibnamefont {Xu}}, \bibinfo
  {author} {\bibfnamefont {T.}~\bibnamefont {Hensgens}}, \bibinfo {author}
  {\bibfnamefont {R.}~\bibnamefont {Versluis}}, \bibinfo {author}
  {\bibfnamefont {H.~W.}\ \bibnamefont {Naus}}, \bibinfo {author}
  {\bibfnamefont {J.~S.}\ \bibnamefont {Clarke}}, \bibinfo {author}
  {\bibfnamefont {M.}~\bibnamefont {Veldhorst}}, \bibinfo {author}
  {\bibfnamefont {F.}~\bibnamefont {Sebastiano}},\ and\ \bibinfo {author}
  {\bibfnamefont {L.~M.}\ \bibnamefont {Vandersypen}},\ }\bibfield  {title}
  {\bibinfo {title} {Spiderweb array: A sparse spin-qubit array},\ }\href
  {https://doi.org/10.1103/PhysRevApplied.18.024053} {\bibfield  {journal}
  {\bibinfo  {journal} {Phys. Rev. Appl.}\ }\textbf {\bibinfo {volume} {18}},\
  \bibinfo {pages} {024053} (\bibinfo {year} {2022})}\BibitemShut {NoStop}%
\bibitem [{\citenamefont {Philips}\ \emph {et~al.}(2022)\citenamefont
  {Philips}, \citenamefont {Madzik}, \citenamefont {Amitonov}, \citenamefont
  {de~Snoo}, \citenamefont {Russ}, \citenamefont {Kalhor}, \citenamefont
  {Volk}, \citenamefont {Lawrie}, \citenamefont {Brousse}, \citenamefont
  {Tryputen}, \citenamefont {Wuetz}, \citenamefont {Sammak}, \citenamefont
  {Veldhorst}, \citenamefont {Scappucci},\ and\ \citenamefont
  {Vandersypen}}]{philips2022}%
  \BibitemOpen
  \bibfield  {author} {\bibinfo {author} {\bibfnamefont {S.~G.~J.}\
  \bibnamefont {Philips}}, \bibinfo {author} {\bibfnamefont {M.~T.}\
  \bibnamefont {Madzik}}, \bibinfo {author} {\bibfnamefont {S.~V.}\
  \bibnamefont {Amitonov}}, \bibinfo {author} {\bibfnamefont {S.~L.}\
  \bibnamefont {de~Snoo}}, \bibinfo {author} {\bibfnamefont {M.}~\bibnamefont
  {Russ}}, \bibinfo {author} {\bibfnamefont {N.}~\bibnamefont {Kalhor}},
  \bibinfo {author} {\bibfnamefont {C.}~\bibnamefont {Volk}}, \bibinfo {author}
  {\bibfnamefont {W.~I.~L.}\ \bibnamefont {Lawrie}}, \bibinfo {author}
  {\bibfnamefont {D.}~\bibnamefont {Brousse}}, \bibinfo {author} {\bibfnamefont
  {L.}~\bibnamefont {Tryputen}}, \bibinfo {author} {\bibfnamefont {B.~P.}\
  \bibnamefont {Wuetz}}, \bibinfo {author} {\bibfnamefont {A.}~\bibnamefont
  {Sammak}}, \bibinfo {author} {\bibfnamefont {M.}~\bibnamefont {Veldhorst}},
  \bibinfo {author} {\bibfnamefont {G.}~\bibnamefont {Scappucci}},\ and\
  \bibinfo {author} {\bibfnamefont {L.~M.~K.}\ \bibnamefont {Vandersypen}},\
  }\bibfield  {title} {\bibinfo {title} {Universal control of a six-qubit
  quantum processor in silicon},\ }\href
  {https://doi.org/10.1038/s41586-022-05117-x} {\bibfield  {journal} {\bibinfo
  {journal} {Nature}\ }\textbf {\bibinfo {volume} {609}},\ \bibinfo {pages}
  {919} (\bibinfo {year} {2022})}\BibitemShut {NoStop}%
\bibitem [{\citenamefont {Geyer}\ \emph {et~al.}(2021)\citenamefont {Geyer},
  \citenamefont {Camenzind}, \citenamefont {Czornomaz}, \citenamefont
  {Deshpande}, \citenamefont {Fuhrer}, \citenamefont {Warburton}, \citenamefont
  {Zumbühl},\ and\ \citenamefont {Kuhlmann}}]{geyer2021}%
  \BibitemOpen
  \bibfield  {author} {\bibinfo {author} {\bibfnamefont {S.}~\bibnamefont
  {Geyer}}, \bibinfo {author} {\bibfnamefont {L.~C.}\ \bibnamefont
  {Camenzind}}, \bibinfo {author} {\bibfnamefont {L.}~\bibnamefont
  {Czornomaz}}, \bibinfo {author} {\bibfnamefont {V.}~\bibnamefont
  {Deshpande}}, \bibinfo {author} {\bibfnamefont {A.}~\bibnamefont {Fuhrer}},
  \bibinfo {author} {\bibfnamefont {R.~J.}\ \bibnamefont {Warburton}}, \bibinfo
  {author} {\bibfnamefont {D.~M.}\ \bibnamefont {Zumbühl}},\ and\ \bibinfo
  {author} {\bibfnamefont {A.~V.}\ \bibnamefont {Kuhlmann}},\ }\bibfield
  {title} {\bibinfo {title} {Self-aligned gates for scalable silicon quantum
  computing},\ }\href {https://doi.org/10.1063/5.0036520} {\bibfield  {journal}
  {\bibinfo  {journal} {Applied Physics Letters}\ }\textbf {\bibinfo {volume}
  {118}},\ \bibinfo {pages} {104004} (\bibinfo {year} {2021})}\BibitemShut
  {NoStop}%
\bibitem [{\citenamefont {Kobayashi}\ \emph {et~al.}(2021)\citenamefont
  {Kobayashi}, \citenamefont {Salfi}, \citenamefont {Chua}, \citenamefont
  {van~der Heijden}, \citenamefont {House}, \citenamefont {Culcer},
  \citenamefont {Hutchison}, \citenamefont {Johnson}, \citenamefont {McCallum},
  \citenamefont {Riemann}, \citenamefont {Abrosimov}, \citenamefont {Becker},
  \citenamefont {Pohl}, \citenamefont {Simmons},\ and\ \citenamefont
  {Rogge}}]{kobayashi2021}%
  \BibitemOpen
  \bibfield  {author} {\bibinfo {author} {\bibfnamefont {T.}~\bibnamefont
  {Kobayashi}}, \bibinfo {author} {\bibfnamefont {J.}~\bibnamefont {Salfi}},
  \bibinfo {author} {\bibfnamefont {C.}~\bibnamefont {Chua}}, \bibinfo {author}
  {\bibfnamefont {J.}~\bibnamefont {van~der Heijden}}, \bibinfo {author}
  {\bibfnamefont {M.~G.}\ \bibnamefont {House}}, \bibinfo {author}
  {\bibfnamefont {D.}~\bibnamefont {Culcer}}, \bibinfo {author} {\bibfnamefont
  {W.~D.}\ \bibnamefont {Hutchison}}, \bibinfo {author} {\bibfnamefont {B.~C.}\
  \bibnamefont {Johnson}}, \bibinfo {author} {\bibfnamefont {J.~C.}\
  \bibnamefont {McCallum}}, \bibinfo {author} {\bibfnamefont {H.}~\bibnamefont
  {Riemann}}, \bibinfo {author} {\bibfnamefont {N.~V.}\ \bibnamefont
  {Abrosimov}}, \bibinfo {author} {\bibfnamefont {P.}~\bibnamefont {Becker}},
  \bibinfo {author} {\bibfnamefont {H.-J.}\ \bibnamefont {Pohl}}, \bibinfo
  {author} {\bibfnamefont {M.~Y.}\ \bibnamefont {Simmons}},\ and\ \bibinfo
  {author} {\bibfnamefont {S.}~\bibnamefont {Rogge}},\ }\bibfield  {title}
  {\bibinfo {title} {Engineering long spin coherence times of spin--orbit
  qubits in silicon},\ }\href {https://doi.org/10.1038/s41563-020-0743-3}
  {\bibfield  {journal} {\bibinfo  {journal} {Nature Materials}\ }\textbf
  {\bibinfo {volume} {20}},\ \bibinfo {pages} {38} (\bibinfo {year}
  {2021})}\BibitemShut {NoStop}%
\bibitem [{\citenamefont {Piot}\ \emph {et~al.}(2022)\citenamefont {Piot},
  \citenamefont {Brun}, \citenamefont {Schmitt}, \citenamefont {Zihlmann},
  \citenamefont {Michal}, \citenamefont {Apra}, \citenamefont {Abadillo-Uriel},
  \citenamefont {Jehl}, \citenamefont {Bertrand}, \citenamefont {Niebojewski},
  \citenamefont {Hutin}, \citenamefont {Vinet}, \citenamefont {Urdampilleta},
  \citenamefont {Meunier}, \citenamefont {Niquet}, \citenamefont {Maurand},\
  and\ \citenamefont {Franceschi}}]{piot2022}%
  \BibitemOpen
  \bibfield  {author} {\bibinfo {author} {\bibfnamefont {N.}~\bibnamefont
  {Piot}}, \bibinfo {author} {\bibfnamefont {B.}~\bibnamefont {Brun}}, \bibinfo
  {author} {\bibfnamefont {V.}~\bibnamefont {Schmitt}}, \bibinfo {author}
  {\bibfnamefont {S.}~\bibnamefont {Zihlmann}}, \bibinfo {author}
  {\bibfnamefont {V.~P.}\ \bibnamefont {Michal}}, \bibinfo {author}
  {\bibfnamefont {A.}~\bibnamefont {Apra}}, \bibinfo {author} {\bibfnamefont
  {J.~C.}\ \bibnamefont {Abadillo-Uriel}}, \bibinfo {author} {\bibfnamefont
  {X.}~\bibnamefont {Jehl}}, \bibinfo {author} {\bibfnamefont {B.}~\bibnamefont
  {Bertrand}}, \bibinfo {author} {\bibfnamefont {H.}~\bibnamefont
  {Niebojewski}}, \bibinfo {author} {\bibfnamefont {L.}~\bibnamefont {Hutin}},
  \bibinfo {author} {\bibfnamefont {M.}~\bibnamefont {Vinet}}, \bibinfo
  {author} {\bibfnamefont {M.}~\bibnamefont {Urdampilleta}}, \bibinfo {author}
  {\bibfnamefont {T.}~\bibnamefont {Meunier}}, \bibinfo {author} {\bibfnamefont
  {Y.-M.}\ \bibnamefont {Niquet}}, \bibinfo {author} {\bibfnamefont
  {R.}~\bibnamefont {Maurand}},\ and\ \bibinfo {author} {\bibfnamefont {S.~D.}\
  \bibnamefont {Franceschi}},\ }\bibfield  {title} {\bibinfo {title} {A single
  hole spin with enhanced coherence in natural silicon},\ }\href
  {https://doi.org/10.1038/s41565-022-01196-z} {\bibfield  {journal} {\bibinfo
  {journal} {Nature Nanotechnology}\ }\textbf {\bibinfo {volume} {17}},\
  \bibinfo {pages} {1072} (\bibinfo {year} {2022})}\BibitemShut {NoStop}%
\bibitem [{\citenamefont {Stano}\ and\ \citenamefont {Loss}(2022)}]{stano2022}%
  \BibitemOpen
  \bibfield  {author} {\bibinfo {author} {\bibfnamefont {P.}~\bibnamefont
  {Stano}}\ and\ \bibinfo {author} {\bibfnamefont {D.}~\bibnamefont {Loss}},\
  }\bibfield  {title} {\bibinfo {title} {Review of performance metrics of spin
  qubits in gated semiconducting nanostructures},\ }\href
  {https://doi.org/10.1038/s42254-022-00484-w} {\bibfield  {journal} {\bibinfo
  {journal} {Nature Reviews Physics}\ }\textbf {\bibinfo {volume} {4}},\
  \bibinfo {pages} {672} (\bibinfo {year} {2022})}\BibitemShut {NoStop}%
\bibitem [{\citenamefont {Zwanenburg}\ \emph {et~al.}(2013)\citenamefont
  {Zwanenburg}, \citenamefont {Dzurak}, \citenamefont {Morello}, \citenamefont
  {Simmons}, \citenamefont {Hollenberg}, \citenamefont {Klimeck}, \citenamefont
  {Rogge}, \citenamefont {Coppersmith},\ and\ \citenamefont
  {Eriksson}}]{zwanenburg2013}%
  \BibitemOpen
  \bibfield  {author} {\bibinfo {author} {\bibfnamefont {F.~A.}\ \bibnamefont
  {Zwanenburg}}, \bibinfo {author} {\bibfnamefont {A.~S.}\ \bibnamefont
  {Dzurak}}, \bibinfo {author} {\bibfnamefont {A.}~\bibnamefont {Morello}},
  \bibinfo {author} {\bibfnamefont {M.~Y.}\ \bibnamefont {Simmons}}, \bibinfo
  {author} {\bibfnamefont {L.~C.~L.}\ \bibnamefont {Hollenberg}}, \bibinfo
  {author} {\bibfnamefont {G.}~\bibnamefont {Klimeck}}, \bibinfo {author}
  {\bibfnamefont {S.}~\bibnamefont {Rogge}}, \bibinfo {author} {\bibfnamefont
  {S.~N.}\ \bibnamefont {Coppersmith}},\ and\ \bibinfo {author} {\bibfnamefont
  {M.~A.}\ \bibnamefont {Eriksson}},\ }\bibfield  {title} {\bibinfo {title}
  {Silicon quantum electronics},\ }\href
  {https://doi.org/10.1103/RevModPhys.85.961} {\bibfield  {journal} {\bibinfo
  {journal} {Rev. Mod. Phys.}\ }\textbf {\bibinfo {volume} {85}},\ \bibinfo
  {pages} {961} (\bibinfo {year} {2013})}\BibitemShut {NoStop}%
\bibitem [{\citenamefont {Scappucci}\ \emph {et~al.}(2021)\citenamefont
  {Scappucci}, \citenamefont {Kloeffel}, \citenamefont {Zwanenburg},
  \citenamefont {Loss}, \citenamefont {Myronov}, \citenamefont {Zhang},
  \citenamefont {De~Franceschi}, \citenamefont {Katsaros},\ and\ \citenamefont
  {Veldhorst}}]{scappucci2021}%
  \BibitemOpen
  \bibfield  {author} {\bibinfo {author} {\bibfnamefont {G.}~\bibnamefont
  {Scappucci}}, \bibinfo {author} {\bibfnamefont {C.}~\bibnamefont {Kloeffel}},
  \bibinfo {author} {\bibfnamefont {F.~A.}\ \bibnamefont {Zwanenburg}},
  \bibinfo {author} {\bibfnamefont {D.}~\bibnamefont {Loss}}, \bibinfo {author}
  {\bibfnamefont {M.}~\bibnamefont {Myronov}}, \bibinfo {author} {\bibfnamefont
  {J.-J.}\ \bibnamefont {Zhang}}, \bibinfo {author} {\bibfnamefont
  {S.}~\bibnamefont {De~Franceschi}}, \bibinfo {author} {\bibfnamefont
  {G.}~\bibnamefont {Katsaros}},\ and\ \bibinfo {author} {\bibfnamefont
  {M.}~\bibnamefont {Veldhorst}},\ }\bibfield  {title} {\bibinfo {title} {The
  germanium quantum information route},\ }\href
  {https://doi.org/10.1038/s41578-020-00262-z} {\bibfield  {journal} {\bibinfo
  {journal} {Nature Reviews Materials}\ }\textbf {\bibinfo {volume} {6}},\
  \bibinfo {pages} {926} (\bibinfo {year} {2021})}\BibitemShut {NoStop}%
\bibitem [{\citenamefont {Vandersypen}\ \emph {et~al.}(2017)\citenamefont
  {Vandersypen}, \citenamefont {Bluhm}, \citenamefont {Clarke}, \citenamefont
  {Dzurak}, \citenamefont {Ishihara}, \citenamefont {Morello}, \citenamefont
  {Reilly}, \citenamefont {Schreiber},\ and\ \citenamefont
  {Veldhorst}}]{vandersypen2017}%
  \BibitemOpen
  \bibfield  {author} {\bibinfo {author} {\bibfnamefont {L.~M.~K.}\
  \bibnamefont {Vandersypen}}, \bibinfo {author} {\bibfnamefont
  {H.}~\bibnamefont {Bluhm}}, \bibinfo {author} {\bibfnamefont {J.~S.}\
  \bibnamefont {Clarke}}, \bibinfo {author} {\bibfnamefont {A.~S.}\
  \bibnamefont {Dzurak}}, \bibinfo {author} {\bibfnamefont {R.}~\bibnamefont
  {Ishihara}}, \bibinfo {author} {\bibfnamefont {A.}~\bibnamefont {Morello}},
  \bibinfo {author} {\bibfnamefont {D.~J.}\ \bibnamefont {Reilly}}, \bibinfo
  {author} {\bibfnamefont {L.~R.}\ \bibnamefont {Schreiber}},\ and\ \bibinfo
  {author} {\bibfnamefont {M.}~\bibnamefont {Veldhorst}},\ }\bibfield  {title}
  {\bibinfo {title} {Interfacing spin qubits in quantum dots and donors---hot,
  dense, and coherent},\ }\href {https://doi.org/10.1038/s41534-017-0038-y}
  {\bibfield  {journal} {\bibinfo  {journal} {npj Quantum Information}\
  }\textbf {\bibinfo {volume} {3}},\ \bibinfo {pages} {34} (\bibinfo {year}
  {2017})}\BibitemShut {NoStop}%
\bibitem [{\citenamefont {Camenzind}\ \emph {et~al.}(2022)\citenamefont
  {Camenzind}, \citenamefont {Geyer}, \citenamefont {Fuhrer}, \citenamefont
  {Warburton}, \citenamefont {Zumbühl},\ and\ \citenamefont
  {Kuhlmann}}]{camenzind2022}%
  \BibitemOpen
  \bibfield  {author} {\bibinfo {author} {\bibfnamefont {L.~C.}\ \bibnamefont
  {Camenzind}}, \bibinfo {author} {\bibfnamefont {S.}~\bibnamefont {Geyer}},
  \bibinfo {author} {\bibfnamefont {A.}~\bibnamefont {Fuhrer}}, \bibinfo
  {author} {\bibfnamefont {R.~J.}\ \bibnamefont {Warburton}}, \bibinfo {author}
  {\bibfnamefont {D.~M.}\ \bibnamefont {Zumbühl}},\ and\ \bibinfo {author}
  {\bibfnamefont {A.~V.}\ \bibnamefont {Kuhlmann}},\ }\bibfield  {title}
  {\bibinfo {title} {A hole spin qubit in a fin field-effect transistor above
  4{\thinspace}kelvin},\ }\href {https://doi.org/10.1038/s41928-022-00722-0}
  {\bibfield  {journal} {\bibinfo  {journal} {Nature Electronics}\ }\textbf
  {\bibinfo {volume} {5}},\ \bibinfo {pages} {178} (\bibinfo {year}
  {2022})}\BibitemShut {NoStop}%
\bibitem [{\citenamefont {Xue}\ \emph {et~al.}(2021)\citenamefont {Xue},
  \citenamefont {Patra}, \citenamefont {van Dijk}, \citenamefont {Samkharadze},
  \citenamefont {Subramanian}, \citenamefont {Corna}, \citenamefont
  {Paquelet~Wuetz}, \citenamefont {Jeon}, \citenamefont {Sheikh}, \citenamefont
  {Juarez-Hernandez}, \citenamefont {Esparza}, \citenamefont {Rampurawala},
  \citenamefont {Carlton}, \citenamefont {Ravikumar}, \citenamefont {Nieva},
  \citenamefont {Kim}, \citenamefont {Lee}, \citenamefont {Sammak},
  \citenamefont {Scappucci}, \citenamefont {Veldhorst}, \citenamefont
  {Sebastiano}, \citenamefont {Babaie}, \citenamefont {Pellerano},
  \citenamefont {Charbon},\ and\ \citenamefont {Vandersypen}}]{xue2021}%
  \BibitemOpen
  \bibfield  {author} {\bibinfo {author} {\bibfnamefont {X.}~\bibnamefont
  {Xue}}, \bibinfo {author} {\bibfnamefont {B.}~\bibnamefont {Patra}}, \bibinfo
  {author} {\bibfnamefont {J.~P.~G.}\ \bibnamefont {van Dijk}}, \bibinfo
  {author} {\bibfnamefont {N.}~\bibnamefont {Samkharadze}}, \bibinfo {author}
  {\bibfnamefont {S.}~\bibnamefont {Subramanian}}, \bibinfo {author}
  {\bibfnamefont {A.}~\bibnamefont {Corna}}, \bibinfo {author} {\bibfnamefont
  {B.}~\bibnamefont {Paquelet~Wuetz}}, \bibinfo {author} {\bibfnamefont
  {C.}~\bibnamefont {Jeon}}, \bibinfo {author} {\bibfnamefont {F.}~\bibnamefont
  {Sheikh}}, \bibinfo {author} {\bibfnamefont {E.}~\bibnamefont
  {Juarez-Hernandez}}, \bibinfo {author} {\bibfnamefont {B.~P.}\ \bibnamefont
  {Esparza}}, \bibinfo {author} {\bibfnamefont {H.}~\bibnamefont
  {Rampurawala}}, \bibinfo {author} {\bibfnamefont {B.}~\bibnamefont
  {Carlton}}, \bibinfo {author} {\bibfnamefont {S.}~\bibnamefont {Ravikumar}},
  \bibinfo {author} {\bibfnamefont {C.}~\bibnamefont {Nieva}}, \bibinfo
  {author} {\bibfnamefont {S.}~\bibnamefont {Kim}}, \bibinfo {author}
  {\bibfnamefont {H.-J.}\ \bibnamefont {Lee}}, \bibinfo {author} {\bibfnamefont
  {A.}~\bibnamefont {Sammak}}, \bibinfo {author} {\bibfnamefont
  {G.}~\bibnamefont {Scappucci}}, \bibinfo {author} {\bibfnamefont
  {M.}~\bibnamefont {Veldhorst}}, \bibinfo {author} {\bibfnamefont
  {F.}~\bibnamefont {Sebastiano}}, \bibinfo {author} {\bibfnamefont
  {M.}~\bibnamefont {Babaie}}, \bibinfo {author} {\bibfnamefont
  {S.}~\bibnamefont {Pellerano}}, \bibinfo {author} {\bibfnamefont
  {E.}~\bibnamefont {Charbon}},\ and\ \bibinfo {author} {\bibfnamefont
  {L.~M.~K.}\ \bibnamefont {Vandersypen}},\ }\bibfield  {title} {\bibinfo
  {title} {Cmos-based cryogenic control of silicon quantum circuits},\ }\href
  {https://doi.org/10.1038/s41586-021-03469-4} {\bibfield  {journal} {\bibinfo
  {journal} {Nature}\ }\textbf {\bibinfo {volume} {593}},\ \bibinfo {pages}
  {205} (\bibinfo {year} {2021})}\BibitemShut {NoStop}%
\bibitem [{\citenamefont {Petit}\ \emph {et~al.}(2022)\citenamefont {Petit},
  \citenamefont {Russ}, \citenamefont {Eenink}, \citenamefont {Lawrie},
  \citenamefont {Clarke}, \citenamefont {Vandersypen},\ and\ \citenamefont
  {Veldhorst}}]{petit2020}%
  \BibitemOpen
  \bibfield  {author} {\bibinfo {author} {\bibfnamefont {L.}~\bibnamefont
  {Petit}}, \bibinfo {author} {\bibfnamefont {M.}~\bibnamefont {Russ}},
  \bibinfo {author} {\bibfnamefont {G.~H. G.~J.}\ \bibnamefont {Eenink}},
  \bibinfo {author} {\bibfnamefont {W.~I.~L.}\ \bibnamefont {Lawrie}}, \bibinfo
  {author} {\bibfnamefont {J.~S.}\ \bibnamefont {Clarke}}, \bibinfo {author}
  {\bibfnamefont {L.~M.~K.}\ \bibnamefont {Vandersypen}},\ and\ \bibinfo
  {author} {\bibfnamefont {M.}~\bibnamefont {Veldhorst}},\ }\bibfield  {title}
  {\bibinfo {title} {Design and integration of single-qubit rotations and
  two-qubit gates in silicon above one kelvin},\ }\href
  {https://doi.org/10.1038/s43246-022-00304-9} {\bibfield  {journal} {\bibinfo
  {journal} {Communications Materials}\ }\textbf {\bibinfo {volume} {3}},\
  \bibinfo {pages} {82} (\bibinfo {year} {2022})}\BibitemShut {NoStop}%
\bibitem [{\citenamefont {Petit}\ \emph {et~al.}(2020)\citenamefont {Petit},
  \citenamefont {Eenink}, \citenamefont {Russ}, \citenamefont {Lawrie},
  \citenamefont {Hendrickx}, \citenamefont {Philips}, \citenamefont {Clarke},
  \citenamefont {Vandersypen},\ and\ \citenamefont {Veldhorst}}]{petit2020-2}%
  \BibitemOpen
  \bibfield  {author} {\bibinfo {author} {\bibfnamefont {L.}~\bibnamefont
  {Petit}}, \bibinfo {author} {\bibfnamefont {H.~G.~J.}\ \bibnamefont
  {Eenink}}, \bibinfo {author} {\bibfnamefont {M.}~\bibnamefont {Russ}},
  \bibinfo {author} {\bibfnamefont {W.~I.~L.}\ \bibnamefont {Lawrie}}, \bibinfo
  {author} {\bibfnamefont {N.~W.}\ \bibnamefont {Hendrickx}}, \bibinfo {author}
  {\bibfnamefont {S.~G.~J.}\ \bibnamefont {Philips}}, \bibinfo {author}
  {\bibfnamefont {J.~S.}\ \bibnamefont {Clarke}}, \bibinfo {author}
  {\bibfnamefont {L.~M.~K.}\ \bibnamefont {Vandersypen}},\ and\ \bibinfo
  {author} {\bibfnamefont {M.}~\bibnamefont {Veldhorst}},\ }\bibfield  {title}
  {\bibinfo {title} {Universal quantum logic in hot silicon qubits},\ }\href
  {https://doi.org/10.1038/s41586-020-2170-7} {\bibfield  {journal} {\bibinfo
  {journal} {Nature}\ }\textbf {\bibinfo {volume} {580}},\ \bibinfo {pages}
  {355} (\bibinfo {year} {2020})}\BibitemShut {NoStop}%
\bibitem [{\citenamefont {Van~Dijk}\ \emph {et~al.}(2020)\citenamefont
  {Van~Dijk}, \citenamefont {Patra}, \citenamefont {Subramanian}, \citenamefont
  {Xue}, \citenamefont {Samkharadze}, \citenamefont {Corna}, \citenamefont
  {Jeon}, \citenamefont {Sheikh}, \citenamefont {Juarez-Hernandez},
  \citenamefont {Esparza}, \citenamefont {Rampurawala}, \citenamefont
  {Carlton}, \citenamefont {Ravikumar}, \citenamefont {Nieva}, \citenamefont
  {Kim}, \citenamefont {Lee}, \citenamefont {Sammak}, \citenamefont
  {Scappucci}, \citenamefont {Veldhorst}, \citenamefont {Vandersypen},
  \citenamefont {Charbon}, \citenamefont {Pellerano}, \citenamefont {Babaie},\
  and\ \citenamefont {Sebastiano}}]{vandijk2020}%
  \BibitemOpen
  \bibfield  {author} {\bibinfo {author} {\bibfnamefont {J.~P.~G.}\
  \bibnamefont {Van~Dijk}}, \bibinfo {author} {\bibfnamefont {B.}~\bibnamefont
  {Patra}}, \bibinfo {author} {\bibfnamefont {S.}~\bibnamefont {Subramanian}},
  \bibinfo {author} {\bibfnamefont {X.}~\bibnamefont {Xue}}, \bibinfo {author}
  {\bibfnamefont {N.}~\bibnamefont {Samkharadze}}, \bibinfo {author}
  {\bibfnamefont {A.}~\bibnamefont {Corna}}, \bibinfo {author} {\bibfnamefont
  {C.}~\bibnamefont {Jeon}}, \bibinfo {author} {\bibfnamefont {F.}~\bibnamefont
  {Sheikh}}, \bibinfo {author} {\bibfnamefont {E.}~\bibnamefont
  {Juarez-Hernandez}}, \bibinfo {author} {\bibfnamefont {B.~P.}\ \bibnamefont
  {Esparza}}, \bibinfo {author} {\bibfnamefont {H.}~\bibnamefont
  {Rampurawala}}, \bibinfo {author} {\bibfnamefont {B.~R.}\ \bibnamefont
  {Carlton}}, \bibinfo {author} {\bibfnamefont {S.}~\bibnamefont {Ravikumar}},
  \bibinfo {author} {\bibfnamefont {C.}~\bibnamefont {Nieva}}, \bibinfo
  {author} {\bibfnamefont {S.}~\bibnamefont {Kim}}, \bibinfo {author}
  {\bibfnamefont {H.-J.}\ \bibnamefont {Lee}}, \bibinfo {author} {\bibfnamefont
  {A.}~\bibnamefont {Sammak}}, \bibinfo {author} {\bibfnamefont
  {G.}~\bibnamefont {Scappucci}}, \bibinfo {author} {\bibfnamefont
  {M.}~\bibnamefont {Veldhorst}}, \bibinfo {author} {\bibfnamefont {L.~M.~K.}\
  \bibnamefont {Vandersypen}}, \bibinfo {author} {\bibfnamefont
  {E.}~\bibnamefont {Charbon}}, \bibinfo {author} {\bibfnamefont
  {S.}~\bibnamefont {Pellerano}}, \bibinfo {author} {\bibfnamefont
  {M.}~\bibnamefont {Babaie}},\ and\ \bibinfo {author} {\bibfnamefont
  {F.}~\bibnamefont {Sebastiano}},\ }\bibfield  {title} {\bibinfo {title} {A
  scalable cryo-cmos controller for the wideband frequency-multiplexed control
  of spin qubits and transmons},\ }\href
  {https://doi.org/10.1109/JSSC.2020.3024678} {\bibfield  {journal} {\bibinfo
  {journal} {IEEE Journal of Solid-State Circuits}\ }\textbf {\bibinfo {volume}
  {55}},\ \bibinfo {pages} {2930} (\bibinfo {year} {2020})}\BibitemShut
  {NoStop}%
\bibitem [{\citenamefont {Liles}\ \emph {et~al.}(2021)\citenamefont {Liles},
  \citenamefont {Martins}, \citenamefont {Miserev}, \citenamefont {Kiselev},
  \citenamefont {Thorvaldson}, \citenamefont {Rendell}, \citenamefont {Jin},
  \citenamefont {Hudson}, \citenamefont {Veldhorst}, \citenamefont {Itoh},
  \citenamefont {Sushkov}, \citenamefont {Ladd}, \citenamefont {Dzurak},\ and\
  \citenamefont {Hamilton}}]{liles2021}%
  \BibitemOpen
  \bibfield  {author} {\bibinfo {author} {\bibfnamefont {S.~D.}\ \bibnamefont
  {Liles}}, \bibinfo {author} {\bibfnamefont {F.}~\bibnamefont {Martins}},
  \bibinfo {author} {\bibfnamefont {D.~S.}\ \bibnamefont {Miserev}}, \bibinfo
  {author} {\bibfnamefont {A.~A.}\ \bibnamefont {Kiselev}}, \bibinfo {author}
  {\bibfnamefont {I.~D.}\ \bibnamefont {Thorvaldson}}, \bibinfo {author}
  {\bibfnamefont {M.~J.}\ \bibnamefont {Rendell}}, \bibinfo {author}
  {\bibfnamefont {I.~K.}\ \bibnamefont {Jin}}, \bibinfo {author} {\bibfnamefont
  {F.~E.}\ \bibnamefont {Hudson}}, \bibinfo {author} {\bibfnamefont
  {M.}~\bibnamefont {Veldhorst}}, \bibinfo {author} {\bibfnamefont {K.~M.}\
  \bibnamefont {Itoh}}, \bibinfo {author} {\bibfnamefont {O.~P.}\ \bibnamefont
  {Sushkov}}, \bibinfo {author} {\bibfnamefont {T.~D.}\ \bibnamefont {Ladd}},
  \bibinfo {author} {\bibfnamefont {A.~S.}\ \bibnamefont {Dzurak}},\ and\
  \bibinfo {author} {\bibfnamefont {A.~R.}\ \bibnamefont {Hamilton}},\
  }\bibfield  {title} {\bibinfo {title} {Electrical control of the $g$ tensor
  of the first hole in a silicon mos quantum dot},\ }\href
  {https://doi.org/10.1103/PhysRevB.104.235303} {\bibfield  {journal} {\bibinfo
   {journal} {Phys. Rev. B}\ }\textbf {\bibinfo {volume} {104}},\ \bibinfo
  {pages} {235303} (\bibinfo {year} {2021})}\BibitemShut {NoStop}%
\bibitem [{\citenamefont {Froning}\ \emph {et~al.}(2021)\citenamefont
  {Froning}, \citenamefont {Camenzind}, \citenamefont {van~der Molen},
  \citenamefont {Li}, \citenamefont {Bakkers}, \citenamefont {Zumb{\"u}hl},\
  and\ \citenamefont {Braakman}}]{froning2021}%
  \BibitemOpen
  \bibfield  {author} {\bibinfo {author} {\bibfnamefont {F.~N.~M.}\
  \bibnamefont {Froning}}, \bibinfo {author} {\bibfnamefont {L.~C.}\
  \bibnamefont {Camenzind}}, \bibinfo {author} {\bibfnamefont {O.~A.~H.}\
  \bibnamefont {van~der Molen}}, \bibinfo {author} {\bibfnamefont
  {A.}~\bibnamefont {Li}}, \bibinfo {author} {\bibfnamefont {E.~P. A.~M.}\
  \bibnamefont {Bakkers}}, \bibinfo {author} {\bibfnamefont {D.~M.}\
  \bibnamefont {Zumb{\"u}hl}},\ and\ \bibinfo {author} {\bibfnamefont {F.~R.}\
  \bibnamefont {Braakman}},\ }\bibfield  {title} {\bibinfo {title} {Ultrafast
  hole spin qubit with gate-tunable spin--orbit switch functionality},\ }\href
  {https://doi.org/10.1038/s41565-020-00828-6} {\bibfield  {journal} {\bibinfo
  {journal} {Nature Nanotechnology}\ }\textbf {\bibinfo {volume} {16}},\
  \bibinfo {pages} {308} (\bibinfo {year} {2021})}\BibitemShut {NoStop}%
\bibitem [{\citenamefont {Michal}\ \emph {et~al.}(2021)\citenamefont {Michal},
  \citenamefont {Venitucci},\ and\ \citenamefont {Niquet}}]{michal2021}%
  \BibitemOpen
  \bibfield  {author} {\bibinfo {author} {\bibfnamefont {V.~P.}\ \bibnamefont
  {Michal}}, \bibinfo {author} {\bibfnamefont {B.}~\bibnamefont {Venitucci}},\
  and\ \bibinfo {author} {\bibfnamefont {Y.-M.}\ \bibnamefont {Niquet}},\
  }\bibfield  {title} {\bibinfo {title} {Longitudinal and transverse electric
  field manipulation of hole spin-orbit qubits in one-dimensional channels},\
  }\href {https://doi.org/10.1103/PhysRevB.103.045305} {\bibfield  {journal}
  {\bibinfo  {journal} {Phys. Rev. B}\ }\textbf {\bibinfo {volume} {103}},\
  \bibinfo {pages} {045305} (\bibinfo {year} {2021})}\BibitemShut {NoStop}%
\bibitem [{\citenamefont {Bosco}\ \emph {et~al.}(2021)\citenamefont {Bosco},
  \citenamefont {Het\'enyi},\ and\ \citenamefont {Loss}}]{bosco2021-2}%
  \BibitemOpen
  \bibfield  {author} {\bibinfo {author} {\bibfnamefont {S.}~\bibnamefont
  {Bosco}}, \bibinfo {author} {\bibfnamefont {B.}~\bibnamefont {Het\'enyi}},\
  and\ \bibinfo {author} {\bibfnamefont {D.}~\bibnamefont {Loss}},\ }\bibfield
  {title} {\bibinfo {title} {Hole spin qubits in $\mathrm{Si}$ finfets with
  fully tunable spin-orbit coupling and sweet spots for charge noise},\ }\href
  {https://doi.org/10.1103/PRXQuantum.2.010348} {\bibfield  {journal} {\bibinfo
   {journal} {PRX Quantum}\ }\textbf {\bibinfo {volume} {2}},\ \bibinfo {pages}
  {010348} (\bibinfo {year} {2021})}\BibitemShut {NoStop}%
\bibitem [{\citenamefont {Hosseinkhani}\ and\ \citenamefont
  {Burkard}(2022)}]{hosseinkhani2022}%
  \BibitemOpen
  \bibfield  {author} {\bibinfo {author} {\bibfnamefont {A.}~\bibnamefont
  {Hosseinkhani}}\ and\ \bibinfo {author} {\bibfnamefont {G.}~\bibnamefont
  {Burkard}},\ }\bibfield  {title} {\bibinfo {title} {Theory of silicon spin
  qubit relaxation in a synthetic spin-orbit field},\ }\href
  {https://doi.org/10.1103/PhysRevB.106.075415} {\bibfield  {journal} {\bibinfo
   {journal} {Phys. Rev. B}\ }\textbf {\bibinfo {volume} {106}},\ \bibinfo
  {pages} {075415} (\bibinfo {year} {2022})}\BibitemShut {NoStop}%
\bibitem [{\citenamefont {Hsiao}\ \emph {et~al.}(2020)\citenamefont {Hsiao},
  \citenamefont {van Diepen}, \citenamefont {Mukhopadhyay}, \citenamefont
  {Reichl}, \citenamefont {Wegscheider},\ and\ \citenamefont
  {Vandersypen}}]{hsiao2020}%
  \BibitemOpen
  \bibfield  {author} {\bibinfo {author} {\bibfnamefont {T.-K.}\ \bibnamefont
  {Hsiao}}, \bibinfo {author} {\bibfnamefont {C.}~\bibnamefont {van Diepen}},
  \bibinfo {author} {\bibfnamefont {U.}~\bibnamefont {Mukhopadhyay}}, \bibinfo
  {author} {\bibfnamefont {C.}~\bibnamefont {Reichl}}, \bibinfo {author}
  {\bibfnamefont {W.}~\bibnamefont {Wegscheider}},\ and\ \bibinfo {author}
  {\bibfnamefont {L.}~\bibnamefont {Vandersypen}},\ }\bibfield  {title}
  {\bibinfo {title} {Efficient orthogonal control of tunnel couplings in a
  quantum dot array},\ }\href
  {https://doi.org/10.1103/PhysRevApplied.13.054018} {\bibfield  {journal}
  {\bibinfo  {journal} {Phys. Rev. Appl.}\ }\textbf {\bibinfo {volume} {13}},\
  \bibinfo {pages} {054018} (\bibinfo {year} {2020})}\BibitemShut {NoStop}%
\bibitem [{\citenamefont {van Diepen}\ \emph {et~al.}(2018)\citenamefont {van
  Diepen}, \citenamefont {Eendebak}, \citenamefont {Buijtendorp}, \citenamefont
  {Mukhopadhyay}, \citenamefont {Fujita}, \citenamefont {Reichl}, \citenamefont
  {Wegscheider},\ and\ \citenamefont {Vandersypen}}]{vandiepen2018}%
  \BibitemOpen
  \bibfield  {author} {\bibinfo {author} {\bibfnamefont {C.~J.}\ \bibnamefont
  {van Diepen}}, \bibinfo {author} {\bibfnamefont {P.~T.}\ \bibnamefont
  {Eendebak}}, \bibinfo {author} {\bibfnamefont {B.~T.}\ \bibnamefont
  {Buijtendorp}}, \bibinfo {author} {\bibfnamefont {U.}~\bibnamefont
  {Mukhopadhyay}}, \bibinfo {author} {\bibfnamefont {T.}~\bibnamefont
  {Fujita}}, \bibinfo {author} {\bibfnamefont {C.}~\bibnamefont {Reichl}},
  \bibinfo {author} {\bibfnamefont {W.}~\bibnamefont {Wegscheider}},\ and\
  \bibinfo {author} {\bibfnamefont {L.~M.~K.}\ \bibnamefont {Vandersypen}},\
  }\bibfield  {title} {\bibinfo {title} {Automated tuning of inter-dot tunnel
  coupling in double quantum dots},\ }\href {https://doi.org/10.1063/1.5031034}
  {\bibfield  {journal} {\bibinfo  {journal} {Applied Physics Letters}\
  }\textbf {\bibinfo {volume} {113}},\ \bibinfo {pages} {033101} (\bibinfo
  {year} {2018})}\BibitemShut {NoStop}%
\bibitem [{\citenamefont {Dong}\ and\ \citenamefont
  {Petersen}(2022)}]{dong2022}%
  \BibitemOpen
  \bibfield  {author} {\bibinfo {author} {\bibfnamefont {D.}~\bibnamefont
  {Dong}}\ and\ \bibinfo {author} {\bibfnamefont {I.~R.}\ \bibnamefont
  {Petersen}},\ }\bibfield  {title} {\bibinfo {title} {Quantum estimation,
  control and learning: Opportunities and challenges},\ }\href
  {https://doi.org/https://doi.org/10.1016/j.arcontrol.2022.04.011} {\bibfield
  {journal} {\bibinfo  {journal} {Annual Reviews in Control}\ }\textbf
  {\bibinfo {volume} {54}},\ \bibinfo {pages} {243} (\bibinfo {year}
  {2022})}\BibitemShut {NoStop}%
\bibitem [{\citenamefont {Liu}\ and\ \citenamefont {Yuan}(2017)}]{liu2017}%
  \BibitemOpen
  \bibfield  {author} {\bibinfo {author} {\bibfnamefont {J.}~\bibnamefont
  {Liu}}\ and\ \bibinfo {author} {\bibfnamefont {H.}~\bibnamefont {Yuan}},\
  }\bibfield  {title} {\bibinfo {title} {Quantum parameter estimation with
  optimal control},\ }\href {https://doi.org/10.1103/PhysRevA.96.012117}
  {\bibfield  {journal} {\bibinfo  {journal} {Phys. Rev. A}\ }\textbf {\bibinfo
  {volume} {96}},\ \bibinfo {pages} {012117} (\bibinfo {year}
  {2017})}\BibitemShut {NoStop}%
\bibitem [{\citenamefont {Degen}\ \emph {et~al.}(2017)\citenamefont {Degen},
  \citenamefont {Reinhard},\ and\ \citenamefont {Cappellaro}}]{degen2017}%
  \BibitemOpen
  \bibfield  {author} {\bibinfo {author} {\bibfnamefont {C.~L.}\ \bibnamefont
  {Degen}}, \bibinfo {author} {\bibfnamefont {F.}~\bibnamefont {Reinhard}},\
  and\ \bibinfo {author} {\bibfnamefont {P.}~\bibnamefont {Cappellaro}},\
  }\bibfield  {title} {\bibinfo {title} {Quantum sensing},\ }\href
  {https://doi.org/10.1103/RevModPhys.89.035002} {\bibfield  {journal}
  {\bibinfo  {journal} {Rev. Mod. Phys.}\ }\textbf {\bibinfo {volume} {89}},\
  \bibinfo {pages} {035002} (\bibinfo {year} {2017})}\BibitemShut {NoStop}%
\bibitem [{\citenamefont {Duan}\ \emph {et~al.}(2020)\citenamefont {Duan},
  \citenamefont {Fogarty}, \citenamefont {Williams}, \citenamefont {Hutin},
  \citenamefont {Vinet},\ and\ \citenamefont {Morton}}]{Duan2020}%
  \BibitemOpen
  \bibfield  {author} {\bibinfo {author} {\bibfnamefont {J.}~\bibnamefont
  {Duan}}, \bibinfo {author} {\bibfnamefont {M.~A.}\ \bibnamefont {Fogarty}},
  \bibinfo {author} {\bibfnamefont {J.}~\bibnamefont {Williams}}, \bibinfo
  {author} {\bibfnamefont {L.}~\bibnamefont {Hutin}}, \bibinfo {author}
  {\bibfnamefont {M.}~\bibnamefont {Vinet}},\ and\ \bibinfo {author}
  {\bibfnamefont {J.~J.~L.}\ \bibnamefont {Morton}},\ }\bibfield  {title}
  {\bibinfo {title} {Remote capacitive sensing in two-dimensional quantum-dot
  arrays},\ }\href {https://doi.org/10.1021/acs.nanolett.0c02393} {\bibfield
  {journal} {\bibinfo  {journal} {Nano Letters}\ }\textbf {\bibinfo {volume}
  {20}},\ \bibinfo {pages} {7123} (\bibinfo {year} {2020})}\BibitemShut
  {NoStop}%
\bibitem [{\citenamefont {Patomäki}\ \emph {et~al.}(2023)\citenamefont
  {Patomäki}, \citenamefont {Williams}, \citenamefont {Berritta},
  \citenamefont {Laine}, \citenamefont {Fogarty}, \citenamefont {Leon},
  \citenamefont {Jussot}, \citenamefont {Kubicek}, \citenamefont {Chatterjee},
  \citenamefont {Govoreanu}, \citenamefont {Kuemmeth}, \citenamefont {Morton},\
  and\ \citenamefont {Gonzalez-Zalba}}]{Patomaki}%
  \BibitemOpen
  \bibfield  {author} {\bibinfo {author} {\bibfnamefont {S.~M.}\ \bibnamefont
  {Patomäki}}, \bibinfo {author} {\bibfnamefont {J.}~\bibnamefont {Williams}},
  \bibinfo {author} {\bibfnamefont {F.}~\bibnamefont {Berritta}}, \bibinfo
  {author} {\bibfnamefont {C.}~\bibnamefont {Laine}}, \bibinfo {author}
  {\bibfnamefont {M.~A.}\ \bibnamefont {Fogarty}}, \bibinfo {author}
  {\bibfnamefont {R.~C.~C.}\ \bibnamefont {Leon}}, \bibinfo {author}
  {\bibfnamefont {J.}~\bibnamefont {Jussot}}, \bibinfo {author} {\bibfnamefont
  {S.}~\bibnamefont {Kubicek}}, \bibinfo {author} {\bibfnamefont
  {A.}~\bibnamefont {Chatterjee}}, \bibinfo {author} {\bibfnamefont
  {B.}~\bibnamefont {Govoreanu}}, \bibinfo {author} {\bibfnamefont
  {F.}~\bibnamefont {Kuemmeth}}, \bibinfo {author} {\bibfnamefont {J.~J.~L.}\
  \bibnamefont {Morton}},\ and\ \bibinfo {author} {\bibfnamefont {M.~F.}\
  \bibnamefont {Gonzalez-Zalba}},\ }\bibfield  {title} {\bibinfo {title} {An
  elongated quantum dot as a distributed charge sensor},\ }\href
  {https://arxiv.org/abs/2301.01650} {\bibfield  {journal} {\bibinfo  {journal}
  {arXiv:2301.01650}\ } (\bibinfo {year} {2023})}\BibitemShut {NoStop}%
\bibitem [{\citenamefont {Bellentani}\ \emph {et~al.}(2021)\citenamefont
  {Bellentani}, \citenamefont {Bina}, \citenamefont {Bonen}, \citenamefont
  {Secchi}, \citenamefont {Bertoni}, \citenamefont {Voinigescu}, \citenamefont
  {Padovani}, \citenamefont {Larcher},\ and\ \citenamefont
  {Troiani}}]{Bellentani2021a}%
  \BibitemOpen
  \bibfield  {author} {\bibinfo {author} {\bibfnamefont {L.}~\bibnamefont
  {Bellentani}}, \bibinfo {author} {\bibfnamefont {M.}~\bibnamefont {Bina}},
  \bibinfo {author} {\bibfnamefont {S.}~\bibnamefont {Bonen}}, \bibinfo
  {author} {\bibfnamefont {A.}~\bibnamefont {Secchi}}, \bibinfo {author}
  {\bibfnamefont {A.}~\bibnamefont {Bertoni}}, \bibinfo {author} {\bibfnamefont
  {S.~P.}\ \bibnamefont {Voinigescu}}, \bibinfo {author} {\bibfnamefont
  {A.}~\bibnamefont {Padovani}}, \bibinfo {author} {\bibfnamefont
  {L.}~\bibnamefont {Larcher}},\ and\ \bibinfo {author} {\bibfnamefont
  {F.}~\bibnamefont {Troiani}},\ }\bibfield  {title} {\bibinfo {title} {Toward
  hole-spin qubits in $\mathrm{Si}$ $p$-mosfets within a planar cmos foundry
  technology},\ }\href {https://doi.org/10.1103/PhysRevApplied.16.054034}
  {\bibfield  {journal} {\bibinfo  {journal} {Phys. Rev. Applied}\ }\textbf
  {\bibinfo {volume} {16}},\ \bibinfo {pages} {054034} (\bibinfo {year}
  {2021})}\BibitemShut {NoStop}%
\bibitem [{\citenamefont {Secchi}\ \emph
  {et~al.}(2021{\natexlab{a}})\citenamefont {Secchi}, \citenamefont
  {Bellentani}, \citenamefont {Bertoni},\ and\ \citenamefont
  {Troiani}}]{Secchi2021a}%
  \BibitemOpen
  \bibfield  {author} {\bibinfo {author} {\bibfnamefont {A.}~\bibnamefont
  {Secchi}}, \bibinfo {author} {\bibfnamefont {L.}~\bibnamefont {Bellentani}},
  \bibinfo {author} {\bibfnamefont {A.}~\bibnamefont {Bertoni}},\ and\ \bibinfo
  {author} {\bibfnamefont {F.}~\bibnamefont {Troiani}},\ }\bibfield  {title}
  {\bibinfo {title} {Interacting holes in si and ge double quantum dots: From a
  multiband approach to an effective-spin picture},\ }\href
  {https://doi.org/10.1103/PhysRevB.104.035302} {\bibfield  {journal} {\bibinfo
   {journal} {Phys. Rev. B}\ }\textbf {\bibinfo {volume} {104}},\ \bibinfo
  {pages} {035302} (\bibinfo {year} {2021}{\natexlab{a}})}\BibitemShut
  {NoStop}%
\bibitem [{\citenamefont {Secchi}\ \emph
  {et~al.}(2021{\natexlab{b}})\citenamefont {Secchi}, \citenamefont
  {Bellentani}, \citenamefont {Bertoni},\ and\ \citenamefont
  {Troiani}}]{Secchi2021b}%
  \BibitemOpen
  \bibfield  {author} {\bibinfo {author} {\bibfnamefont {A.}~\bibnamefont
  {Secchi}}, \bibinfo {author} {\bibfnamefont {L.}~\bibnamefont {Bellentani}},
  \bibinfo {author} {\bibfnamefont {A.}~\bibnamefont {Bertoni}},\ and\ \bibinfo
  {author} {\bibfnamefont {F.}~\bibnamefont {Troiani}},\ }\bibfield  {title}
  {\bibinfo {title} {Inter- and intraband coulomb interactions between holes in
  silicon nanostructures},\ }\href
  {https://doi.org/10.1103/PhysRevB.104.205409} {\bibfield  {journal} {\bibinfo
   {journal} {Phys. Rev. B}\ }\textbf {\bibinfo {volume} {104}},\ \bibinfo
  {pages} {205409} (\bibinfo {year} {2021}{\natexlab{b}})}\BibitemShut
  {NoStop}%
\bibitem [{\citenamefont {Venitucci}\ \emph {et~al.}(2018)\citenamefont
  {Venitucci}, \citenamefont {Bourdet}, \citenamefont {Pouzada},\ and\
  \citenamefont {Niquet}}]{Venitucci18}%
  \BibitemOpen
  \bibfield  {author} {\bibinfo {author} {\bibfnamefont {B.}~\bibnamefont
  {Venitucci}}, \bibinfo {author} {\bibfnamefont {L.}~\bibnamefont {Bourdet}},
  \bibinfo {author} {\bibfnamefont {D.}~\bibnamefont {Pouzada}},\ and\ \bibinfo
  {author} {\bibfnamefont {Y.-M.}\ \bibnamefont {Niquet}},\ }\bibfield  {title}
  {\bibinfo {title} {Electrical manipulation of semiconductor spin qubits
  within the $g$-matrix formalism},\ }\href
  {https://doi.org/10.1103/PhysRevB.98.155319} {\bibfield  {journal} {\bibinfo
  {journal} {Phys. Rev. B}\ }\textbf {\bibinfo {volume} {98}},\ \bibinfo
  {pages} {155319} (\bibinfo {year} {2018})}\BibitemShut {NoStop}%
\bibitem [{\citenamefont {Luttinger}\ and\ \citenamefont
  {Kohn}(1955)}]{Luttinger55}%
  \BibitemOpen
  \bibfield  {author} {\bibinfo {author} {\bibfnamefont {J.~M.}\ \bibnamefont
  {Luttinger}}\ and\ \bibinfo {author} {\bibfnamefont {W.}~\bibnamefont
  {Kohn}},\ }\bibfield  {title} {\bibinfo {title} {Motion of electrons and
  holes in perturbed periodic fields},\ }\href
  {https://doi.org/10.1103/PhysRev.97.869} {\bibfield  {journal} {\bibinfo
  {journal} {Phys. Rev.}\ }\textbf {\bibinfo {volume} {97}},\ \bibinfo {pages}
  {869} (\bibinfo {year} {1955})}\BibitemShut {NoStop}%
\bibitem [{\citenamefont {Voon}\ and\ \citenamefont
  {Willatzen}(2009)}]{Voon09}%
  \BibitemOpen
  \bibfield  {author} {\bibinfo {author} {\bibfnamefont {L.~C. L.~Y.}\
  \bibnamefont {Voon}}\ and\ \bibinfo {author} {\bibfnamefont {M.}~\bibnamefont
  {Willatzen}},\ }\href {https://doi.org/10.1007/978-3-540-92872-0} {\emph
  {\bibinfo {title} {The k p Method}}}\ (\bibinfo  {publisher} {Springer
  Berlin},\ \bibinfo {address} {Heidelberg},\ \bibinfo {year}
  {2009})\BibitemShut {NoStop}%
\bibitem [{\citenamefont {Chao}\ and\ \citenamefont {Chuang}(1992)}]{Chao92}%
  \BibitemOpen
  \bibfield  {author} {\bibinfo {author} {\bibfnamefont {C.~Y.-P.}\
  \bibnamefont {Chao}}\ and\ \bibinfo {author} {\bibfnamefont {S.~L.}\
  \bibnamefont {Chuang}},\ }\bibfield  {title} {\bibinfo {title}
  {Spin-orbit-coupling effects on the valence-band structure of strained
  semiconductor quantum wells},\ }\href
  {https://doi.org/10.1103/PhysRevB.46.4110} {\bibfield  {journal} {\bibinfo
  {journal} {Phys. Rev. B}\ }\textbf {\bibinfo {volume} {46}},\ \bibinfo
  {pages} {4110} (\bibinfo {year} {1992})}\BibitemShut {NoStop}%
\bibitem [{\citenamefont {Dunlap}\ and\ \citenamefont
  {Watters}(1953)}]{Dunlap53}%
  \BibitemOpen
  \bibfield  {author} {\bibinfo {author} {\bibfnamefont {W.~C.}\ \bibnamefont
  {Dunlap}}\ and\ \bibinfo {author} {\bibfnamefont {R.~L.}\ \bibnamefont
  {Watters}},\ }\bibfield  {title} {\bibinfo {title} {Direct measurement of the
  dielectric constants of silicon and germanium},\ }\href
  {https://doi.org/10.1103/PhysRev.92.1396} {\bibfield  {journal} {\bibinfo
  {journal} {Phys. Rev.}\ }\textbf {\bibinfo {volume} {92}},\ \bibinfo {pages}
  {1396} (\bibinfo {year} {1953})}\BibitemShut {NoStop}%
\bibitem [{\citenamefont {Hogg}\ \emph {et~al.}(2023)\citenamefont {Hogg},
  \citenamefont {Pakkiam}, \citenamefont {Gorman}, \citenamefont {Timofeev},
  \citenamefont {Chung}, \citenamefont {Gulati}, \citenamefont {House},\ and\
  \citenamefont {Simmons}}]{Hogg23a}%
  \BibitemOpen
  \bibfield  {author} {\bibinfo {author} {\bibfnamefont {M.}~\bibnamefont
  {Hogg}}, \bibinfo {author} {\bibfnamefont {P.}~\bibnamefont {Pakkiam}},
  \bibinfo {author} {\bibfnamefont {S.}~\bibnamefont {Gorman}}, \bibinfo
  {author} {\bibfnamefont {A.}~\bibnamefont {Timofeev}}, \bibinfo {author}
  {\bibfnamefont {Y.}~\bibnamefont {Chung}}, \bibinfo {author} {\bibfnamefont
  {G.}~\bibnamefont {Gulati}}, \bibinfo {author} {\bibfnamefont
  {M.}~\bibnamefont {House}},\ and\ \bibinfo {author} {\bibfnamefont
  {M.}~\bibnamefont {Simmons}},\ }\bibfield  {title} {\bibinfo {title}
  {Single-shot readout of multiple donor electron spins with a gate-based
  sensor},\ }\href {https://doi.org/10.1103/PRXQuantum.4.010319} {\bibfield
  {journal} {\bibinfo  {journal} {PRX Quantum}\ }\textbf {\bibinfo {volume}
  {4}},\ \bibinfo {pages} {010319} (\bibinfo {year} {2023})}\BibitemShut
  {NoStop}%
\bibitem [{\citenamefont {Helstrom}(1969)}]{Helstrom1969}%
  \BibitemOpen
  \bibfield  {author} {\bibinfo {author} {\bibfnamefont {C.~W.}\ \bibnamefont
  {Helstrom}},\ }\bibfield  {title} {\bibinfo {title} {Quantum detection and
  estimation theory},\ }\href {https://doi.org/10.1007/BF01007479} {\bibfield
  {journal} {\bibinfo  {journal} {Journal of Statistical Physics}\ }\textbf
  {\bibinfo {volume} {1}},\ \bibinfo {pages} {231} (\bibinfo {year}
  {1969})}\BibitemShut {NoStop}%
\bibitem [{\citenamefont {Bae}\ and\ \citenamefont {Kwek}(2015)}]{Bae2015a}%
  \BibitemOpen
  \bibfield  {author} {\bibinfo {author} {\bibfnamefont {J.}~\bibnamefont
  {Bae}}\ and\ \bibinfo {author} {\bibfnamefont {L.-C.}\ \bibnamefont {Kwek}},\
  }\bibfield  {title} {\bibinfo {title} {Quantum state discrimination and its
  applications},\ }\href {https://doi.org/10.1088/1751-8113/48/8/083001}
  {\bibfield  {journal} {\bibinfo  {journal} {Journal of Physics A:
  Mathematical and Theoretical}\ }\textbf {\bibinfo {volume} {48}},\ \bibinfo
  {pages} {083001} (\bibinfo {year} {2015})}\BibitemShut {NoStop}%
\bibitem [{\citenamefont {Troiani}\ \emph {et~al.}(2020)\citenamefont
  {Troiani}, \citenamefont {Rotunno}, \citenamefont {Frabboni}, \citenamefont
  {Ravelli}, \citenamefont {Peters}, \citenamefont {Karimi},\ and\
  \citenamefont {Grillo}}]{Troiani20a}%
  \BibitemOpen
  \bibfield  {author} {\bibinfo {author} {\bibfnamefont {F.}~\bibnamefont
  {Troiani}}, \bibinfo {author} {\bibfnamefont {E.}~\bibnamefont {Rotunno}},
  \bibinfo {author} {\bibfnamefont {S.}~\bibnamefont {Frabboni}}, \bibinfo
  {author} {\bibfnamefont {R.~B.~G.}\ \bibnamefont {Ravelli}}, \bibinfo
  {author} {\bibfnamefont {P.~J.}\ \bibnamefont {Peters}}, \bibinfo {author}
  {\bibfnamefont {E.}~\bibnamefont {Karimi}},\ and\ \bibinfo {author}
  {\bibfnamefont {V.}~\bibnamefont {Grillo}},\ }\bibfield  {title} {\bibinfo
  {title} {Efficient molecule discrimination in electron microscopy through an
  optimized orbital angular momentum sorter},\ }\href
  {https://doi.org/10.1103/PhysRevA.102.043510} {\bibfield  {journal} {\bibinfo
   {journal} {Phys. Rev. A}\ }\textbf {\bibinfo {volume} {102}},\ \bibinfo
  {pages} {043510} (\bibinfo {year} {2020})}\BibitemShut {NoStop}%
\bibitem [{\citenamefont {Paris}(2009)}]{Paris09}%
  \BibitemOpen
  \bibfield  {author} {\bibinfo {author} {\bibfnamefont {M.~G.~A.}\
  \bibnamefont {Paris}},\ }\bibfield  {title} {\bibinfo {title} {Quantum
  estimation for quantum technology},\ }\href
  {https://doi.org/10.1142/S0219749909004839} {\bibfield  {journal} {\bibinfo
  {journal} {International Journal of Quantum Information}\ }\textbf {\bibinfo
  {volume} {07}},\ \bibinfo {pages} {125} (\bibinfo {year} {2009})}\BibitemShut
  {NoStop}%
\bibitem [{\citenamefont {Nielsen}\ and\ \citenamefont
  {Chuang}(2016)}]{nielsen}%
  \BibitemOpen
  \bibfield  {author} {\bibinfo {author} {\bibfnamefont {M.~A.}\ \bibnamefont
  {Nielsen}}\ and\ \bibinfo {author} {\bibfnamefont {I.~L.}\ \bibnamefont
  {Chuang}},\ }\href
  {https://www.cambridge.org/de/academic/subjects/physics/quantum-physics-quantum-information-and-quantum-computation/quantum-computation-and-quantum-information-10th-anniversary-edition?format=HB}
  {\emph {\bibinfo {title} {Quantum Computation and Quantum Information}}}\
  (\bibinfo  {publisher} {Cambridge University Press},\ \bibinfo {year}
  {2016})\BibitemShut {NoStop}%
\bibitem [{\citenamefont {Connors}\ \emph {et~al.}(2022)\citenamefont
  {Connors}, \citenamefont {Nelson}, \citenamefont {Edge},\ and\ \citenamefont
  {Nichol}}]{Connors22a}%
  \BibitemOpen
  \bibfield  {author} {\bibinfo {author} {\bibfnamefont {E.~J.}\ \bibnamefont
  {Connors}}, \bibinfo {author} {\bibfnamefont {J.}~\bibnamefont {Nelson}},
  \bibinfo {author} {\bibfnamefont {L.~F.}\ \bibnamefont {Edge}},\ and\
  \bibinfo {author} {\bibfnamefont {J.~M.}\ \bibnamefont {Nichol}},\ }\bibfield
   {title} {\bibinfo {title} {Charge-noise spectroscopy of si/sige quantum dots
  via dynamically-decoupled exchange oscillations},\ }\href
  {https://doi.org/10.1038/s41467-022-28519-x} {\bibfield  {journal} {\bibinfo
  {journal} {Nature Communications}\ }\textbf {\bibinfo {volume} {13}},\
  \bibinfo {pages} {940} (\bibinfo {year} {2022})}\BibitemShut {NoStop}%
\bibitem [{\citenamefont {Sakurai}(1994)}]{Sakurai}%
  \BibitemOpen
  \bibfield  {author} {\bibinfo {author} {\bibfnamefont {J.~J.}\ \bibnamefont
  {Sakurai}},\ }\href {https://cds.cern.ch/record/1167961} {\emph {\bibinfo
  {title} {{Modern quantum mechanics}}}}\ (\bibinfo  {publisher}
  {Addison-Wesley},\ \bibinfo {address} {Reading, MA},\ \bibinfo {year}
  {1994})\BibitemShut {NoStop}%
\bibitem [{\citenamefont {Ramsey}(1950)}]{ramsey1950}%
  \BibitemOpen
  \bibfield  {author} {\bibinfo {author} {\bibfnamefont {N.~F.}\ \bibnamefont
  {Ramsey}},\ }\bibfield  {title} {\bibinfo {title} {A molecular beam resonance
  method with separated oscillating fields},\ }\href
  {https://doi.org/10.1103/PhysRev.78.695} {\bibfield  {journal} {\bibinfo
  {journal} {Phys. Rev.}\ }\textbf {\bibinfo {volume} {78}},\ \bibinfo {pages}
  {695} (\bibinfo {year} {1950})}\BibitemShut {NoStop}%
\bibitem [{\citenamefont {Wang}\ \emph {et~al.}(2022)\citenamefont {Wang},
  \citenamefont {Bushmakin}, \citenamefont {Stein}, \citenamefont {Schell},\
  and\ \citenamefont {Gerhardt}}]{wang2022}%
  \BibitemOpen
  \bibfield  {author} {\bibinfo {author} {\bibfnamefont {Y.}~\bibnamefont
  {Wang}}, \bibinfo {author} {\bibfnamefont {V.}~\bibnamefont {Bushmakin}},
  \bibinfo {author} {\bibfnamefont {G.~A.}\ \bibnamefont {Stein}}, \bibinfo
  {author} {\bibfnamefont {A.~W.}\ \bibnamefont {Schell}},\ and\ \bibinfo
  {author} {\bibfnamefont {I.}~\bibnamefont {Gerhardt}},\ }\bibfield  {title}
  {\bibinfo {title} {Optical ramsey spectroscopy on a single molecule},\ }\href
  {https://doi.org/10.1364/OPTICA.443727} {\bibfield  {journal} {\bibinfo
  {journal} {Optica}\ }\textbf {\bibinfo {volume} {9}},\ \bibinfo {pages} {374}
  (\bibinfo {year} {2022})}\BibitemShut {NoStop}%
\bibitem [{\citenamefont {Singh}\ and\ \citenamefont {Singh}(1989)}]{Singh89}%
  \BibitemOpen
  \bibfield  {author} {\bibinfo {author} {\bibfnamefont {S.~B.}\ \bibnamefont
  {Singh}}\ and\ \bibinfo {author} {\bibfnamefont {C.~A.}\ \bibnamefont
  {Singh}},\ }\bibfield  {title} {\bibinfo {title} {Extensions of the
  feynman–hellman theorem and applications},\ }\href
  {https://doi.org/10.1119/1.15842} {\bibfield  {journal} {\bibinfo  {journal}
  {American Journal of Physics}\ }\textbf {\bibinfo {volume} {57}},\ \bibinfo
  {pages} {894} (\bibinfo {year} {1989})}\BibitemShut {NoStop}%
\end{thebibliography}%

\end{document}